\documentclass{aa}
\usepackage[dvips]{rotating}
\usepackage{amsmath}
\usepackage{amssymb}
\usepackage{txfonts}
\usepackage{multirow}
\usepackage{longtable}
\usepackage{graphicx}
\usepackage{epsfig}
\usepackage{url}
\usepackage{xspace}

\usepackage[round]{natbib}
\bibpunct{(}{)}{;}{a}{}{,}
\usepackage{aas_macros} 

\newcommand{\survey}{\textit{400d}}
\newcommand{\megacam}{\textsc{Megacam}}
\newcommand{\theli}{\texttt{THELI}}
\newcommand{\tr}{\operatorname{tr}}
\newcommand{\median}{\operatorname{median}}
\def \cl00 {CL0030+2618}

\begin{document}

\title{The \textit{400d} Galaxy Cluster Survey Weak Lensing Programme:\\
I: MMT/Megacam Analysis of CL0030+2618 at $z\!=\!0.50$\thanks{
Observations reported here were obtained at the MMT Observatory, a joint facility of the Smithsonian Institution and the University of Arizona.
Our MMT observations were supported in part by a donation from the
F.~H. Levinson Fund of the Peninsula Community Foundation to the 
University of Virginia. In addition, MMT observations used for this project 
were granted by the Smithsonian Astrophysical Observatory and by NOAO,
through the Telescope System Instrumentation Program (TSIP). TSIP is
funded by NSF.}}
\titlerunning{The \textit{400d} Weak Lensing Survey I}

\author{Holger Israel\inst{1} \and Thomas Erben\inst{1} \and Thomas H. Reiprich\inst{1}
   \and Alexey Vikhlinin\inst{2}
   \and Hendrik Hildebrandt\inst{3}
   \and Daniel S. Hudson\inst{1}
   \and Brian A. McLeod\inst{2}
   \and Craig L. Sarazin\inst{4}
   \and Peter Schneider\inst{1}
   \and Yu-Ying Zhang\inst{1}
    }
\institute{Argelander-Institut f\"ur Astronomie, Auf dem H\"ugel 71, 53121 Bonn,
Germany
 \and Harvard-Smithsonian Center for Astrophysics, 60 Garden Street, Cambridge, MA 02138, USA
 \and Leiden Observatory, Leiden University, Niels Bohrweg 2, 2333 CA Leiden, The Netherlands
 \and Department of Astronomy, University of Virginia, 530 McCormick Road, Charlottesville, VA 22904, USA}

\date{Received / Accepted }

\abstract{Cosmological structure formation offers insights on all of the 
universe's components: baryonic matter, dark matter, and, notably, dark energy.
The mass function of galaxy clusters at high redshifts is a particularly
useful probe to learn about the history of structure formation and constrain
cosmological parameters.}
{We aim at deriving reliable masses for a high-redshift, high-luminosity sample
of clusters of galaxies selected from the \survey ~survey of X-ray selected
clusters. 
Weak gravitational lensing allows a comparison of the mass estimates derived
from X-rays, forming an independent test.  
Here, we will focus on a particular object, \cl00 ~at $z\!=\!0.50$.}
{Using deep imaging in three passbands with the \megacam ~instrument at MMT,
we show that \megacam ~is well-suited for measuring gravitational shear, i.e.\
the shapes of faint galaxies. A catalogue of background galaxies is constructed
by analysing photometric properties of galaxies in the $g'r'i'$ bands.}
{We detect the weak lensing signal of \cl00 ~at $5.8\sigma$ significance,
using the aperture mass technique. Furthermore, we find 
significant tangential alignment of
galaxies out to $\sim\!10\arcmin$ or $>\!2~r_{200}$ 
distance from the cluster centre. The weak
lensing centre of \cl00 ~agrees with several X-ray measurements and the 
position of the brightest cluster galaxy. Finally, we infer a weak lensing
virial mass of 
$M_{200}\!=\!7.2^{+3.6+2.3}_{-2.9-2.5}\times 10^{14}\mathrm{M}_{\sun}$ for \cl00 .}
{Despite complications by a tentative foreground galaxy group in the line of
sight, the X-ray and weak lensing estimates for \cl00 ~are in remarkable
agreement. 
}


\maketitle


\section{Introduction} \label{sec:intro}
The mass function $n(M,z)$
of galaxy clusters is a sensitive probe to both cosmic
expansion and evolution of structure by gravitational collapse
\citep[cf.\ e.g.,][]{2002ARA&A..40..539R,2005RvMP...77..207V,2005RvMA...18...76S}.
Therefore, mass functions derived from statistically well-understood 
cluster samples can be and are frequently used to determine cosmological
parameters like $\Omega_{\mathrm{m}}$, the total matter density of the universe
in terms of the critical density, and $\sigma_{8}$, the dispersion of the matter
density contrast.
In addition, measurements of the mass function 
at different redshifts have the potential to
constrain the possible evolution of the dark energy component of the universe
\citep{2006ewg3.rept.....P,2006astro.ph..9591A},
expressed by the value and 
change with time of the equation-of-state parameter 

Because the abundance and mass function of clusters are sensitive functions
of these cosmological parameters, they have been studied intensively both
theoretically 
\citep{1974ApJ...187..425P,1999MNRAS.308..119S,2001MNRAS.321..372J,2008ApJ...688..709T} and observationally.

Several methods to measure cluster masses have been employed to
determine their mass function. Assuming hydrostatic equilibrium of the
intracluster medium (ICM), the mass of a cluster can be computed, once its 
X-ray gas density and temperature profiles are known.
If the quality of the X-ray data does not allow the determination of profiles
for individual clusters, for instance at high redshift, X-ray
\emph{scaling relations} of the X-ray luminosity 
\citep[$L_{\mathrm{X}}$--$M_{\mathrm{tot}}$; e.g.,][]{2002ApJ...567..716R,2008MNRAS.387.1179M}, temperature ($T_{\mathrm{X}}$--$M_{\mathrm{tot}}$),
gas mass ($M_{\mathrm{gas}}$--$M_{\mathrm{tot}}$) or 
$Y_{\mathrm{X}}\!=\!T_{\mathrm{X}}M_{\mathrm{gas}}$
with the total mass are used as proxies \citep{2009ApJ...692.1033V}.
Simultaneously constraining cosmological parameters and X-ray cluster scaling
relations, \citet{2009arXiv0909.3098M} found 
$w_{\mathrm{DE}}\!=\!-1.01\!\pm\!0.20$ for the dark energy equation-of-state
parameter, compiling data from a large sample of galaxy clusters.

Weak gravitational lensing provides a completely independent probe of a
cluster's mass, as it is sensitive to baryonic and dark matter alike,
not relying on assumptions of the thermodynamic state of the gas. 
The fact that sources of systematic errors in the lensing and X-ray
methods are unrelated opens the possibility to compare and cross-calibrate 
X-ray and weak lensing masses.

Several studies have been recently undertaken, comparing X-ray and weak lensing
cluster observables: 
\citet{2006ApJ...653..954D} found the weak lensing mass to scale with X-ray
luminosity like $\propto~L_{\mathrm{X}}^{1.04\pm0.46}$ and constrain a
combination of $\Omega_{\mathrm{m}}$ and $\sigma_{8}$.
\citet{2007MNRAS.379..317H} established a proportionality between the weak
lensing mass $M_{\mathrm{wl}}$ within the radius $r_{2500}$
inside which the density 
exceeds the critical density by a factor $2500$ and $T_{\mathrm{X}}^{\alpha}$, 
with an exponent $\alpha\!=\!1.34^{+0.30}_{-0.28}$.
For the same radius, \citet{2008MNRAS.384.1567M} quoted a ratio
$M_{\mathrm{X}}/M_{\mathrm{wl}}\!=\!1.03\pm0.07$ along with 
a decrease towards smaller radii. 
Directly aiming at the ratio of weak lensing to X-ray mass, 
\citet{2008A&A...482..451Z} measured at a radius of 
$M_{\mathrm{wl}}/M_{\mathrm{X}}\!=\!1.09\!\pm\!0.08$, for a radius $r_{500}$,
and later confirmed this value and the trend with radius \citep{2010arXiv1001.0780Z}. 
\citet{2009MNRAS.396..315C} investigated the role of the mass estimator on 
the systematics of the weak lensing mass function.


In order to 
make progress, it is particularly important to determine the masses
of more clusters to a high accuracy, especially at high 
($z\!>\!0.3$) redshifts.
To date, only a few studies have been undertaken in this regime which provides
the strongest leverage on structure formation and is thus crucial for tackling
the problem of dark energy.

In the redshift range $0.3\!\lesssim\!z\!\lesssim\!0.8$,
for two reasons, weak gravitational lensing becomes more
advantageous over methods relying on the X-ray emission from the ICM with
increasing redshift.
First, X-ray temperature profiles are becoming increasingly difficult to
determine with cluster redshift.
Second, the fraction of clusters undergoing a merger increases with $z$, in
accordance with the paradigm of hierarchical structure formation
\citep[e.g.][]{2005APh....24..316C}, rendering the assumption of hydrostatic
equilibrium more problematic at higher $z$. 

With this paper, we report the first results of the largest weak lensing 
follow-up of an X-ray selected cluster sample at high redshifts.

\section{Observations}

\subsection{The 400d Survey and Cosmological Sample}

The 400 Square Degree Galaxy Cluster Survey 
(abbreviated as \survey) comprises all clusters of galaxies detected
serendipitously in an analysis of (nearly) all suitable \textit{ROSAT} PSPC 
pointings \citep{2007ApJS..172..561B}. 

The survey's name derives from the total area of $397\degr^{2}$ on the sky
covered by these pointings.
The sample is flux-limited, using a threshold of 
$1.4\times 10^{-13}\,\mbox{erg s}^{-1}\,\mbox{cm}^{-2}$ in the 
$0.5$--$2.0~\mbox{keV}$ band. 

The analysis of the \textit{ROSAT} data is described in detail in 
\citet{2007ApJS..172..561B}.
All clusters in the \survey~sample are confirmed by the identification of
galaxy overdensities in optical images. Their redshifts have been determined by
optical spectroscopy of sample galaxies.

The final \survey~catalogue contains 242 objects in the redshift range
$0.0032<z<0.888$. 
In order to be able to accurately constrain the mass function of galaxy clusters
at $z\!\approx\!0.5$, specifications of the \survey ~sample were devised such
that the cluster catalogue provides a sample of 
``typical'' clusters in the $0.3<z<0.8$ range.

This work is based on the \emph{\survey ~Cosmological Sample},
a carefully selected subsample of 
high-redshift and X-ray luminous \survey~clusters. 
It has been defined and published in \citet[Table~1]{2009ApJ...692.1033V}
and comprises 36 clusters. 
This cosmological or high-redshift sample is drawn from the \survey~catalogue by
selecting all clusters both at a redshift $z\!\geq\!0.350$, as given by 
\citet{2007ApJS..172..561B}, and with a \textit{ROSAT} luminosity exceeding
\begin{equation} \label{eq:lxmin}
L_{\mathrm{X,min}}=4.8\times 10^{43} (1+z)^{1.8}\,\mbox{erg s}^{-1}\quad .
\end{equation}
\citet{2007ApJS..172..561B} assume a $\Lambda$CDM cosmology with Hubble parameter $h\!=\!0.72$ and density
parameters $\Omega_{\mathrm{m}}\!=\!0.30$ and $\Omega_{\Lambda}\!=\!0.70$ 
for matter and dark energy, respectively. For consistency, the same
cosmology is adopted throughout this paper.

The X-ray luminosity limit in Eq.\ (\ref{eq:lxmin}) has been chosen in such 
a way that it, given an 
$L_{\mathrm{X}}$-$M_{\mathrm{X}}$-scaling relation valid for low-$z$ clusters, 
corresponds to a mass of  $\approx 10^{14} \mathrm{M}_{\sun}$.
This renders our cosmological sample nearly mass-limited.

All clusters in the cosmological sample have been reobserved with 
\textsc{Chandra} within the framework of the \textit{Chandra Cluster Cosmology
Project} (CCCP) and constraints on cosmological parameters derived from these
X-ray data have been published in 
\citet{2009ApJ...692.1033V,2009ApJ...692.1060V}.

With this paper, we report first results of a weak lensing follow-up survey 
of the galaxy clusters in the \survey~cosmological sample. 
%
Here, we focus on one particular object, \cl00 ,
which, as we will see below, represents an exceptionally interesting case. 
Also, we demonstrate in detail the methods we use for data reduction and
analysis as this is the first weak lensing study using \megacam ~at MMT.

The cluster \cl00 ~is listed at a redshift of $z\!=\!0.500$ 
both in \citet{2007ApJS..172..561B}
(designation BVH~002) as well as in the precursor to the \survey , termed the
\textit{160d survey} \citep{1998ApJ...502..558V} as VMF~001. 
It was first identified as a cluster of galaxies by \citet{1997MNRAS.285..511B}
conducting a spectroscopic follow-up to \textsc{Rosat} observations in the visual
wavelength range. These authors assigned
the designation CRSS J0030.5+2618 and measured a redshift of $z\!=\!0.516$.
\citet{2000AJ....119.2349B} observed the field of \cl00 ~with \textsc{Chandra}
during its calibration phase, studying faint hard X-ray sources in the
vicinity of the cluster.
\citet{2008ApJS..176..374H} confirm the redshift of $z\!=\!0.500$ for the
cluster with their designation WARP J0030.5+2618 in their X-ray selected
survey of \textsl{ROSAT} clusters, but point out a possible
contamination of the X-ray signal by a line-of-sight structure at the lower
redshift of $z\!\approx\!0.27$.

Additional \textsc{Chandra} observations were conducted as part of the CCCP
\citep{2009ApJ...692.1033V,2009ApJ...692.1060V}.
Its X-ray emission as detected by \textsc{Rosat} is centred at 
$\alpha_{\mathrm{J}2000}=00^{\mathrm{h}}30^{\mathrm{m}}33^{\mathrm{s}}\!\!.6$, 
$\delta_{\mathrm{J}2000}=+26^{\circ}18'16''$. The analysis of \textsc{Chandra}
data by \citet{2009ApJ...692.1033V}
results in a luminosity in the $0.5\ldots 2.0~\text{keV}$-band of
$L_{\mathrm{X}}\!=\!1.57\times 10^{44}\,\mbox{erg s}^{-1}$ 
and a ICM temperature of
$k_{\mathrm{B}}T_{\mathrm{X}}\!=\!(5.63\pm 1.13)\,\mbox{keV}\,$. \cl00 ~has been
classified as a possibly merging by \citet{2009ApJ...692.1033V} based on its 
X-ray morphology.

As \cl00 ~has not been studied with deep optical imaging before, and there are 
no observations with large optical telescopes in the major public archives,
we present the first such study of this cluster.
(The SEGUE observations used in Sect.~\ref{sec:ccc} 
for cross-calibration have some overlap with our
\megacam ~imaging south of \cl00 , but do not contain the object itself.)
%

\subsection{The \megacam ~Instrument at MMT } \label{sec:mmt}

The observations discussed in this work have been obtained using the \megacam
~36-chip camera at the 6.5~m MMT telescope, located at Fred Lawrence Whipple
Observatory on Mt. Hopkins, Arizona 
\citep{2000fdso.conf...11M,2006sda..conf..337M}.
\megacam ~is a wide-field imaging instrument with a field-of-view of 
$\sim\!24\arcmin\times 24\arcmin$, resulting from a mosaic of $4\times 9$ CCDs, 
each consisting of $2048\times 4608$ pixels which corresponds to
a very small pixel size of $0.08\arcsec\,\mbox{px}^{-1}$.
Each chip has two read-out circuits and amplifiers, each reading out half a
chip (cf. Sect.~\ref{sec:amp} in the Appendix). 
The gaps between the chips measure $6\arcsec$ in the direction corresponding
to declination using the default derotation and $33\arcsec$, $5\arcsec$, and
$33\arcsec$ in the direction associated with right ascension.
We use \megacam ~in the default $2\times 2$ binning mode.

A system of $u'g'r'i'z'$ filters, similar to but subtly different from
their namesakes in the Sloan Digital Sky Survey \citep{1996AJ....111.1748F}
is used
for \megacam. The relations between the \megacam ~and SDSS filter systems are
described in detail in Sect.~\ref{sec:pcrel} and visualised in 
Fig.~\ref{fig:filters}.

None of the previous studies with \megacam ~\citep[e.g.:][]{2008ApJ...675.1233H,2008ApJ...688..245W}
is related to gravitational lensing. Thus, in this paper 
we will show that \megacam ~indeed is suitable for weak lensing work.

\subsection{Observing Strategy}

\begin{figure*}
\includegraphics[width=0.85\textwidth]{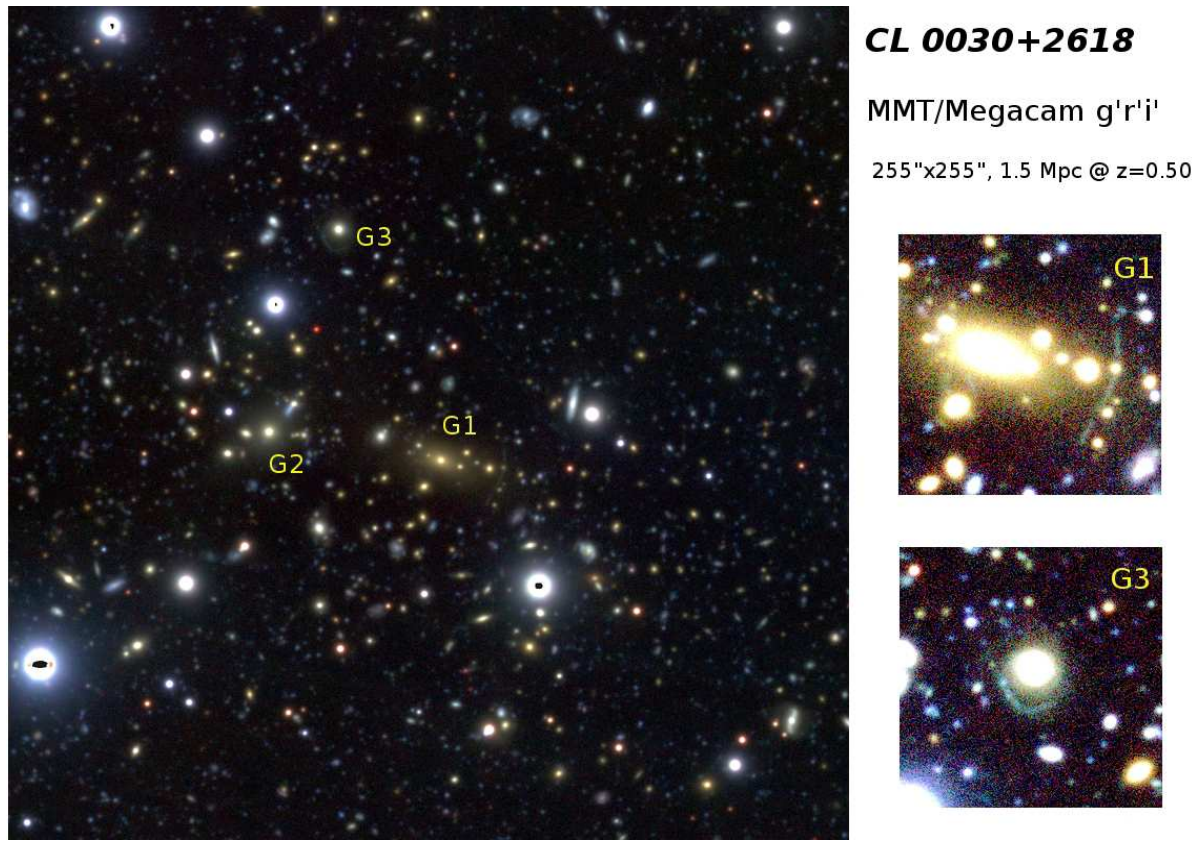}
 \caption{Three-colour composite of \cl00 , prepared from the \megacam ~$g'r'i'$ coadded images.
          The main image shows a cut-out of the central region of \cl00 , with an edge length of
          $\approx\!255\arcsec$ ($1600\,\mbox{px}$), corresponding to the virial radius of 
          $\approx\!1.5\,\mbox{Mpc}$ at the cluster redshift of $z\!=\!0.50$.
          North is up and East is to the left.
          The tentative luminous arcs near the galaxies G1 and G3 (Table~\ref{tab:gal} and 
          Sect.~\ref{sec:arcs}) are emphasized in the two smaller, zoomed 
          ($40\arcsec\times40\arcsec$) images.}
 \label{fig:img}
\end{figure*}

In principle, the small distortions of background sources which we want to
measure are achromatic. In practice, however, the optimal passband for weak
lensing observations is determined by the signal-to-noise ratio which can be
obtained in a given amount of time and depends on seeing and instrumental
throughput.
To maximise the number of high signal-to-noise, background galaxies whose
shapes can be determined reliably for a given exposure time, we choose
the $r'$-band as the default lensing band. Aiming at a limiting magnitude of 
$r'_{\mathrm{lim}}\!\approx\!26$ for $T_{\mathrm{exp}}\!\approx\!3\,\mbox{h}$,
we obtain a sufficient number of high-quality shape sources
($n_{\mathrm{gal}}\!\gtrsim\!15\,\mbox{arcmin}^{-2}$) in the final 
catalogue.\footnote{Eq.~(\ref{eq:maglim}) gives our definition of ``limiting magnitude'',
while the actual values for \cl00 ~are listed in Table~\ref{tab:data}.}

Lensing effects depend on the relative distances between source and deflector
\citep[Chap.\ 4.3]{2001PhR...340..291B}.
Ideally, we would like to determine a \emph{photometric redshift}
estimate for each galaxy in our lensing catalogue
\citep[e.g.:][]{2000ApJ...536..571B,2000A&A...363..476B,2001A&A...365..660W,2006A&A...457..841I,2008A&A...480..703H}.
However, this is observationally expensive as deep imaging in 
$\gtrsim\! 5$ passbands is necessary to obtain accurate photometric redshifts.

On the other hand, using only one filter (the lensing band) and a simple 
magnitude cut 
for a rough separation of background from foreground galaxies needs
a minimum of observing time but neglects the galaxies' intrinsic distribution in
magnitude. 
We are following an intermediary approach here, using three filters from which
we construct colour-colour-diagrammes of the detected galaxies and use this 
information to achieve a more accurate background selection than using the
simplistic magnitude cut. This method has successfully been applied to
weak-lensing galaxy cluster data by e.g.\
\citet{2002A&A...395..385C,2005A&A...437...49B,2007A&A...471...31K}.
\megacam 's $g'$ and $i'$ passbands 
straddle the Balmer break, the most distinctive feature in an elliptical
galaxy's optical spectrum in the redshift range $z\!\approx\!0.5$
in which we are interested. Therefore, we use the $g'r'i'$ filters and
resulting colours to identify foreground and cluster objects in our catalogues.

In order 
to obtain a high level of homogeneity in data quality over the field-of-view
despite the gaps between \megacam's chips, we stack dithered exposures.
Our dither pattern consists of $5\!\times\!5$ positions in a square array with
$40\arcsec$ distance between neighbouring points, inclined by $10\degr$ with
respect to the right ascension axis on which the chips normally are aligned. 
We find this pattern to be robust against missing frames (exposures
which couldn't be used in the final stack for whatever reason). 

\subsection{The Data} \label{sec:thedata}

The data presented in this paper have been collected in five nights distributed
over two observing runs on October 6th and 7th, 2004 and October 30th, 
October 31st, and November 1st of 2005.
In these two observing runs, a total of four \survey ~Cosmological Sample 
clusters have been
observed. In the first phase of data reduction, the so-called 
\emph{run processing}, these data have been processed in a consistent fashion.
A weak lensing analysis of the three clusters besides \cl00 , namely
CL0159+0030, CL0230+1836, and CL0809+2811, 
will be the topic of an upcoming paper in this series.
In this work, we make use of part of these data for the photometric calibration
of \cl00 ~images (see Sect.~\ref{sec:filters}).

In Table~\ref{tab:data}, we give the exposure times, seeings, and
related information for the final image stacks on which we base all
further analysis.
Fig.~\ref{fig:img} shows a three-colour composite image prepared from
the stacked \megacam ~$g'r'i'$ observations of \cl00 .

\section{Outline of Data Reduction} \label{sec:datared}

\begin{figure*}
\includegraphics[width=\textwidth]{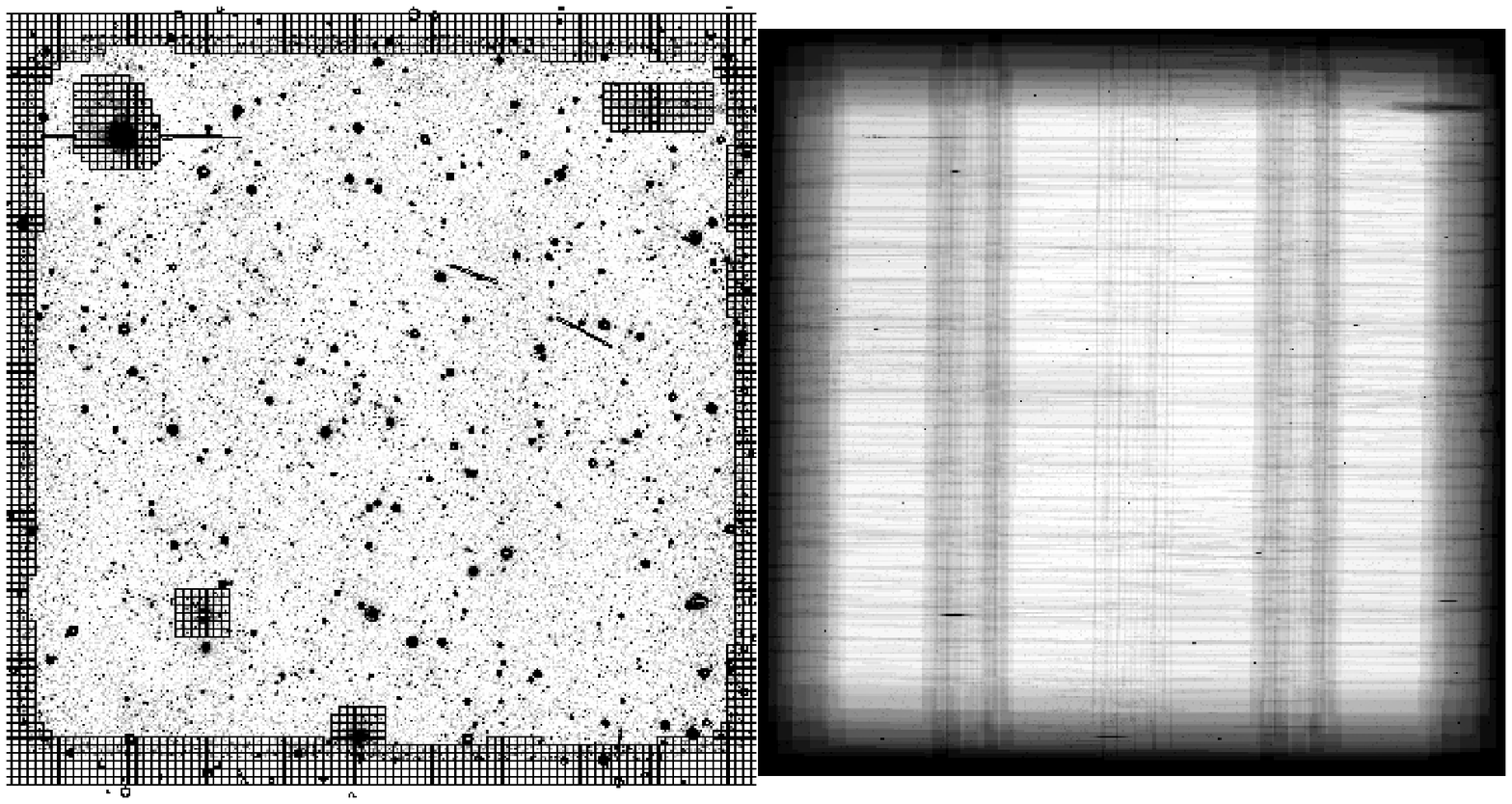}
 \caption{Left: The coadded $r'$-band image of \cl00 , overlaid with
   its final masks. The target cluster is located in the frame
   centre. Small square masks cover regions masked because of their 
   source counts strongly deviating from the average in the field 
   (Sect.~\ref{sec:post}).
   The elongated masks enclose tracks
   of slowly moving objects (asteroids) which had to be identified on
   the image by visual inspection.  The small octagonal masks are
   saturated stars found from the \texttt{USNO B1} catalogue and
   manually.  Right: The $r'$-band weight image of \cl00 . 
   Pixels lying in the interior of the chips get significantly higher weights
   than those which fall into an intra-chip gap in some of the
   dithered exposures.}
\label{fig:coadd}
\end{figure*}
\begin{table*}
 \centering
 \caption{Specifications of the coadded images for \cl00 }
 \label{tab:data}
 \begin{center} 
 \begin{tabular}{cccccc} \hline\hline
 Filter & Observation Dates & $T_{\mathrm{exp}}^{\mathrm{fin}} [s]$& Seeing &
 Calib. Method & $m_{\mathrm{lim}}$ \\ \hline
 r' & 2004-10-06/7 & 6600 & 0.82
& stellar colours & 25.9\\
 g' & 2005-10-30/1, 2005-11-01 &  7950 & 0.87 & SDSS standards & 26.8\\
 i' & 2005-10-31 & 5700 & 1.03 & SDSS standards & 25.1\\ \hline\hline
 \end{tabular}
 \end{center}
\end{table*}
%
%
The reduction of the data presented in this paper relies on the THELI
pipeline originally 
designed and tested on
observations obtained using the Wide-Field Imager (WFI) on ESO's La~Silla 2.2~m
telescope \citep{2005AN....326..432E}.
While the reduction follows the procedure detailed in
\citet{2005AN....326..432E}, some important changes had to be made to
adapt the THELI pipeline to work on MMT \megacam\ data. In the
following discussion of the subsequent individual reduction routines,
special emphasis has been given to those adaptations arising from
the fact that MMT \megacam ~is a ``new'' camera with a small field-of-view per 
chip ($325\arcsec\!\times\!164\arcsec$ instead of 
$853\arcsec\!\times379\!\arcsec$
for MegaPrime at CFHT, or a factor $1/6$ in field-of-view), using a larger
telescope.

The THELI pipeline distinguishes two stages of data reduction: run processing
and set processing. During \textit{run processing}, the first phase, all frames
taken during an observation run in a particular filter are treated in the same
way. Run processing comprises the removal of instrumental signatures, e.g.\
de-biasing and flatfielding.
In \textit{set processing} the data are re-ordered according to their celestial
coordinates rather than their date of observation. 
Astrometric and photometric calibration lead to a resulting ``coadded''
(stacked) image for each set.
In Fig.~\ref{fig:coadd}, we present the coadded $r'$-band image of \cl00 , 
overlaid with its final masks in the left and its weight image in
the right part of the plot.

\subsection{Coaddition ``Post Production''} \label{sec:post}

\begin{figure*}
 \sidecaption
 \includegraphics[width=12cm,angle=90]{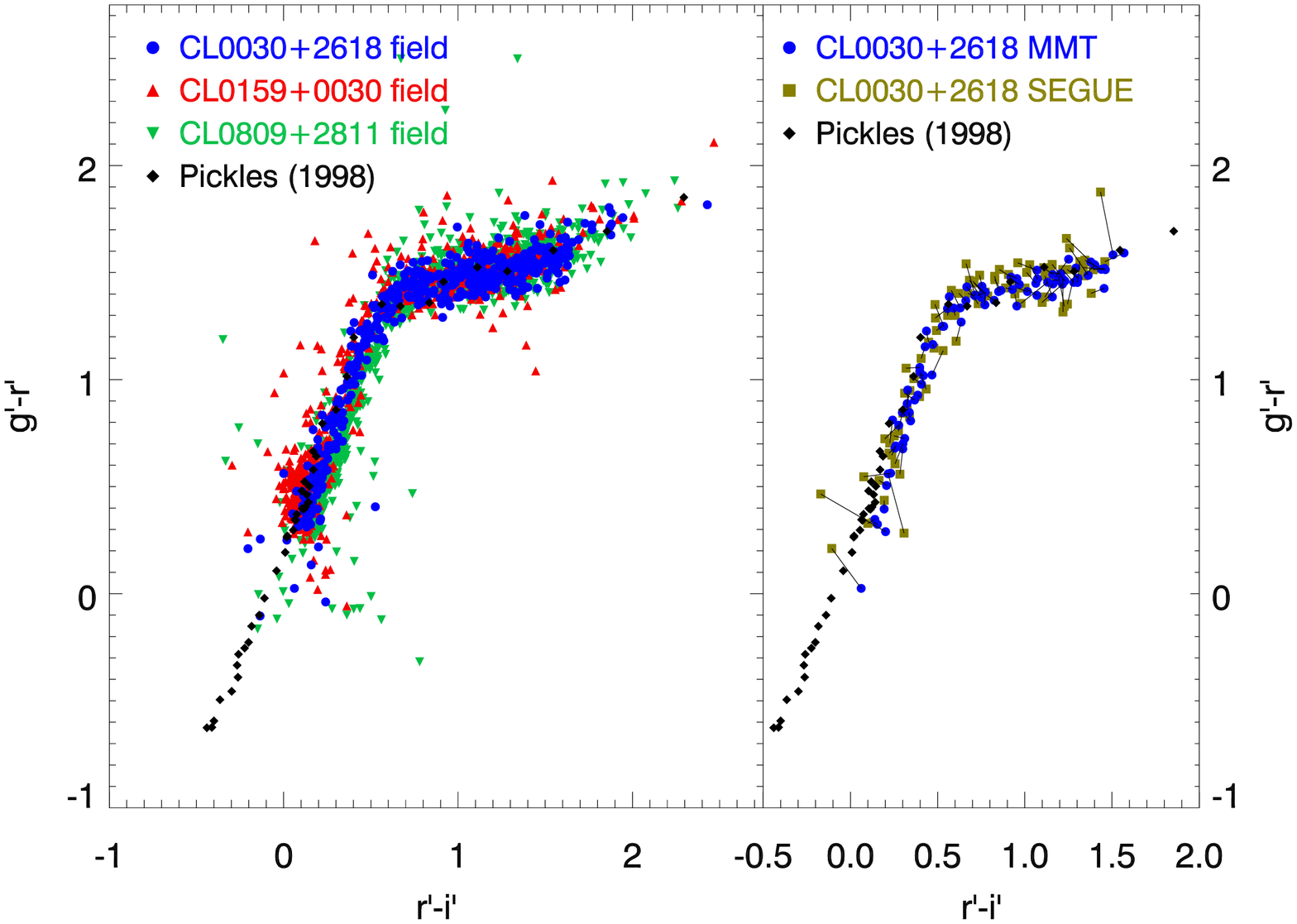}
 \caption{Photometric calibration by stellar colours: 
 \emph{Left panel:} plotted here are the
$g'\!-\!r'$ vs.\ $r'\!-\!i'$ colours of sources identified as stars in
three galaxy cluster fields observed with \megacam . For two of these fields,
CL0159+0030 (upward triangles) and CL0809+2811 (downward triangles), 
absolute photometric calibration with SDSS standards could be performed.
For \cl00 , 
results for recalibrated $r'$-band zeropoints are shown 
(dots; details see main text)
The colours in all three fields agree with the 
colours of main sequence stars from the 
\citet{1998PASP..110..863P} spectral library (diamonds). 
\emph{Right panel:} The $g'\!-\!r'$ vs. $r'\!-\!i'$ colours of stars in the
\megacam ~images of \cl00 ~(dots) 
which could also be identified in the partially
overlapping SEGUE strip \citep{2003AAS...20311211N} and shown here as squares
are both consistent with each other as well as with the 
\citet{1998PASP..110..863P} colours (diamonds). 
Each pair of measurements of one individual source is connected.}  
 \label{fig:pickles}
\end{figure*}

The final stage of the data reduction is to mask problematic regions in the
coadded images, applying the methods presented in \citet{2007A&A...470..821D}.
By subdividing the image into grid cells of a suitable size and counting 
\textsl{SExtractor} 
detections within those, we identify regions whose source density strongly
deviate from the average
as well as those with large gradients in source density.
This method not only detects the image's borders but also masks, effectively, 
zones of increased background close to bright stars, galaxies or defects.

In a similar fashion, we mask bright and possibly saturated stars
which are likely to introduce spurious objects in catalogues created
with \textsl{SExtractor}. We place a mask at each position of such
sources as drawn from the \texttt{USNO B1} catalogue. 
The method in which the size of the mask is
scaled according to the star's magnitude has been described in some
detail in \citet{2009A&A...493.1197E}. A small number of objects per
field which are missing in the \texttt{USNO B1} catalogue has to be
masked manually while masks around catalogue positions where no source
can be found have to be removed.\footnote{This last manual
step can be largely avoided by also automatically masking objects
drawn from the Hubble Space Telescope Guide Star Catalog (GSC),
as demonstrated by \citet{2009A&A...493.1197E}.}

Conforming with \citet{2005A&A...441..905H}, we compute the limiting
magnitudes in the coadded images for a $5\sigma$-detection in a
$2\arcsec$ aperture as
\begin{equation}
m_{\mathrm{lim}}\!=\!Z_{f}-2.5\log{(5\sqrt{N_{\mathrm{pix}}}\sigma_{\mathrm{sky}})}
\label{eq:maglim}
\end{equation}
with $Z_{f}$ being the photometric zeropoint in the respective filter
$f$, $N_{\mathrm{pix}}$ the number of pixels within the $2\arcsec$ aperture
and $\sigma_{\mathrm{sky}}$ the RMS sky-background variation measured
from the image.

Obtaining accurate colours for objects from CCD images is not as
trivial as it might seem. Next to the photometric calibration (see
Sect.~\ref{sec:photo}), aperture effects have to be taken into
account. Our approach is to measure \textsl{SExtractor} isophotal
(\texttt{ISO}) magnitudes from seeing-equalised images in our three
bands.  We perform a simplistic PSF matching based on the assumption
of Gaussian PSFs as it has been described in
\citet{2007A&A...462..865H}: The width of the filter with which to
convolve the $k$-th image is given as
\begin{equation}
\sigma_{\mathrm{filter,k}}=\sqrt{\sigma_{\mathrm{worst}}^{2}-\sigma_{\mathrm{k}}^{2}}
\end{equation}
with $\sigma_{\mathrm{k}}$ and $\sigma_{\mathrm{worst}}$ being the
widths of the best fitting Gaussians to the PSFs measured from the
$k$-th and the worst seeing image.

\subsection{Photometric Calibration} \label{sec:photo}

\subsubsection{Calibration Pipeline} \label{sec:filters}

The photometric calibration of our data is largely based on the method
developed by \citet{2006A&A...452.1121H} but using AB magnitudes and
SDSS-like filters together with the SDSS Data Release Six
\citep[DR6]{2008ApJS..175..297A} as our calibration catalogue.

As mentioned in Sect.~\ref{sec:thedata}, the \cl00 ~data we
present here have been observed together with three other clusters
from the \survey\ sample. Two of these, CL0159+0030 and CL0809+2811,
are situated within the SDSS DR6 footprint.
In addition to these, numerous
\citet{2000PASP..112..925S} standard fields were observed along with
the science data of which some provide further SDSS photometry.

The MMT/\megacam\ filter system is based on the SDSS one but not identical to 
it (see Fig.~\ref{fig:filters}); 
the comparison between the systems is detailed in
Sect.~\ref{sec:pcrel} in the Appendix.
Therefore, relations between instrumental magnitudes and calibrated magnitudes
in the SDSS system have to take into account colour terms.

In order to determine the photometric solution, we 
use \textsl{SExtractor} to draw catalogues from all
science and standard frames having SDSS overlap. 
Using the \citet{2006A&A...452.1121H} pipeline,
we then match these catalogues with a photometric catalogue 
assembled from the SDSS archives, serving as indirect photometric standards.

%
The relation between \megacam ~instrumental magnitudes 
$m_{\mathrm{inst}}$ and catalogue magnitudes $m_{\mathrm{SDSS}}$ for a filter
$f$ can be fitted as a linear function
of airmass $a$ and a first-order expansion with respect to the colour index,
simultaneously:
\begin{equation} \label{eq:photofit}
m_{\mathrm{f,inst}}-m_{\mathrm{f,SDSS}}\!=\!
\beta_{\mathrm{f}}(m_{\mathrm{f,SDSS}}-m_{\mathrm{f',SDSS}})+
\gamma_{\mathrm{f}}a+Z_{\mathrm{f}}
\end{equation}
where $m_{\mathrm{f,SDSS}}-m_{\mathrm{f',SDSS}}$ is a general colour index 
w.r.t.\ another filter $f'$, $\beta_{\mathrm{f}}$ the corresponding colour term,
and $Z_{\mathrm{f}}$ the photometric zeropoint
in which we are mainly interested.
For the fit, we select objects of intermediate magnitude 
which are neither saturated nor show a too large scatter in $m_{\mathrm{inst}}$ 
given a certain $m_{\mathrm{SDSS}}$.
Following the model of \citet{2006A&A...452.1121H}, 
we account for the variable photometric quality of our data by fitting
$\beta_{\mathrm{f}}$, $\gamma_{\mathrm{f}}$, and $Z_{\mathrm{f}}$
simultaneously under the best conditions, fixing $\gamma_{\mathrm{f}}$ in
intermediate, and fixing $\gamma_{\mathrm{f}}$ and $\beta_{\mathrm{f}}$ in even 
poorer conditions.
The fixed extinctions and colour terms are set to default values which are
discussed in Sect.~\ref{sec:pcrel}
in the Appendix.

\subsubsection{Colour-Colour-Diagrammes} \label{sec:ccc}

Comparing the zeropoints for different nights and fields, we conclude that
the nights on which the $r'$-band observations of \cl00
~were performed, were not entirely photometric but show a thin, uniform cirrus.
Therefore, an indirect re-calibration method is needed here.
To this end, we fitted the position in the $r'\!-\!i'$ vs. $g'\!-\!r'$ 
colour-colour-diagramme of the stars identified in the
\cl00 ~field to those found in two other, fully calibrated, galaxy cluster
fields, CL0159+0030 and CL0809+2811.

In the left panel of 
Fig.~\ref{fig:pickles}, we plot the $g'\!-\!r'$ versus $r'\!-\!i'$
colours of stars identified in these two fields and compare them to theoretical
spectra of main-sequence stars from the \citet{1998PASP..110..863P}
spectral library to
find good agreement between both the two observed sequences and the predicted
stellar colours.


As we find reliable absolute photometric calibrations for the $g'$- and
$i'$-bands of \cl00 , the location of the stellar main sequence for this
field is determined up to a shift along the main diagonal of the $g'\!-\!r'$ 
versus $r'\!-\!i'$ diagramme, corresponding to the $r'$
zeropoint. We fix this parameter by shifting the main sequence of \cl00 ~on top
of the other observed main sequences as well as the \citet{1998PASP..110..863P}
sequence.
We go in steps of 0.05~magnitudes, assuming this to be the best accuracy we
can reach adopting this rather qualitative method and settle for the
best-fitting test value (see Table~\ref{tab:data}).
The dots in Fig.~\ref{fig:pickles} 
show the best match with the CL0159+0030 and CL0809+2811 stellar
colours obtained by the re-calibration of the \cl00 ~$r'$-band.

After the photometric calibration, we became aware of a field observed in the
SEGUE project \citep{2003AAS...20311211N} using the SDSS telescope and filter
system which became publicly available along with the Sixth Data Release of
SDSS \citep{2008ApJS..175..297A} and partially overlaps with the \cl00 
~\megacam ~observations.
Thus, we are able to directly validate the indirect calibration by comparing
the colours of stars in the overlapping region. The right panel of  
Fig.~\ref{fig:pickles} shows the good agreement between the two independent
photometric measurements and the \citet{1998PASP..110..863P} templates
from which we conclude that our calibration holds to a good quality.

For comparison we also calibrated the \cl00 ~$r'$-band by comparing its
source counts to the ones in the CL0159+0030 and CL0809+2811 fields for the
same filter, but discard this calibration as we find a discrepancy of the
resulting main sequence in $g'\!-\!r'$ versus $r'\!-\!i'$ with the
theoretical \citet{1998PASP..110..863P} models mentioned earlier.

\section{The Shear Signal}

\begin{figure}
 \resizebox{\hsize}{!}{\includegraphics[angle=90,width=12cm]{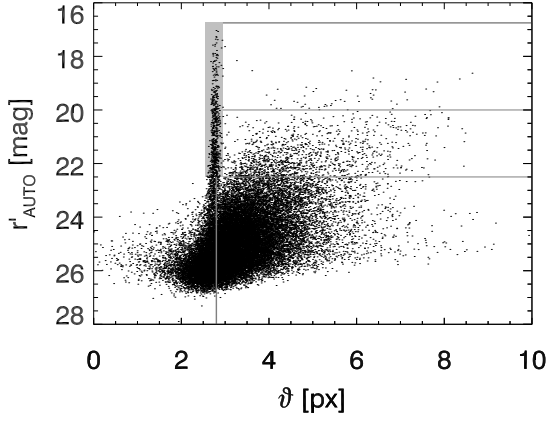}}
 \caption{The distribution of sources in the apparent size -- magnitude --
 space. Plotted are the \textsl{SExtractor} magnitudes 
 $r'_{\mathrm{AUTO}}$ against the half-light radii $\vartheta$ of all
 sources in the ``reliable'' catalogue.
 The stellar locus is visible prominently and is defined for further
 analysis by 
 $2.55~\mbox{px}\!<\vartheta\!<\! 2.95~\mbox{px}$ and
 $16.75~\mbox{mag}\!<\!r'_{\mathrm{AUTO}}\!<\! 22.5~\mbox{mag}$
 (the grey shaded area).
 The sources to the right and lower sides of the thick lines
 $r'_{\mathrm{AUTO}}\!<\!16.75~\mbox{mag}$ and 
 $\vartheta\!>\! 2.8~\mbox{px}$ for sources fainter than $22.5~\mbox{mag}$
 we treat as unsaturated galaxies. 
 In addition, the magnitude limits 
 $m_{\mathrm{bright}}$ and $m_{\mathrm{faint}}$ defined in 
 Sect.~\ref{sec:colcolsel} are given as thin lines for illustrative purposes. 
}
\label{fig:rhmag}
\end{figure}

Gravitational lensing leads to a distortion of images of distant sources by
tidal gravitational fields of intervening masses.
Here, we describe the method to measure this shear while referring to
\citet{2006glsw.book.....S} for the basic concepts and notation.

\subsection{KSB analysis}

The analysis of the weak lensing data is based on the 
\citet[KSB]{1995ApJ...449..460K} 
algorithm. The reduction pipeline we use was adapted from
the ``TS'' implementation presented in \citet{2006MNRAS.368.1323H}
and explored further in \citet{2007A&A...468..823S,2009arXiv0901.3269H}.
Its basic concepts are already outlined in \citet{2001A&A...366..717E}.
Thus, in this section, we will focus more on the properties of our data than on
the methods themselves since they are well documented in the above references.

The KSB algorithm confronts the problem of reconstructing the shear signal from
measured galactic ellipticities; therefore, it has to disentangle the shear
$\gamma$ from the intrinsic ellipticities\footnote{In this study, we 
adopt the following definition of ellipticity: if $r\!\leq\!1$ is an ellipse's
axis ratio its ellipticity is described by a two-component (polar) quantity
$e$ with $|e|\!=\!(1-r^{2})/(1+r^{2})$
which we represent as a complex number with Cartesian components 
$e\!=\!e_{1}+\mathrm{i}e_{2}$.}  
of the galaxies and from PSF effects.
The simultaneous effects of shear dilution by the PSF and the superposition of
the incoming ellipticity with the anisotropic PSF component can be isolated by
tracing the shapes of sources we can identify as stars, bearing neither
intrinsic ellipticity nor lensing shear.

\begin{figure*}
\includegraphics[width=12cm,angle=90]{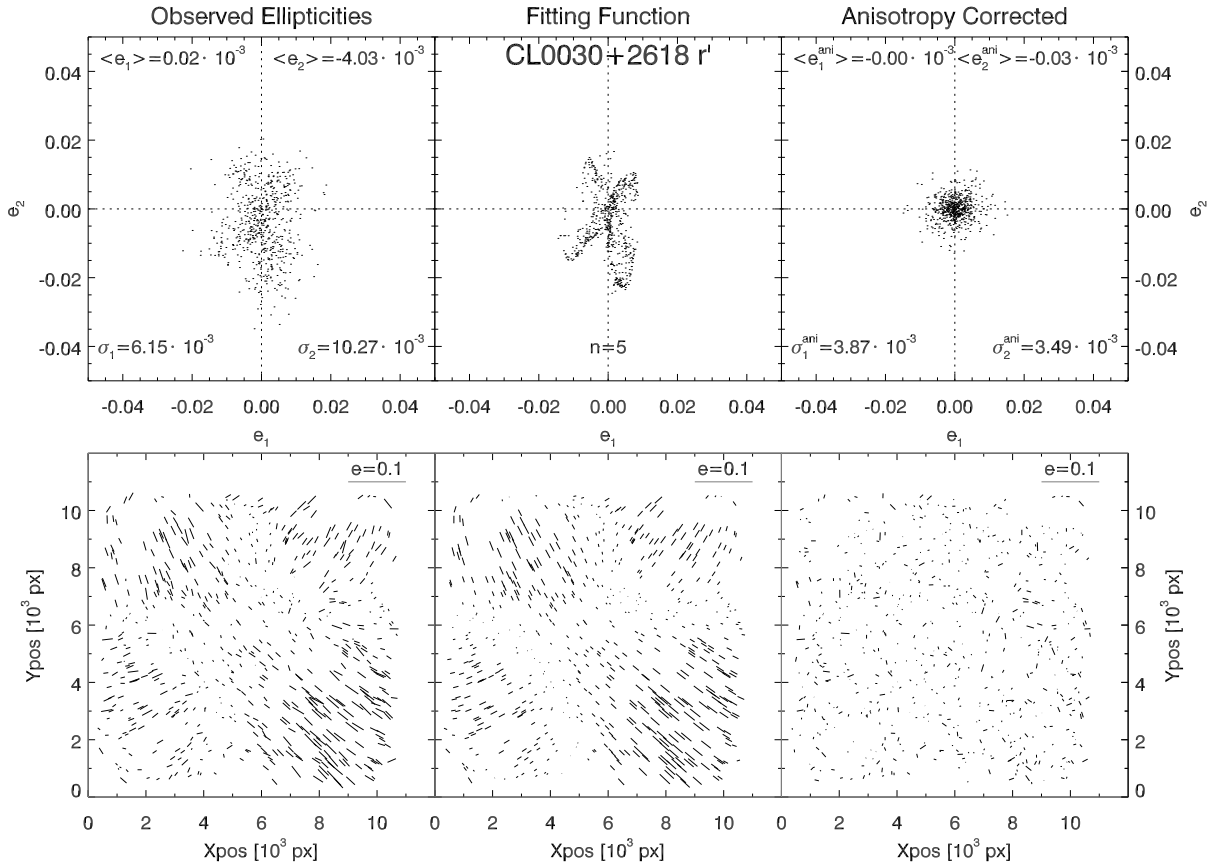}
\caption{Correction of PSF anisotropy of the \cl00 ~$r'$ band as 
  used in the analysis. The upper panel shows the distribution of the 
  ellipticity components $e_{1,2}$ of the stars identified in the
  field, together with the numeric values of their dispersions 
  $\sigma_{1,2}\!:=\!\sigma(e_{1,2})$. The ``whisker plots'' 
  in the lower panel show the size and orientation of
  PSF anisotropy as they vary as a function of the spatial coordinates $x$ and
  $y$. On the left, the situation before correction, i.e.\ the ellipticities as
  measured in the stars are depicted. The middle two plots give the fit by a
  global fifth order polynomial in $x$ and $y$. Residuals of this correction 
  are presented in the plots on the right.}
\label{fig:aniso}
\end{figure*}

The complete correction yielding a direct, and ideally unbiased, estimator 
$\varepsilon$ 
of the (reduced) shear $g$ exerted on a galaxy in our catalogue reads:
  \begin{equation} \label{eq:ksb}
   \varepsilon_{\alpha}=\left(\tens{P}^{\mathrm{g}}\right)^{-1}_{\alpha\beta}
  \left[ e_{\beta}-\tens{P}^{\mathrm{sm}}_{\beta\gamma}
  \left(\left(\tens{P}^{\mathrm{sm*}}\right)^{-1}_{\gamma\delta}
  e^{*}_{\delta}\right)\right]
  \end{equation}
Here, small Greek indices denote either of the two components of the complex
ellipticity. Quantities with asterisks are measured from stellar sources.
The $e_{\beta}$ are the ellipticities of galaxies as measured from the input
``shape'' image while the $e^{*}_{\delta}$ are the ellipticities of stars tracing
the PSF (plotted in the left two panels of Fig.~\ref{fig:aniso}).
The $2\times 2$ matrices $\tens{P}^{\mathrm{g}}$ and 
$\tens{P}^{\mathrm{sm}}$ give the transformation of ellipticities under the
influences of gravitational shear fields and the presence of an (anisotropic)
PSF, respectively. See, e.g., \citet{2006glsw.book.....S} 
on how these quantities are determined from
higher-order moments of the measured brightness distributions.


\subsection{The KSB and Galaxy Shape Catalogues} \label{sec:galcat}

Catalogues are created from the images using the \textsl{SExtractor} double
detection mode: Sources are identified on the lensing band image in its
original seeing. Photometric quantities (fluxes, magnitudes) are determined
at these coordinates from the measurement images in the three bands 
$g'r'i'$ convolved to the worst seeing (found in the $i'$-band).

The photometric properties determined from the three bands are merged
into one catalogue based on the detection image. From those catalogues
problematic sources are removed. 
Those are sources near the boundaries of the field-of-view
or blended with other sources, as well
as objects whose flux radii do not fall in the range 
$\vartheta^{*}\!<\!\vartheta_{\mathrm{g}}\!<\!10~\mbox{px}$,
for which KSB works safely,
with $\vartheta^{*}$ the angular size of unsaturated stars.
For the remaining objects, the shapes can now be determined.

Note that the \emph{KSB catalogue} 
presented in Fig.~\ref{fig:rhmag} and all
catalogues discussed hereafter only contain those objects for which a
half-light radius $\vartheta$ could be determined by our implementation 
of the \citet{2001A&A...366..717E} method.
Objects for which the measurements on the (noisy) data yield negative fluxes,
semi major axes, or second-order brightness moments or which lie to close to
the image border are removed from the catalogue, reducing its size by 
$\sim\!3$~\%.

Figure~\ref{fig:rhmag} shows the distribution of the sources in the
``reliable'' catalogue in apparent size -- magnitude space. The
prominent stellar locus enables us to define a sample of stars by
$\vartheta^{*}_{\mathrm{min}}\!\!<\!\vartheta\!\!<\!\vartheta^{*}_{\mathrm{max}}$
and
$r'^{*}_{\mathrm{min}}\!<\!r'_{\mathrm{AUTO}}\!<\!r'^{*}_{\mathrm{max}}$
with $\vartheta^{*}_{\mathrm{min}}\!\!=\!2.55~\mbox{px}$,
$\vartheta^{*}_{\mathrm{max}}\!\!=\!2.95~\mbox{px}$,
$r'^{*}_{\mathrm{min}}\!\!=\!16.75~\mbox{mag}$, and
$r'^{*}_{\mathrm{max}}\!\!=\!22.5~\mbox{mag}$
(the shaded area in Fig.~\ref{fig:rhmag}) 
from which the PSF anisotropy $e^{*}_{\delta}$
in Eq.~(\ref{eq:ksb}) is determined.

In the transition to the \emph{galaxy shape catalogue}, we 
regard as unsaturated galaxies all objects 
$r'_{\mathrm{AUTO}}\!>\!r'^{*}_{\mathrm{min}}$ (i.e.\ fainter than the brightest 
unsaturated point sources) and more extended than
$\vartheta\!\!>\!\vartheta^{*}_{\mathrm{max}}$ for 
$r'_{\mathrm{AUTO}}\!\!<\!r'^{*}_{\mathrm{max}}$ or 
$\vartheta\!\!>\! 2.8~\mbox{px}$ for 
$r'_{\mathrm{AUTO}}\!\!>\!r'^{*}_{\mathrm{max}}$,
respectively. The latter is justified by the fact that while for bright sources
it is easy to distinguish galaxies from point sources, there is a significant
population of faint galaxies for which a very small radius is measured by the
\textsl{SExtractor} algorithm. Thus, we relax the radius criterion by
$5$~\% for sources fainter than $r'^{*}_{\mathrm{max}}$.

However, we notice that among those small objects there is a population of
faint stars, not distinguishable from poorly resolved galaxies using an
apparent size -- magnitude diagramme alone and resulting in a dilution of the
lensing signal compared to a perfect star -- galaxy distinction.
Our decision to nevertheless include these small sources into our catalogue is
based on the resulting higher cluster signal as compared to a more 
conservative criterion 
(e.g.\ $\vartheta/\vartheta^{*}_{\mathrm{max}}\!\leq\!1.10$ 
for the galaxies fainter than $r'^{*}_{\mathrm{max}}$).
We call ``galaxy shape catalogue'' the list of objects that both pass
this galaxy selection and the cuts for signal quality discussed in 
Sect.~\ref{sec:veri}. This important catalogue yields the final ``lensing catalogue''
by means of the \emph{background selection} discussed in Sect.~\ref{sec:bgsel}.

\subsection{PSF Anisotropy of \megacam } \label{sec:aniso}

\begin{figure}
\resizebox{\hsize}{!}{\includegraphics[width=\textwidth,angle=90]{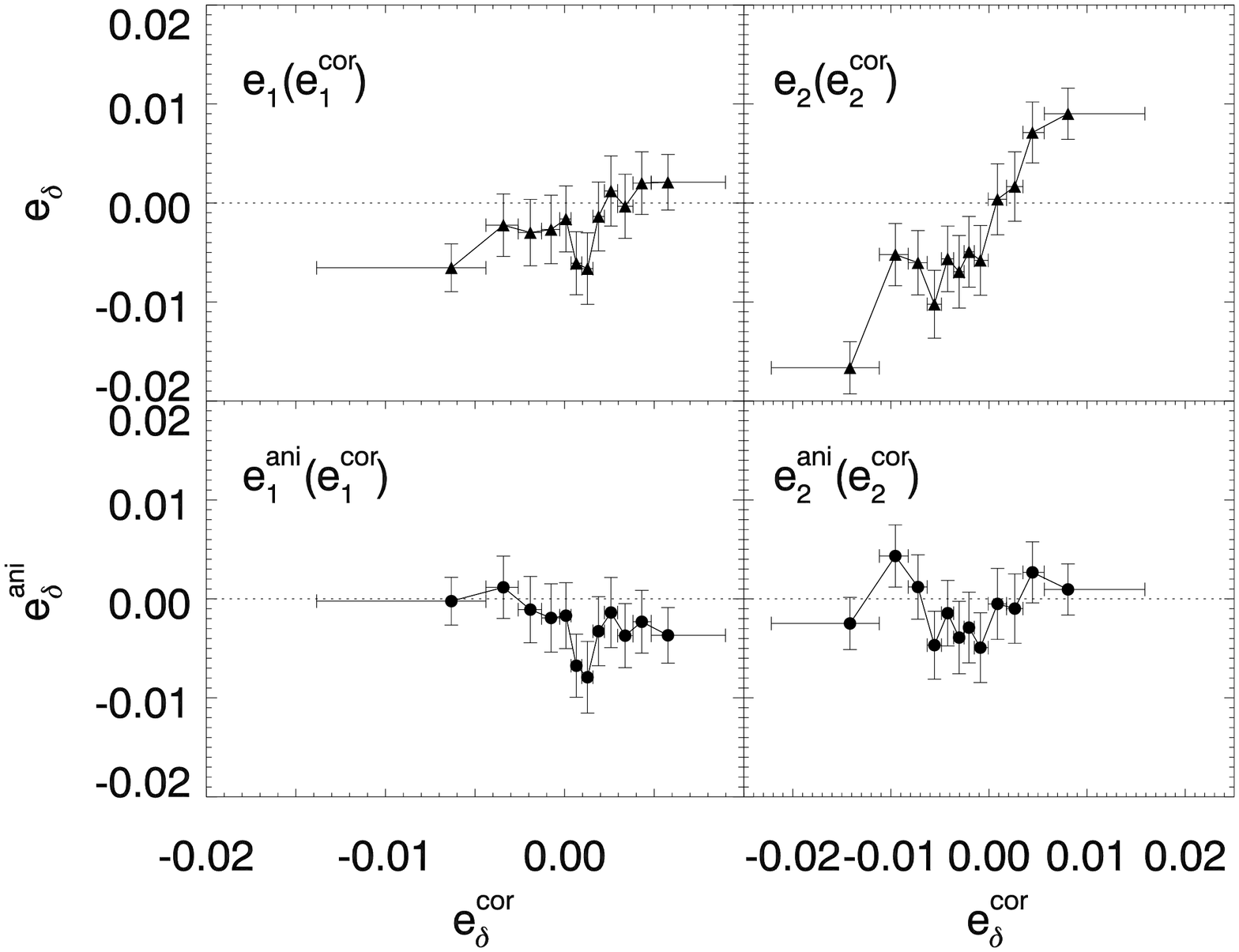}}
\caption{The effect of the polynomial correction for the PSF anisotropy
   on the ellipticities of galaxies averaged in equally populated bins. 
   As a function of the amount of correction
   $e^{\mathrm{cor}}$ applied to the components $\delta\!=\!1$ (left panels) and
   $\delta\!=\!2$ (right panels), we show the raw ellipticities \emph{before}
   correction in the upper panels and the PSF-corrected ellipticities 
   $\varepsilon$ in the lower panels. 
   The bars in the abscissa and ordinate denote the range of the bin and the 
   standard deviation
   of the ellipticity in this bin, respectively. }
\label{fig:corrpol}
\end{figure}

In the KSB pipeline, 
we fit a model $e^{\mathrm{cor},*}(x,y)$ in the pixel coordinates $x$ and $y$
to the measured ellipticities $e^{*}$ of stars such that the residual 
anisotropies $e^{\mathrm{ani},*}\!=e^{*}\!-\!e^{\mathrm{cor},*}$ 
of stellar images should effectively be zero. 
Figure~\ref{fig:aniso} shows the effect of PSF anisotropy correction: 
The raw ellipticities of the tracing stars as presented
in the left two panels are modelled by a polynomial 
$e^{\mathrm{cor}}\!=\!\sum_{k=0}^{n}{\sum_{\ell=0}^{n-k}{p_{k\ell}x^{k}y^{\ell}}}$
defined globally over the whole field-of-view. 
The best-fitting solution for the case
$n\!=\!5$ we adopt here is shown in the middle panels of Fig.~\ref{fig:aniso}
while the residual ellipticities of the stars 
$e^{\mathrm{ani}}$ are displayed in the panels to the right.

Simultaneously aiming at reducing both the mean 
$\langle e^{\mathrm{ani,*}}_{\delta}\rangle$ of the residual ellipticities 
and their dispersion $\sigma(e^{\mathrm{ani,*}}_{\delta})$ we find that a polynomial
order as high as $n\!=\!5$ is necessary to effectively correct for the
distinctive quadrupolar pattern in the spatial distribution of the 
``raw'' stellar ellipticities (see lower left and middle panels of 
Fig.~\ref{fig:aniso}).
There is no obvious relation between the zones of preferred orientation
of the PSF ellipticity in Fig.~\ref{fig:aniso} and the $4\times 9$ chip 
detector layout of \megacam .
See Sect.~\ref{sec:veri} for further details.

For stacking in the lensing band, we select only those frames which show
moderate PSF ellipticity in the first place (see Sect.~\ref{sec:stack} in the
Appendix for details). 
Thus, we ensure the images used for lensing analysis to be isotropic to a high
degree even before any corrections are applied. By stacking images in which the
PSF anisotropy is different in magnitude and orientation 
(cf.\ Figs.~\ref{fig:framesel} and \ref{fig:indiv}), we further reduce the
ellipticity owing to the imaging system. 
The total amount of PSF anisotropy present in our \megacam ~data is small:
Before correction, we measure $\langle e_{1}\rangle\!=\!1.77\times 10^{-5}$,
$\langle e_{2}\rangle\!=\!-4.03\times 10^{-3}$,  
$\langle |e|\rangle=\!1.10\times 10^{-2}$, and 
$\sigma(|e|)\!=\!6.19\times 10^{-3}$, reducing after the correction to
$\langle e^{\mathrm{ani}}_{1}\rangle\!=\!-5.60\times 10^{-7}$,
$\langle e^{\mathrm{ani}}_{2}\rangle\!=\!-2.60\times 10^{-5}$,
$\langle |e^{\mathrm{ani}}|\rangle=\!4.34\times 10^{-3}$, and 
$\sigma(|e^{\mathrm{ani}}|)\!=\!2.90\times 10^{-3}$.
Note that the very small averages for the individual components result from
partial cancellation of anisotropies from different parts of the field-of-view.
Thus, MMT/\megacam ~shows a similar degree of PSF anisotropy as other 
instruments from which lensing signals have been measured successfully,
e.g.\ \textsc{MegaPrime/Megacam} on CFHT \citep{2006A&A...452...51S}
or Subaru's SuprimeCam \citep{2008PASJ...60..345O}.
The latter authors measured, as an RMS average of seven galaxy cluster fields,
$\langle e_{1}\rangle\!=\!1.41\times 10^{-2}$, 
$\langle e_{2}\rangle\!=\!1.63\times 10^{-2}$, and
$\sigma(|e|)\!=\!2.32\times 10^{-2}$ before correction with larger values 
for the anisotropy components but a simpler spatial pattern.

Although we find small-scale changes in the PSF
ellipticity which have to be modelled by a polynomial of relatively high order,
the more important point is that the PSF anisotropy varies smoothly as a
function of the position on the detector surface in every \emph{individual} 
exposure, showing a simpler pattern than Fig.~\ref{fig:aniso}.
(See Fig.~\ref{fig:indiv} for example exposures at both small and large
values of overall PSF anisotropy induced by the tracking behaviour of MMT.) 
Consequently, it \emph{can} be modelled by a smooth function which is a
necessary prerequisite for using the instrument with the current weak lensing
analysis pipelines. Thus, we have shown that weak lensing work is feasible
using MMT \megacam .

\subsection{Selection of lensed background galaxies} \label{sec:bgsel}

Before we proceed with the details of our lensing analysis, we explain
how we arrive from the galaxy shape catalogue at the ``lensing catalogue'' of objects
we classify as background galaxies w.r.t.\ to \cl00 . This 
\emph{background selection}, as we will call it from now on=
is based on their $g'r'i'$ photometry.
While unlensed objects remaining in the catalogue dilute the shear signal,
rejection of actual background galaxies reduce it as well.
Note that a sensible foreground removal is especially important for
relatively distant objects like the \survey ~Cosmology Sample clusters.

We introduce two free parameters in our analysis: the magnitude limit 
$m_{\mathrm{faint}}$ below which all fainter galaxies are included in the shear 
catalogue, regardless of their $g'\!-\!r'$ and $r'\!-\!i'$ 
colour indices, and the magnitude $m_{\mathrm{bright}}$
above which all brighter galaxies will be considered foreground objects and
discarded. Only in the intermediary interval 
$m_{\mathrm{bright}}\!<\!r'\!<\!m_{\mathrm{faint}}$ the selection of galaxies
based on their position in the colour-colour-diagramme will take place.
In these terms, a simple magnitude cut would
correspond to $m_{\mathrm{bright}}\!=\!m_{\mathrm{faint}}$. 
We vary these parameters in order to optimise the detection of \cl00 
~and find $m_{\mathrm{bright}}\!=\!20.0$ and $m_{\mathrm{faint}}\!=\!22.5$.
For details of the colour-colour-diagramme method, see 
Sect.~\ref{sec:colcolsel}. The photometric cuts reduce the catalogue
size by $6.0$~\%, leaving us with a lensing catalogue of 
$N_{\mathrm{cat}}\!=\!14813$ objects, corresponding to a galaxy 
surface density of $n_{\mathrm{cat}}\!=\!21.2\,\mbox{arcmin}^{-2}$.

\subsection{Aperture mass and lensing detection}

The weak lensing analysis we conduct is a two-step process. First, we 
confirm the presence of a cluster signal by constructing \emph{aperture mass}
maps of the field which will provide us with a position for the cluster centre
and the corresponding significance. In the second step, building on this
position for \cl00 , the tangential shear profile can be determined and 
fitted, leading to the determination of the cluster mass.

More precisely, we use the so-called $S$-statistics, corresponding to the 
signal-to-noise ratio of the aperture mass estimator which for any given centre
$\pmb{\theta}_{\mathrm{c}}$ is a weighted sum over the tangential ellipticities
of all lensing catalogue galaxies within a circular aperture of radius 
$\theta_{\mathrm{out}}$. The estimator
can be written analytically as \citep{1996MNRAS.283..837S}:
\begin{equation} \label{eq:s-stat}
S_{\theta_{\mathrm{out}}}(\pmb{\theta}_{\mathrm{c}})
\!=\!\frac{\sqrt{2}}{\sigma_{\varepsilon}}
\frac{\sum_{i}{\varepsilon_{\mathrm{t},i}Q_{i}(|\pmb{\theta}_{\mathrm{i}}\!-\!\pmb{\theta}_{\mathrm{c}}|)}}
{\sqrt{\sum_{i}{Q_{i}^{2}(|\pmb{\theta}_{\mathrm{i}}\!-\!\pmb{\theta}_{\mathrm{c}}|)}}}
\end{equation}
where $\varepsilon_{\mathrm{t},i}$ denotes the measured shear 
component tangential with respect to the
centre for the galaxy at position $\pmb{\theta}_{\mathrm{i}}$.
As filter function 
$Q(x\!:=\!|\pmb{\theta}_{\mathrm{i}}\!-\!\pmb{\theta}_{\mathrm{c}}|/\theta_{\mathrm{out}})$,
we apply the hyperbolic tangent filter introduced by
\citet{2007A&A...462..875S}:
\begin{equation}
Q_{\mathrm{TANH}}(x)\!=\!\frac{1}{1+\mathrm{e}^{a-bx}+\mathrm{e}^{c+dx}}
\frac{\tanh{(x/x_{\mathrm{c}})}}{x/x_{\mathrm{c}}}
\end{equation}
with the width of the filter determined by $x_{\mathrm{c}}\!=\!0.15$ and 
the shape of its exponential cut-offs for small and large $x$ given by the 
default values $\{a,b,c,d\}\!=\!\{6,150,47,50\}$.
The $S$-statistics includes as a noise term the
intrinsic source ellipticity, calculated from the data galaxies as
$\sigma_{\varepsilon}\!=\!\langle\varepsilon_{1}^{2}\!+\!
\varepsilon_{2}^{2}\rangle^{1/2}$ with typical values
$\sigma_{\varepsilon}\!\approx\!0.38$.

The value of $\theta_{\mathrm{out}}$ in Eq.~(\ref{eq:s-stat}) is also 
fixed such that it maximises $S_{\theta_{\mathrm{out}}}(\pmb{\theta}_{\mathrm{c}})$
which strongly depends on the filtering size used.
Exploring the parameter space spanned by $\theta_{\mathrm{out}}$ and the
photometric parameters $m_{\mathrm{bright}}$ and $m_{\mathrm{faint}}$,
we find, independent of the latter two, the
highest $S$-values with $14\arcmin\!<\!\theta_{\mathrm{out}}\!<\!15\arcmin$.
The behaviour of $S$ as a function of $\theta_{\mathrm{out}}$ (at a fixed
$\pmb{\theta}_{\mathrm{c}}$) is in good general agreement with the results of
\citet{2007A&A...462..875S} for the same filter function $Q_{\mathrm{TANH}}(x)$.
Thus, we fix $\theta_{\mathrm{out}}\!=\!14\farcm5$ for the further analysis,
noting this number's agreement with the size of our \megacam ~images
(cf. Fig.~\ref{fig:overlay}).
%
%
We also tested the influence of the parameter $x_{\mathrm{c}}$ in the 
$Q_{\mathrm{TANH}}$ filter and find that, with all other parameters kept fixed,
in the $0.15\!\leq\!x_{\mathrm{c}}\!\leq0.6$
interval the maximum $S$-value changes by less than
0.5~\% but decreases more steeply for smaller values of $x_{\mathrm{c}}$.

Applying these parameters and measuring $S$ on a reference grid of $60\arcsec$
mesh size, we detect \cl00 ~at the level of $5.8\sigma$ in a grid cell
centred at a distance of $34\arcsec$ from the \textsl{ROSAT}
position at $\alpha_{\mathrm{J2000}}\!=\!00^{\mathrm{h}}30^{\mathrm{m}}33\fs6, 
\delta_{\mathrm{J2000}}\!=\!26\degr18\arcmin16\arcsec$, i.e. smaller than the grid resolution.
We will investigate further into the cluster position in Sect.~\ref{sec:cent}.

\subsection{Verification of the shear signal} \label{sec:veri}

\begin{figure}
\resizebox{\hsize}{!}{\includegraphics[width=\textwidth,angle=90]{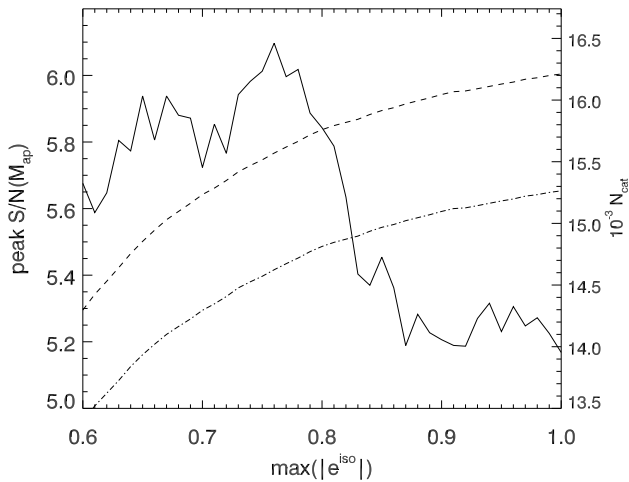}}
\caption{The $S$-statistics
 (solid line) as a function of the maximum value of the
 ellipticity estimator $\varepsilon$ resulting we include in
 the galaxy shape catalogue. The dashed and dash-dotted lines show the sizes of the
 resulting catalogue before and after background selection 
 (see Sect.~\ref{sec:colcolsel}), respectively.}
\label{fig:abseiso}
\end{figure}
\begin{table*}
 \caption{Notable galaxies in the field of \cl00 .}
 \begin{center}
 \begin{tabular}{ccccccccc} \hline\hline
 Galaxy & $\alpha_{\mathrm{J2000}}$ & $\delta_{\mathrm{J2000}}$ & $r'^{\mathrm{MMT}}_{\mathrm{AUTO}}$ & $g'^{\mathrm{MMT}}_{\mathrm{AUTO}}$ & $i'^{\mathrm{MMT}}_{\mathrm{AUTO}}$ & $z$ & Note & See Figure\\ \hline
 G1 & $00^{\mathrm{h}}30^{\mathrm{m}}34\fs0$ & $26\degr18\arcmin09\arcsec$ & $19.20$ & $21.14$ & $18.31$ & $0.516^{\dagger}$ & dominant in \cl00 & Fig.~1 \\
 G2 & $00^{\mathrm{h}}30^{\mathrm{m}}37\fs9$ & $26\degr18\arcmin18\arcsec$ & $18.82$ & $20.23$ & $18.27$ & n.a. & dominant in foreground group & Fig.~1 \\
 G3 & $00^{\mathrm{h}}30^{\mathrm{m}}36\fs3$ & $26\degr19\arcmin20\arcsec$ & $19.46$ & $20.76$ & $18.95$ & n.a. & strong lensing feature & Fig.~1 \\
 G4 & $00^{\mathrm{h}}30^{\mathrm{m}}39\fs5$ & $26\degr20\arcmin56\arcsec$ & $17.23$ & $17.98$ & $16.94$ & $0.493^{\dagger}$ & QSO & Fig.~9 \\ 
\hline \hline \label{tab:gal}

 \end{tabular}
 \begin{minipage}{100mm}
  \smallskip
  $^{\dagger}$ Redshift taken from \citet{1997MNRAS.285..511B}.
 \end{minipage}
 \end{center}
\end{table*}
In this subsection, we summarise the consistency tests performed on the data
to validate the galaxy shape measurements giving rise to the shear signal
discussed below.

\begin{itemize}

 \item{\textbf{Correction of PSF anisotropy:} We assess the performance of the
       correction polynomial by analysing the PSF-corrected ellipticities
       $e^{\mathrm{ani}}_{\mathrm{gal},\delta}$ 
       of galaxies as a function of the amount of correction 
       $e^{\mathrm{cor}}_{\delta}$ that has been applied to them 
       by fitting a polynomial to the anisotropy distribution of star images
       (see Sect.~\ref{sec:aniso}).
       Theoretically, the expected positive correlation between the
       \emph{uncorrected} ellipticities 
       and the correcting polynomial should
       be removed and $e^{\mathrm{ani}}_{\mathrm{gal}}(e^{\mathrm{cor}})$ 
       thus scatter around zero.
       We note that most anisotropy is found in the $\delta\!=\!2$ component
       from the beginning (Fig.~\ref{fig:corrpol}). This is removed in the
       corrected ellipticities, with 
      $\langle e^{\mathrm{ani}}_{\mathrm{gal,2}}\rangle\!=\!-0.0010\!\pm\!0.0010$
       marginally consistent with zero in the standard deviation. In the
       $\delta\!=\!1$ component, we measure a residual anisotropy of
      $\langle e^{\mathrm{ani}}_{\mathrm{gal,1}}\rangle\!=\!-0.0026\!\pm\!0.0010$
       which is one order of magnitude smaller than the lensing signal we
       are about to measure.

       Alternatively to the $n\!=\!5$ polynomial correction to the entire image,
       we consider a piecewise solution based on the pattern of preferred 
       orientation in Fig.~\ref{fig:aniso}. Dividing the field into four regions
       at $y\!=\!6100\,\mbox{px}$ and at $x\!=\!4300\,\mbox{px}$ for
       $y\!<\!6100\,\mbox{px}$ and $x\!=\!5800\,\mbox{px}$ for 
       $y\!>\!6100\,\mbox{px}$ with a polynomial degree up to $n\!=\!5$ we do
       not find a significant improvement in terms of 
       $\langle e^{\mathrm{ani,*}}\rangle$, $\sigma(e^{\mathrm{ani,*}})$, 
       or $e^{\mathrm{ani}}_{\mathrm{gal}}(e^{\mathrm{cor}})$ over the simpler 
       model defined over the whole field.
}
 \item{\textbf{Maximum shear:} Due to the inversion of the noisy matrix
       $\tens{P}^{\mathrm{g}}$ in Eq.~(\ref{eq:ksb}), resulting values for
       the estimator $|\varepsilon|$ are not bound from above while
       ellipticities are confined to $0\!\leq\!e\!\leq\!1$.
       Thus, attempting to measure weak lensing using the KSB method, we need
       to define an upper limit $\max{(|\varepsilon|)}$ of the shear 
       estimates we consider reliable.
       We evaluate the influence of the choice for $\max{(|\varepsilon|)}$
       on the $S$-statistics (Eq.~\ref{eq:s-stat}) by varying it while
       keeping the other parameters, like $\min{(\tr{\tens{P}^{\mathrm{g}}})}$,
       the minimum $\min{(\nu)}$ of the
       signal-to-noise ratio $\nu$ of the individual galaxy
       detection determined by the KSB code, and the photometric parameters
       $m_{\mathrm{bright}}$ and $m_{\mathrm{faint}}$ defined in 
       Sect.~\ref{sec:colcolsel} fixed.
       In the range $0.6\!\lesssim\!\max{(|\varepsilon|)}\!\lesssim\!0.8$,
       we find an increasing shear signal due to the higher number of galaxies in
       the catalogues using less restrictive cuts (Fig.~\ref{fig:abseiso}). For
       $\max{(|\varepsilon|)}\!\gtrsim\!0.8$, we see a sharp decline of the
       lensing signal which we explain as an effect of galaxies entering the
       catalogue whose ellipticity estimate is dominated by noise.
       We fix $\max{(|\varepsilon|)}\!=\!0.8$, 
       $\min{(\tr{\tens{P}^{\mathrm{g}}})}\!=\!0.1$, and $\min{(\nu)}\!=\!4.5$
       simultaneously to their given values.
       We note that, while optimising the $S$-statistics, this might introduce
       a bias in the mass estimate as a cut in $\max{(|\varepsilon|)}\!=\!0.8$
       directly affects the averaging process yielding the shear.
}
 \item{\textbf{Shear calibration: } 
       We can account for this bias by scaling the shear estimates with a
       \emph{shear calibration factor} $f$ such that 
       $\varepsilon\!\rightarrow \!f_{0} \varepsilon$ 
       to balance biases like the effect of $\max{(|\varepsilon|)}$.
       The question how gravitational shear can be measured unbiased and 
       precisely 
       has been identified as the crucial challenge to future weak lensing 
       experiments \citep[see e.g.,][]{2006MNRAS.368.1323H,2007MNRAS.376...13M,2009arXiv0908.0945B}.
       The ``TS'' KSB method employed here has 
       been studied extensively and is well understood in many aspects.
       In order to correct for the biased shear measurements, found by testing
       the KSB pipeline with the simulated data in \citet{2006MNRAS.368.1323H},
       the shear calibration factor has been introduced and studied 
       subsequently \citep{2007A&A...468..823S,2009arXiv0901.3269H}.
       As pointed out by these authors, the calibration bias depends on both 
       the strength of the shear signal under inspection, as well as on the 
       details of the implementation and galaxy selection for the shear 
       catalogue.
       In the absence of detailed shape measurement simulations under 
       cluster lensing conditions, we chose a fiducial 
       $f_{0}\!=\!1.08$ from \citet{2009arXiv0901.3269H} and assign an error 
       of $\sigma_{\mathrm{f_{0}}}\!=\!0.05$ to it, 
       covering a significant part of the discussed interval.
}
 \item{\textbf{Complementary catalogue:} We check the efficacy of the
       set of parameters we adopted by reversing the selection of galaxies and
       calculating the $S$-statistics from those galaxies excluded in our
       normal procedure. 
        Reversing the background selection, i.e.\ only keeping those galaxies
        regarded as cluster or foreground sources,
        we find from $10^{5}$ bootstrap realisations of the
        complentary catalogue an aperture mass significance of
        $S\!=\!-0.83\pm1.06$. From the consistency with zero, we conclude these
        cuts to effectively select the signal-carrying galaxies.
       As the background selection removes 
       $f_{\mathrm{ph}}\!=\!N_{\mathrm{complem}}/N_{\mathrm{cat}}\!=\!6.0$~\% of
       the sources in the catalogue, we only expect a small bias 
       $\approx\!f_{\mathrm{ph}}S_{\mathrm{complem}}/S_{\mathrm{cat}}\!\approx\!-0.8$~\% 
       resulting from the background selection.
}

\end{itemize}

\begin{table}
 \caption{Colours of prominent galaxies observed in the \cl00 ~field compared to
          colours computed from CWW80 elliptical templates at $z\!=\!0.50$
          and $z\!=\!0.25$.}
 \begin{center}
 \begin{tabular}{ccccc} \hline\hline
    Galaxy & $z$ & $g'\!-\!i'$ & $g'\!-\!r'$ & $r'\!-\!i'$ \\ \hline
    G1 & $0.516$ & $2.83$ & $1.94$ & $0.89$ \\
    G2 & n.a. & $1.96$ & $1.41$ & $0.55$ \\ \hline
    CWW80 Ell $z\!=\!0.50$ & $0.50$ & $2.782$ & $1.876$ & $0.906$ \\
    CWW80 Ell $z\!=\!0.25$ & $0.25$ & $2.098$ & $1.554$ & $0.544$ \\
\hline \hline \label{tab:galcww}
 \end{tabular}
 \end{center}
\end{table}
\section{The Multi-Wavelength View of \cl00 }

\subsection{Identifying  the BCG of \cl00 } \label{sec:colours} \label{sec:bcg}

Figure \ref{fig:img} shows two candidates for the brightest cluster galaxy
of \cl00 , galaxies with extended cD-like haloes and similar $i'$-magnitudes 
(Table~\ref{tab:gal}). The galaxy G1, closer to the \textsc{Rosat} and 
\textsc{Chandra} centres of \cl00 , was attributed to a cluster by 
\citet{1997MNRAS.285..511B}, measuring a spectroscopic redshift of 
$z_{\mathrm{G1}}\!=\!0.516$, while three of their six spectro-$z$s are 
$z\!\approx\!0.25$.

We note that G1 and G2 show different colours in Fig.~\ref{fig:img}, each being
similar to their fainter immediate neighbours. As very extended sources, G1
and G2 are flagged early-on in the pipeline but \emph{are} included in the raw 
\textsl{SExtractor} catalogues. Aware of their larger uncertainties, we use 
these magnitudes\footnote{Here, we use \textsl{SExtractor AUTO} instead of \texttt{ISO} 
magnitudes, known to be more robust at the expense of less accurate colour 
measurements. Nevertheless, we find only small differences between the two 
apertures, allowing for cautious direct comparison.}
for G1, G2, and two other interesting extended galaxies (Table~\ref{tab:gal}).

The observed $g'\!-\!r'$, $r'\!-\!i'$, and $g'\!-\!i'$colours are compared to 
the ones predicted for a typical BCG at $z\!=\!0.50$ and $z\!=\!0.25$, using
the \citet[CWW80]{1980ApJS...43..393C} elliptical galaxy template
(Table \ref{tab:galcww}).
Nicely consistent with its spectroscopic redshift, we find the colours of G1
to be similar to the $z\!=\!0.50$ template, while G2's bluer colours resemble 
the CWW template at $z\!=\!0.25$. We conclude that G1, located close to the
X-ray centres, is a member of \cl00 , and indeed its BCG. On the other hand,
G2 can be considered the brightest member of a \emph{foreground group} at 
$z\!\approx\!0.25$. The existence of such foreground structure is corroborated
by the broad $g'\!-\!i'$ distribution (Fig.~\ref{fig:cmdccd}). Its implications
are discussed in Sect.~\ref{sec:cent} and \ref{sec:fits}.

\subsection{Comparing Centres of \cl00 } \label{sec:cent}

\begin{figure*}
\includegraphics[width=0.9\textwidth,angle=90]{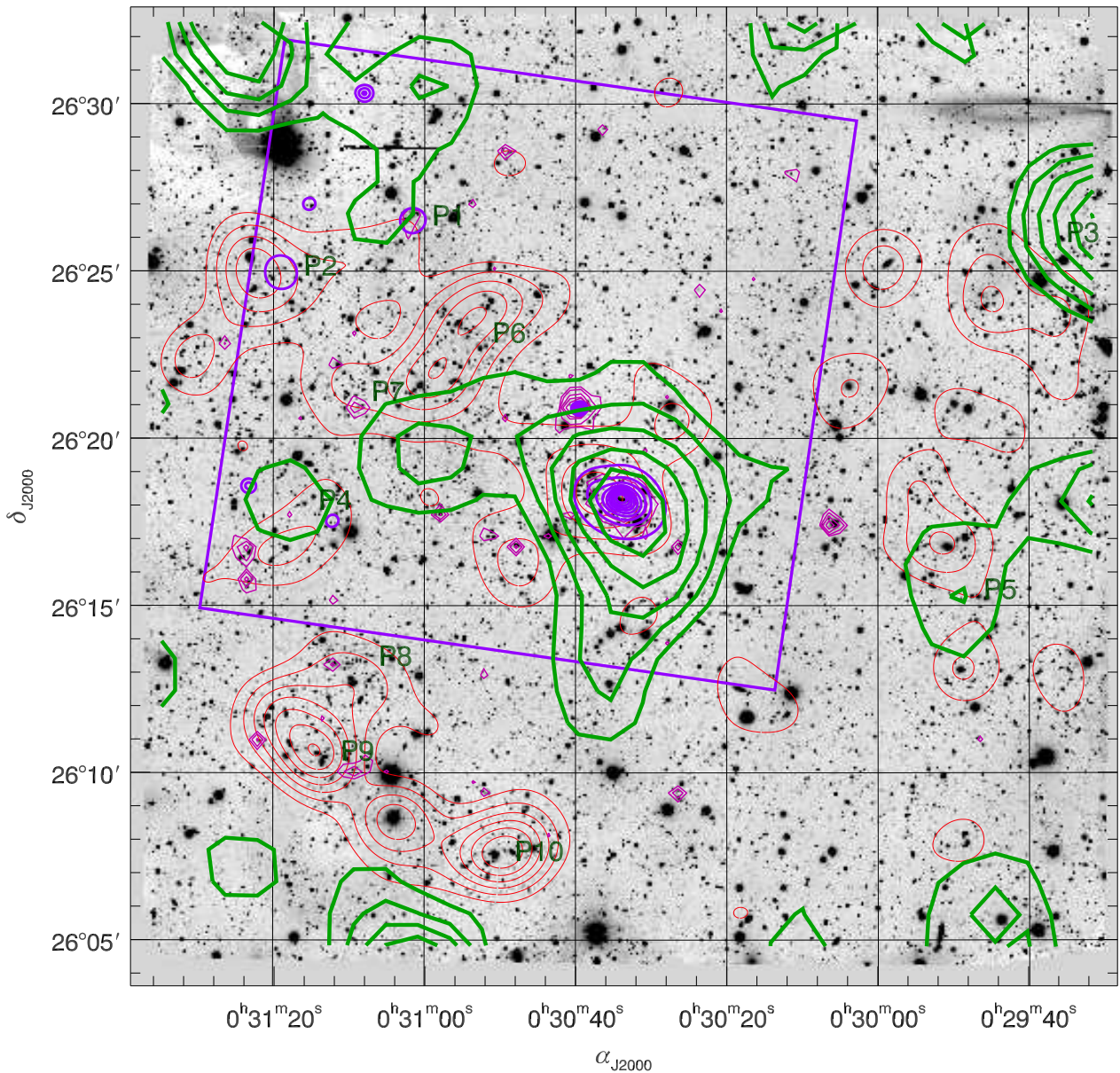}
\caption{The $r'$-band image of \cl00 , overlaid with $r'$-band galaxy light 
(thin, red), \textsc{Chandra} (medium-thick, blue; within the smaller square 
footprint), and \textsc{XMM-Newton} (medium-thin, magenta), 
and lensing surface mass density contours (thick, green). 
We show X-ray surface brightness levels in multiples of 
$5\times 10^{-9} \mbox{cts}\,\mbox{cm}^{-2}\mbox{s}^{-1}\mbox{arcmin}^{-2}$ in the 
$0.5\ldots 2.\,\mbox{keV}$ band. 
The $r'$-band flux density contours (thin red lines) start from $0.015$ flux 
units per pixel, in intervals of $0.005$ flux units. 
Lensing convergence contour levels were obtained smoothing the
shear field $\gamma(\vec{\theta})$ with a Gaussian filter of $2\arcmin$ 
width and are linearly spaced in intervals of $\Delta\kappa\!=\!0.01$, 
starting at $\kappa\!=\!0.01$. \textsc{XMM-Newton} 
contours show MOS2 counts smoothed
by an adaptive Gaussian kernel in logarithmic spacing.
The labels ``P1'' to ``P10'' designate the peaks discussed in 
Sect.~\ref{sec:others}.
}
\label{fig:overlay}
\end{figure*}

\begin{figure*}
\includegraphics[width=0.9\textwidth,angle=90]{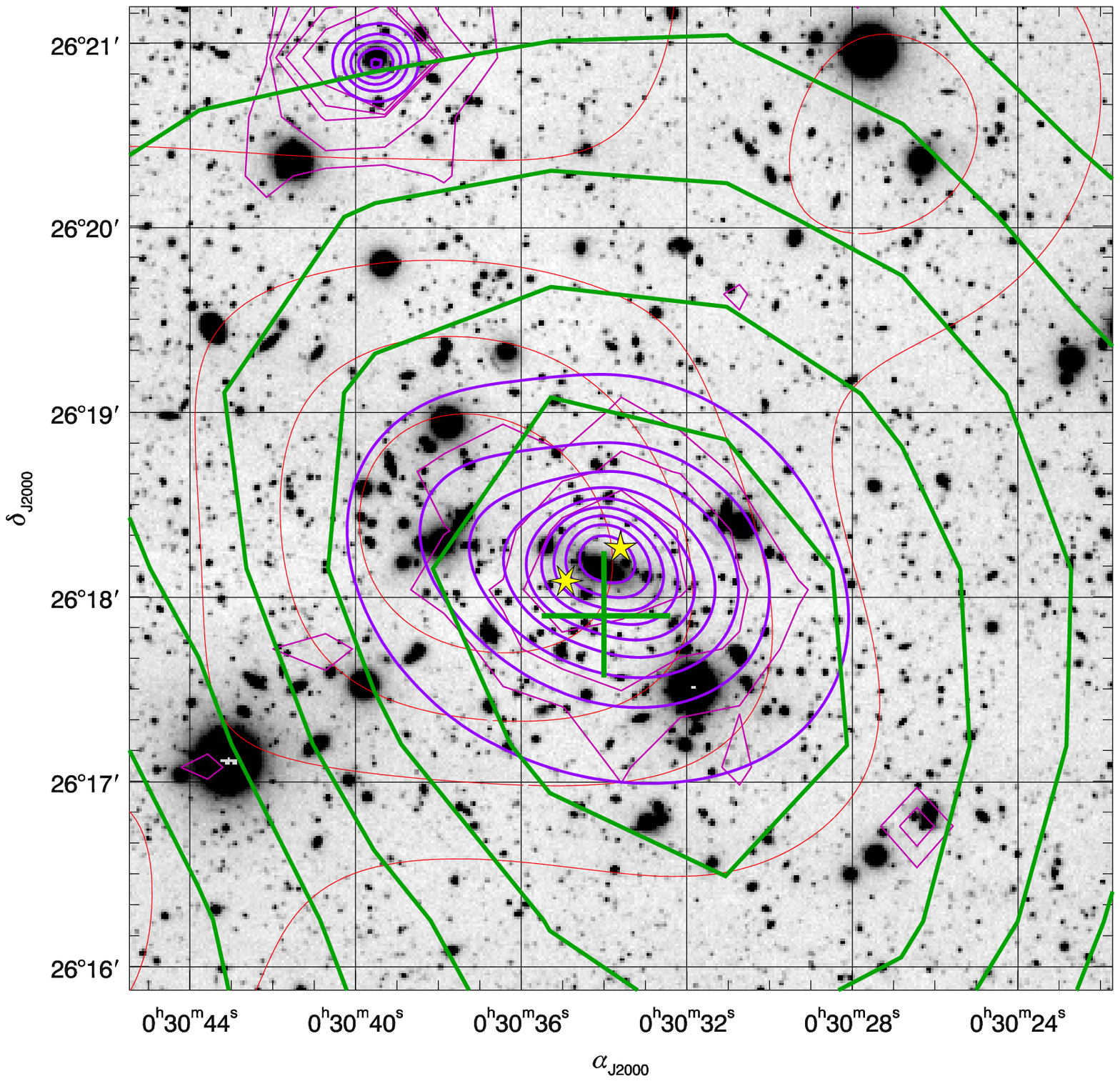}
\caption{Zoomed version of Fig.~\ref{fig:overlay}, showing only the central 
region of \cl00 .
The cross gives the position and $1\sigma$ uncertainty of
the centre position of the $S$-peak obtained by bootstrapping
(cf.\ Sect.~\ref{sec:cent}) while the 
star symbols with five and six points denote the X-ray centres found with
\textsc{ROSAT} and \textsc{Chandra}, respectively.
The source on the northern edge on which strong X-ray emission is centred
is named G4 in Table~\ref{tab:gal}.
}
\label{fig:zoomin}
\end{figure*}
\paragraph{The $S$-statistics lensing centre}
We determine the centre of the \cl00 ~lensing signal and its accuracy
by bootstrap resampling of the galaxy catalogue 
of $N_{\mathrm{cat}}$ galaxies used
in the measurements of the $S$-statistics. From the basic catalogue we draw
$10^{5}$ realisations each containing $N_{\mathrm{cat}}$ sources. 
For each realisation, we determine the $S$-statistics in the central region of
$3\arcmin\!\times\!3\arcmin$ side length ($\sim\!\!1 \mbox{Mpc}$ or roughly the
virial radius of \cl00 ) using a gridsize of $15\arcsec$ and record
the highest $S$-value found on the grid and the grid cell in which it occurs.  

Re-running $10^{5}$ bootstrap realisations of the lensing catalogue with the
centre fixed to the lensing centre, we calculate a detection significance of
$S(\pmb{\theta}^{\mathrm{WL}})\!=\!5.94\pm0.98$. 
\paragraph{Weak Lensing Mass Reconstruction}
In order to get an impression of the (total) mass distribution in \cl00 ,
we perform a finite field mass reconstruction \citep{2001A&A...374..740S}.
This method directly aims at the two-dimensional mass distribution
$\kappa(\pmb{\theta})$ and breaks the \emph{mass-sheet degeneracy},
i.e.\ the fact that the reduced shear, our observable, is invariant under a
transformation $\kappa\!\rightarrow\!\kappa(1-\lambda)+\lambda$
with an arbitrary scalar $\lambda$, 
by assuming $\kappa\!=\!0$ along the border of the field.

The resulting mass map, derived by smoothing the shear field with a scale of 
$2\arcmin$ is shown in Fig.~\ref{fig:overlay}, and a zoomed version displaying 
the central region of \cl00 ~as Fig.~\ref{fig:zoomin}.
The thick contours give the surface mass density.\footnote{The small 
surface mass densities, in contrast to the fact that \cl00 ~likely
has strong lensing arcs (see Sect.~\ref{sec:arcs}), hinting at 
$\kappa\!\gtrsim\!1$ locally, is due to smoothing.} 
Beside the clear main peak of \cl00 , we find a number of smaller additional
peaks whose significance we are going to discuss in the following section.
\paragraph{\textsc{Chandra} and \textsc{XMM-Newton}}
We compare these lensing results to detections by two X-ray observatories,
\textsc{Chandra} and \textsc{XMM-Newton}. For \textsc{Chandra}, we use a
surface brightness map produced from the $58\,\mbox{ks}$ ACIS exposure by
\citet{2009ApJ...692.1033V}
(medium-thick, blue lines in Figs.~\ref{fig:overlay} and \ref{fig:zoomin}).
Using the \citet{2010arXiv1001.0780Z} method, we find the
flux-weighted \textsc{Chandra} centre at 
$00^{\mathrm{h}}30^{\mathrm{m}}34\fs9$, $26\degr18\arcmin05\arcsec$,
slightly off the flux peak at $00^{\mathrm{h}}30^{\mathrm{m}}34\fs0$, 
$26\degr18\arcmin13\arcsec$.

For \textsc{XMM-Newton}, we show detections in the EPIC-MOS2 chip, binned in
$64\times 64$ pixels and smoothed with an adaptive $6\sigma$ Gaussian kernel.
Therefore, the respective contours (medium-thick, magenta
lines in Figs.~\ref{fig:overlay} and \ref{fig:zoomin}) appear more jagged.
\paragraph{Lensing and X-ray Centres}
As can be seen from the cross in Fig.~\ref{fig:zoomin}, the cluster centre
determined with the aperture mass technique falls within the
most significant ($\kappa\!>\!0.05$) convergence contour and is, 
within its $1\sigma$ error ellipse of $24\arcsec\times21\arcsec$,
in good agreement with the flux-weighted \textsc{Chandra} centre of \cl00 , 
separated by $17\arcsec$.
Similar, it is consistent with the \textsc{Rosat} centre just outside the confidence
ellipse and the \textsc{XMM-Newton} contours (~\ref{fig:zoomin}). 
All of these cluster centres are, in turn, within $<\!20\arcsec$ distance from
G1, the BCG.
\paragraph{Optical Galaxy Light}
In addition, we determine the distribution of $r'$-band light from galaxies
by adding the fluxes of all unflagged sources in the \textsl{SExtractor} 
catalogue whose magnitudes and flux radii are consistent with the criteria 
defined for the galaxy catalogue in Sect.~\ref{sec:galcat} and 
Fig.~\ref{fig:rhmag}.\footnote{In absence of a usable half-light radius 
$\vartheta$ 
for the more extended galaxies, we have to substitute flux radii 
$r_{\mathrm{g}}$ here. Using the observed relation between $\vartheta$ and
$r_{\mathrm{g}}$ in our dataset, we consider as galaxies objects with
$r_{\mathrm{g}}\!>\!3.5\,\mbox{px}$ at $16.75\!<\!r'_{\mathrm{AUTO}}\!<\!22.75$
and $r_{\mathrm{g}}\!>\!3.2\,\mbox{px}$ at $r'_{\mathrm{AUTO}}\!>\!22.75$.}
We do so for each pixel of an auxiliary grid, then smoothing it with a 
Gaussian of $60\arcsec$ full-width half-maximum.
In Figs.~\ref{fig:overlay} and \ref{fig:zoomin}, the $r'$-band flux 
density is given in isophotal flux units per \megacam ~pixel, with a flux
of one corresponding to a $r'\!=\!25.8$ galaxy assigned to that pixel
(thin red contours).
There is, amongst others, a discernible $r'$-band flux peak centred between
the galaxies G1 and G2 (Fig.~\ref{fig:zoomin}). 

\subsection{Secondary Peaks} \label{sec:others}

\begin{table}
 \caption{The additional shear, X-ray, and optical flux peaks discussed in 
    Sect.~\ref{sec:others}.}
 \begin{center}
 \begin{tabular}{cccc} \hline\hline
 Peak & $\alpha_{\mathrm{J2000}}$ & $\delta_{\mathrm{J2000}}$ & detected by\\ \hline
 P1 & $00^{\mathrm{h}}31^{\mathrm{m}}02^{\mathrm{s}}$ & $26\degr26\arcmin30\arcsec$ & X-ray, optical \\
 P2 & $00^{\mathrm{h}}31^{\mathrm{m}}19^{\mathrm{s}}$ & $26\degr25\arcmin0\arcsec$ & X-ray, optical \\
 P3 & $00^{\mathrm{h}}29^{\mathrm{m}}31^{\mathrm{s}}$ & $26\degr26\arcmin$ & shear\\
 P4 & $00^{\mathrm{h}}31^{\mathrm{m}}17^{\mathrm{s}}$ & $26\degr18\arcmin$ & shear\\
 P5 & $00^{\mathrm{h}}29^{\mathrm{m}}49^{\mathrm{s}}$ & $26\degr15\arcmin20\arcsec$ & shear\\
 P6 & $00^{\mathrm{h}}30^{\mathrm{m}}54^{\mathrm{s}}$ & $26\degr23\arcmin$ & optical\\
 P7 & $00^{\mathrm{h}}31^{\mathrm{m}}10^{\mathrm{s}}$ & $26\degr21\arcmin15\arcsec$ & optical\\
 P8 & $00^{\mathrm{h}}31^{\mathrm{m}}09^{\mathrm{s}}$ & $26\degr13\arcmin20\arcsec$ & optical\\
 P9 & $00^{\mathrm{h}}31^{\mathrm{m}}14^{\mathrm{s}}$ & $26\degr10\arcmin30\arcsec$ & optical\\
 P10 & $00^{\mathrm{h}}30^{\mathrm{m}}51^{\mathrm{s}}$ & $26\degr07\arcmin30\arcsec$ & optical\\
\hline \hline \label{tab:peaks}
 \end{tabular}
 \end{center}
\end{table}

 The shear peak clearly associated with \cl00 ~is the most dominant signal in 
 the \megacam ~field-of-view, in the lensing $\kappa$-map as
 well as in the X-rays, which can be seen from the \textsc{XMM-Newton} 
 count distribution.
 In the smoothed $r'$-band light distribution, \cl00 ~shows up as a significant
 but not the most prominent peak.
 We have to stress that the background selection using the
 $m_{\mathrm{bright}}$ and $m_{\mathrm{faint}}$ parameters optimise the lensing
 signal for \cl00 , with the likely effect that cluster signals at other 
 redshifts and hence with different photometric properties will be suppressed.
 Keeping this in mind, we compare secondary peaks in the $\kappa$-map to 
 apparent galaxy overdensities, as indicated by the smoothed distribution of 
 $r'$-band light, and to the X-ray detections.

The galaxy listed as G4 in Table~\ref{tab:gal}, a strong X-ray emitter detected
with a high signal by both \textsc{Chandra} and \textsc{XMM-Newton}, 
is identified as a QSO at
redshift $z\!=\!0.493$ by \citet{1997MNRAS.285..511B} and confirmed to be at
$z\!=\!0.492$ by \citet{2001ApJ...548..624C} who found a
significant overdensity of $0.5\ldots2\,\mbox{keV}$ \textsc{Chandra} sources
in the vicinity of \cl00 . Regarding its redshift, it is thus a likely member
of \cl00 .

The \textsc{Chandra} analysis finds two additional sources of 
extended X-ray emission at low surface brightness  
One of them, 
``P1'' in Fig.~\ref{fig:overlay}, (see Table~\ref{tab:peaks} for coordinates
of this and all following peaks) 
is also detected by \textsc{XMM-Newton} and has been identified as a probable 
high-redshift galaxy cluster by \citet{2002A&A...396..397B}
(his candidate \#1 at 
$\alpha_{\mathrm{J2000}}\!=\!00^{\mathrm{h}}31^{\mathrm{m}}01\fs3$, 
$\delta_{\mathrm{J2000}}\!=\!26\degr26\arcmin39\arcsec$) in a deep 
survey for galaxy clusters using pointed \textsc{Chandra} observations.
In the $\kappa$ map, contours near the north-eastern corner of \megacam 's
field-of-view extend close to the position of this cluster, but their 
significance near this corner and close to the bright star BD+25 65 is
doubtful.
The \megacam ~images show a small grouping\footnote{Not visible in
Fig.~\ref{fig:overlay} due to its binning.} of $r'\!\approx\!21$
galaxies with similar colour in the three-colour composite at the position of
``P1''.

The other \textsc{Chandra} peak, ``P2'' 
is located near a prominent peak in the $r'$-band light,
but with a strong contribution from a single bright galaxy within its 
$60\arcsec$ smoothing radius. It
does not correspond to a tabulated source in either
NED\footnote{NASA-IPAC Extragalatic Database:\\ \texttt{http://nedwww.ipac.caltech.edu/}}
or SIMBAD\footnote{\texttt{http://simbad.u-strasbg.fr/simbad/}}.
We do not notice a significant surface mass density from lensing at this
position, but have to stress again that a possible signal might have been
downweighted by the catalogue selection process. 

Most peaks in the $\kappa$ map, apart from the one associated with \cl00 ,
are located at a distance smaller than the $2\arcmin$ smoothing scale from the 
edges of the field, likely due to noise amplification by missing information.  
Amongst them, only the second strongest $\kappa$ peak, ``P3'' 
seems possibly associated with an overdensity of galaxies, but the coverage is
insufficient to draw further conclusions.

For a shear peak ``P4'' 
close to several 
\textsc{Chandra} and \textsc{XMM-Newton} peaks, there also is an enhancement 
in $r'$-band  flux, while galaxies do not appear concentrated. 
Likewise, the high flux density 
close to a possible shear peak ``P5'' 
seems to be caused by a single, bright galaxy.

On the other hand, we notice agglomerations of galaxies 
(``P6'' to ``P8'') with a cluster-like or
group-like appearance that show neither X-ray nor lensing signal. 
For ``P7'', the nearby \textsc{XMM-Newton} signal is the distant quasar named 
I3 by \citet{2000AJ....119.2349B}.
The two strong $r'$-band flux overdensities 
``P9'' and ``P10'' in the south-east corner of the \megacam ~image
appear to be poor, nearby groups of galaxies.

\subsection{Arc-like Features in \cl00 } \label{sec:arcs}

We note that, being a massive cluster of galaxies, \cl00 ~is a probable strong
gravitational lens, leading to the formation of giant arcs.
Indeed, we identify two tentative strong lensing features in our deep \megacam
~exposures. 
The first is a very prominent, highly elongated arc 
$\sim\!20\arcsec$ west from the BCG (Fig.~\ref{fig:img}). 
Its centre is at 
$\alpha_{\mathrm{J2000}}\!=\!00^{\mathrm{h}}30^{\mathrm{m}}32\fs7$ and 
$\delta_{\mathrm{J2000}}\!=\!26\degr18\arcmin05\arcsec$; its length is $>\!20\arcsec$.
The giant arc is not circular but apparently bent around a nearby galaxy.

The second feature possibly due to strong lensing is located near galaxy G3
which appears to be an elliptical. With the centre of the tentative arc at
$\alpha_{\mathrm{J2000}}\!=\!00^{\mathrm{h}}30^{\mathrm{m}}36\fs5$ and
$\delta_{\mathrm{J2000}}\!=\!26\degr19\arcmin14\arcsec$, it is bent around the centre of the
galaxy forming the segment of a circle with $\sim\!6\arcsec$ radius.
Thus, an alternative explanation might be that the arc-like feature corresponds
to a spiral arm of the close-by galaxy. However, this seems less likely 
given its appearance in the \megacam ~images.
If this arc is due to gravitational lensing it is likely to be strongly 
influenced by the gravitational field of the aforementioned galaxy as it is
opening to the opposite side of the cluster centre.

Whether these two candidate arcs are indeed strong lensing features in \cl00 
~will have to be confirmed by spectroscopy.


\section{Mass Determination and Discussion} \label{sec:disc}

We analyse the tangential shear profile 
$g_{\mathrm{t}}(\theta)\!=\!\langle \varepsilon_{\mathrm{t}}(\theta)\rangle$, i.e., 
the averaged tangential component
of $\varepsilon$ with respect to the weak lensing centre of \cl00 ~found in
Sect.~\ref{sec:cent} as a function of the separation $\theta$ to this centre.
At this point, we also introduce the shear calibration factor, $f_{0}\!=\!1.08$,
an empirical correction to the shear recovery by our KSB method and catalogue
selection (cf.\ Sect.~\ref{sec:veri}), and the contamination correction factor
$f_{1}(\theta)$ we will specify in Sect.~\ref{sec:fcg}, thus replacing 
$\varepsilon$ by $f_{0}f_{1}(\theta)\varepsilon$. 
First, the \citet[NFW]{1997ApJ...490..493N} shear profile will be introduced.

\begin{figure}
\resizebox{\hsize}{!}{\includegraphics[angle=90]{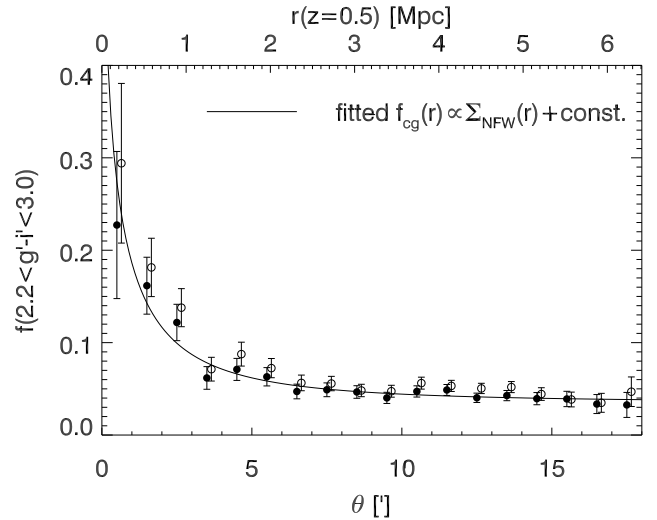}}
\caption{The fraction of ``red sequence-like'' galaxies 
 $2.2\!<\!g'\!-\!i'\!<\!3.0$ as a function of
 clustercentric distance before (open symbols) and after (filled symbols)
 background selection. The solid line denotes the best fitting sum 
 $f_{\mathrm{cg}}$ of a NFW surface mass profile and a constant to the latter.
 We use $f_{1}\!=\!f_{\mathrm{cg}}\!+\!1$ as a correction factor for cluster
 contamination in Sect.~\ref{sec:res}.}
\label{fig:fcg}
\end{figure}

\begin{figure}[ht!]
\sidecaption
\resizebox{\hsize}{!}{\includegraphics[width=0.25\textwidth,angle=90]{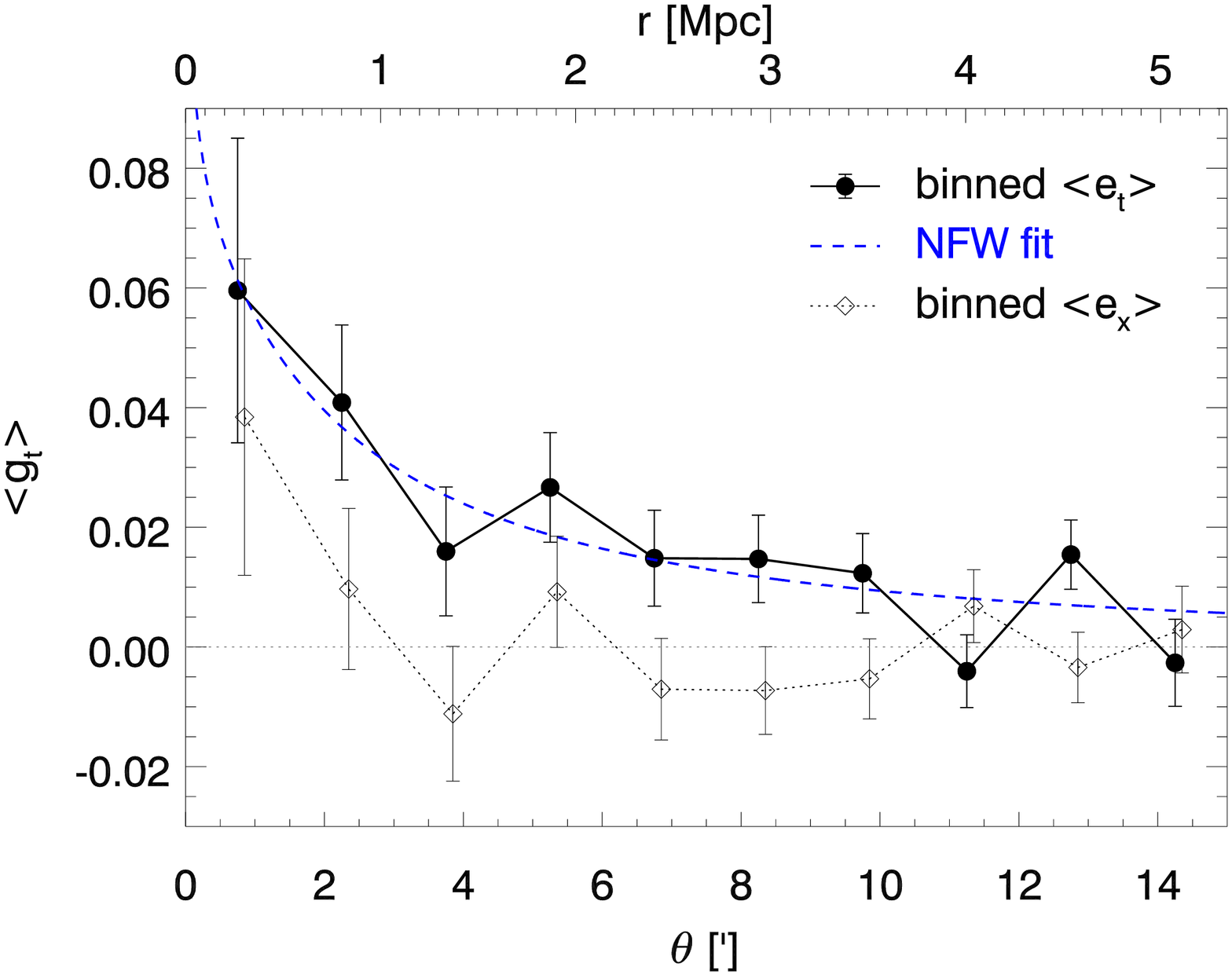}}
\resizebox{\hsize}{!}{\includegraphics[width=0.25\textwidth,angle=90]{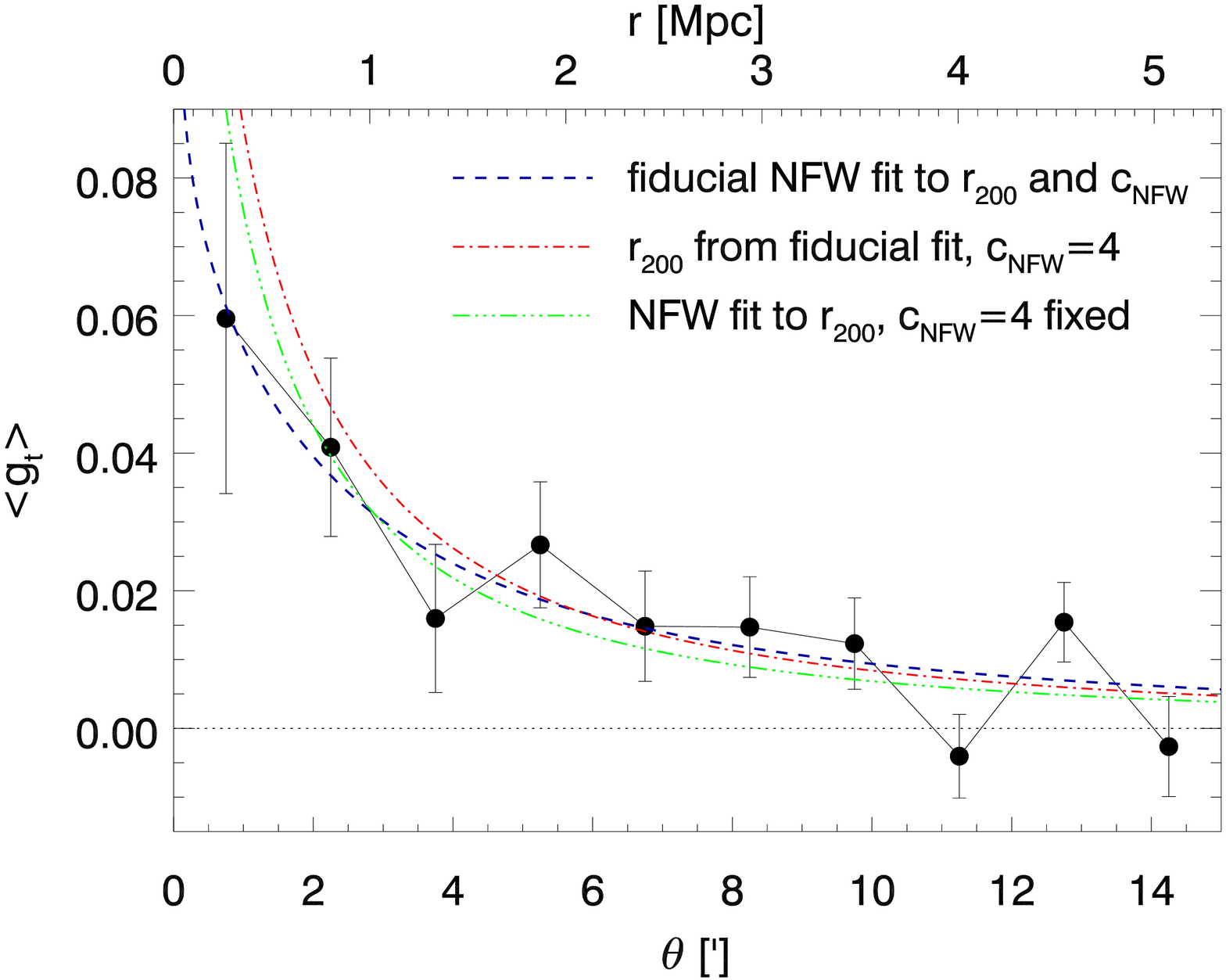}}
\resizebox{\hsize}{!}{\includegraphics[width=0.25\textwidth,angle=90]{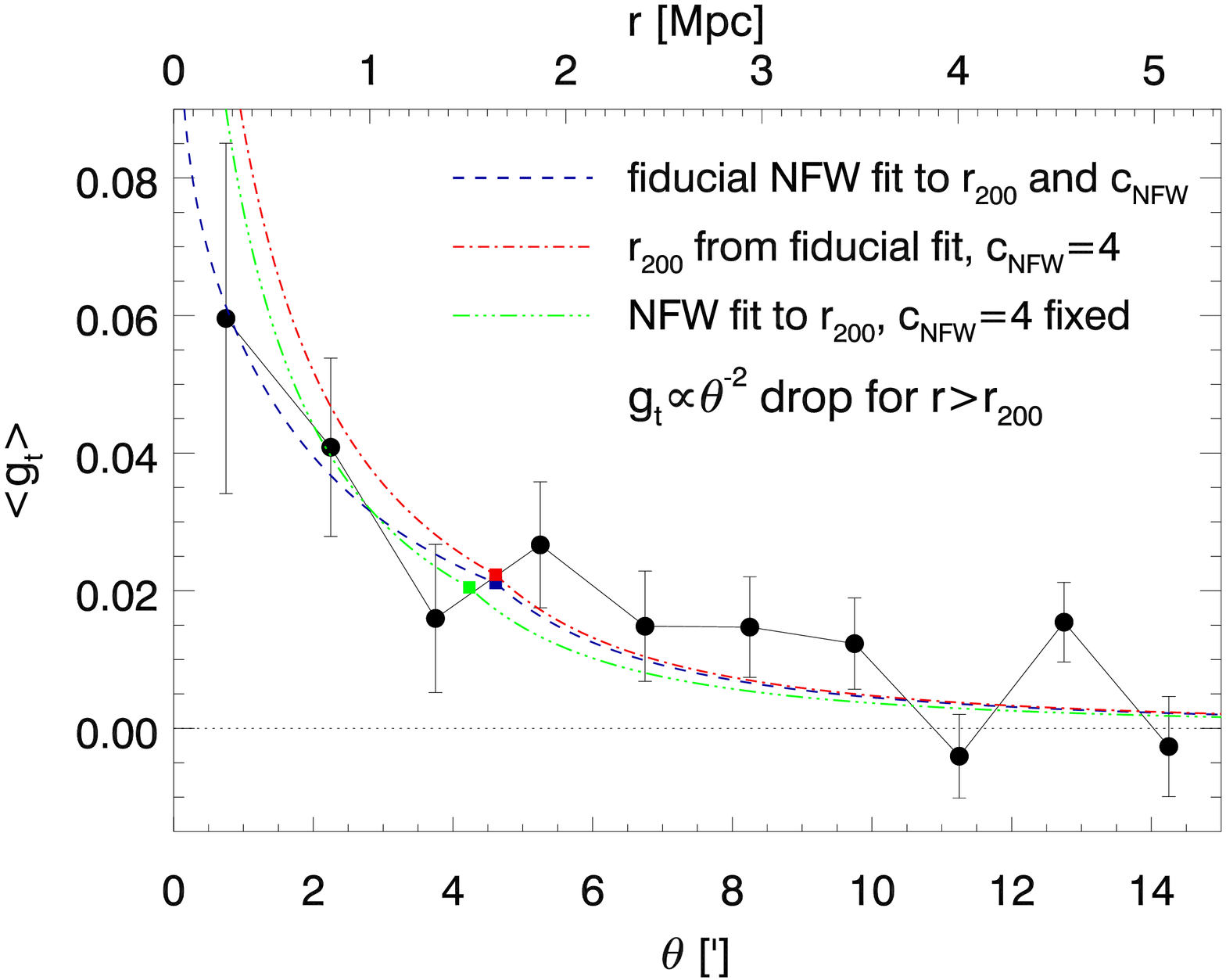}}
\caption{The tangential shear profile of \cl00 , averaged in bins of 
$90\arcsec$ width (solid line with dots).
\textit{Upper panel: }
The best fitting NFW model in the fiducial case (see text; dashed line) and the
binned cross-component $g_{\times}$ of the measured shear
(dotted line with diamonds).
Error bars give the standard deviation of measured values in the
resp. bin.
\textit{Middle panel: } NFW models with $r_{200}$ from the fiducial fit and
concentration set to $c_{\mathrm{NFW}}\!=\!4.0$ (dot-dashed line) and fitting 
to $r_{200}$ only keeping $c_{\mathrm{NFW}}\!=\!4.0$ fixed (triple dot-dashed 
line), compared to the fiducial fit (dashed line).
\textit{Lower panel: } The same models as in the middle panel, but all truncated
at $r_{200}$, with a $g_{\mathrm{t}}\!\propto\theta^{-2}$ drop-off at larger radii.
}
\label{fig:profile}
\end{figure}

\subsection{The NFW model} 

To derive an estimator for the mass of \cl00 ~from the weak lensing data,
we fit the tangential shear profile $g_{\mathrm{t}}(\theta)$ with a 
NFW profile \citep[e.g.][]{1996A&A...313..697B,2000ApJ...534...34W}. 
The NFW density profile 
has two free parameters\footnote{While 
\citet{1997ApJ...490..493N} originally designed their profile as a 
single-parameter model, we follow the usual approach in weak lensing studies 
of expressing the NFW profile in terms of two independent parameters.}, 
the radius $r_{200}$ inside
which the mean density of matter exceeds the critical
mass density $\rho_{\mathrm{c}}$
by a factor of $200$ 
and the concentration parameter $c_{\mathrm{NFW}}$
from which the characteristic overdensity $\delta_{\mathrm{c}}$ can be computed.

The overdensity radius $r_{200}$ being an estimator for the cluster's virial
radius, we define as the mass of the cluster the mass enclosed within 
$r_{200}$, given by:
\begin{equation}
M_{200}\!=\!200 \frac{4\pi}{3} \rho_{\mathrm{c}} r_{200}^{3}  .
 \label{eq:mass}
\end{equation}
%
The reduced shear observable is:
\begin{equation} \label{eq:gnfw}
g_{\mathrm{NFW}}(u)\!=\!\frac{\gamma_{\mathrm{NFW}}(u)}{1-\kappa_{\mathrm{NFW}}(u)}
\end{equation}
where the dimensionless radial distance 
$u\!=\!c_{\mathrm{NFW}} D_{\mathrm{d}}\theta r_{200}^{-1}$ contains the
angular separation $\theta$ and the angular diameter distance $D_{\mathrm{d}}$
between lens and observer. 
The $\gamma_{\mathrm{NFW}}(u)$ and $\kappa_{\mathrm{NFW}}(u)$ profiles are given
in \citet{2000ApJ...534...34W}.
The critical surface mass density 
\begin{equation}
\Sigma_{\mathrm{c}}\!=\!\frac{c^{2}}{4\pi G D_{\mathrm{d}}}
\Bigg\langle\frac{D_{\mathrm{ds}}}{D_{\mathrm{s}}}\Bigg\rangle^{-1}
\label{eq:sigmac}
\end{equation}
depends on $D_{\mathrm{d}}$ and the mean ratio 
$\langle D_{\mathrm{ds}}/D_{\mathrm{s}}\rangle$
of angular diameter distances between source and observer and source and lens.
%

\subsection{Contamination by Cluster Galaxies} \label{sec:fcg}

In addition to the background selection based on $g'\!-\!r'$ and
$r'\!-\!i'$ colours we estimate the remaining fraction of cluster galaxies in
the catalogue using the $g'\!-\!i'$ index. We will use this to devise a
correction factor accounting for the shear dilution by (unsheared) cluster
members.
As discussed in Sect.~\ref{sec:redseq}, the colour-magnitude diagramme of the
\cl00 ~field (Fig.~\ref{fig:cmdccd}) does not show a clear-cut cluster red
sequence, but a broad distribution in $g'\!-\!i'$, indicating two redshift
components. We therefore define a wide region $2.2\!<\!g'\!-\!i'\!<\!3.0$
of \emph{possible} red sequence sources, including galaxies with colours
similar to the $z\!=\!0.50$ CWW elliptical template but redder than the
$z\!=\!0.25$ one (cf.\ Table~\ref{tab:galcww}). As this definition of 
``red sequence-like'' galaxies is meant to encompass all early-type cluster 
members, it will also contain 
background systems, giving an
\emph{upper limit} for the actual contamination in the catalogue.

Figure~\ref{fig:fcg} shows the fraction of sources $2.2\!<\!g'\!-\!i'\!<\!3.0$
in the galaxy catalogue before (open symbols) and after (filled symbols) the 
final cut based on $m_{\mathrm{bright}}$ and $m_{\mathrm{faint}}$ has been applied
as a function of distance to the centre of \cl00 ~as determined by lensing
(Sect.~\ref{sec:cent}). 
Error bars give the propagated Poissonian uncertainties in the counts.
We note a strong increase of the number of ``red sequence-like'' systems
compared to the overall number of galaxies towards the cluster centre, 
indicating that a large fraction of those are indeed cluster members.
Most intriguingly, the background selection seems to remove only few of these
tentative cluster members, with the fractions before and after selection
consistent within their mutual uncertainties at all radii.
This finding can be explained to a large extent by galaxies too faint to be
removed by the background selection criterion:
If background selection is extended to the faintest magnitudes
($m_{\mathrm{faint}}\!=\!29$),
no significant overdensity of ``red sequence-like'' galaxies at the position
of \cl00 ~is detected. Although using a different selection method, this modest
effect of background selection is in agreement with \citet{2007MNRAS.379..317H}.

By repeating this analysis centred on several random position in our field
and not finding a significant increase of the ``red sequence-like'' fraction
towards these positions we show that the peak around the position of \cl00 ~is
indeed caused by concentration of these galaxies towards the cluster centre.

We find the residual contamination to be well represented by the sum 
$f_{\mathrm{cg}}(\theta)\!=\!f_{\mathrm{cg}}^{\mathrm{NFW}}(\theta)\!+\!f_{\mathrm{cg}}^{0}$
of a NFW surface mass profile and a constant
(solid line in Fig.~\ref{fig:fcg}).
We follow the approach of \citet{2007MNRAS.379..317H} and define a radially
dependent factor $f_{1}(\theta)\!=\!f_{\mathrm{cg}}^{\mathrm{NFW}}(\theta)\!+\!1$ correcting 
for the residual contamination.
Here we take into account only the NFW component $f_{\mathrm{cg}}^{\mathrm{NFW}}(\theta)$
of the fit, as the offset $f_{\mathrm{cg}}^{0}$ represents a population of field galaxies,
and not diluting cluster members.
This correction factor scales up the shear
estimates close to the cluster centre, counterweighing the dilution by the
larger number of cluster members there.

\subsection{Mass Modelling of \cl00 } \label{sec:res}

\subsubsection{Fits to the Ellipticity Profile} \label{sec:fits}
\begin{table}
 \centering
 \caption{Properties of the fiducial model combining the parameter values and
  assumptions going into the NFW modelling.}
 \begin{tabular}{ccc}\hline\hline
   Parameter & Value & see Sect. \\ \hline
   $\max(|\varepsilon|)$ & $0.8$ & \ref{sec:veri} \\
   $\min(\nu)$ & $4.5$ & \ref{sec:veri} \\
   $\min(\tr\tens{P}^{\mathrm{g}})$ & $0.1$ & \ref{sec:veri} \\
   $m_{\mathrm{bright}}$ & $20.0$ & \ref{sec:colcolsel} \\
   $m_{\mathrm{faint}}$ & $22.5$ & \ref{sec:colcolsel} \\
   centre & from $S$-statistics & \ref{sec:cent} \\
   radial fit range & $0\arcmin\!<\!\theta\!<\!15\arcmin$ & \ref{sec:fits} \\ \hline
   $f_{0}$ & $1.08\pm0.05$ & \ref{sec:veri} \\
   $f_{1}(\theta)$ & $f_{\mathrm{cg}}^{\mathrm{NFW}}(\theta)+1$ & \ref{sec:fcg} \\
   $\langle D_{\mathrm{ds}}/D_{\mathrm{s}}\rangle$ & $0.33\pm0.03$ & \ref{sec:ddsds} \\ \hline \hline
 \end{tabular}
 \label{tab:fiducial}
\end{table} 

In Fig.~\ref{fig:profile}, there
is a discernible positive tangential alignment signal extending out
to $\sim\!\!10\arcmin$ or $\sim\!\!3.5\,\mbox{Mpc}$) from the cluster centre.
(The solid line and dots in all panels give the
shear averaged in bins of $90\arcsec$ width.) 
In order to validate that this tangential alignment is indeed caused by 
gravitational shear of a cluster-like halo,
we fit the NFW reduced shear profile given in Eq.~(\ref{eq:gnfw})
to the measured shear estimates, probing the range 
$0\arcmin\!<\!\theta\!<\!15\arcmin$. 

We define a \emph{fiducial model} using the preferred
parameter values presented in Table~\ref{tab:fiducial}.
The table also lists references to the sections where these values are 
justified. 
In order to determine $r_{200}$ and $c_{\mathrm{NFW}}$, we fit an NFW model to 
the shear estimates of the \emph{lensing catalogue} galaxies, defined by the 
parameters above the vertical line in Table~\ref{tab:fiducial}.
Parameters below the line do not affect the catalogue but influence the
relation between shear and cluster mass.

The fitting is done by minimising $\chi^{2}$ using 
an \texttt{IDL} implementation of the Levenberg-Marquardt algorithm
\citep{1978LNM..630..105M,2009ASPC..411..251M} 
and returning $r_{200}^{\mathrm{fit}}\!=\!1.64\!\pm\!0.16\,\mbox{Mpc}$ and 
$c_{\mathrm{NFW}}^{\mathrm{fit}}\!=\!2.1\!\pm\!1.1$ for the free parameters of the model.
Comparing the best-fitting NFW model (dashed curve in the upper and middle 
panels of Fig.~\ref{fig:profile}) to the data, we find
the shear profile to be reasonably well-modelled by an NFW profile: we measure
$\chi^{2}/\nu_{\mathrm{dof}}\!=\!13404/13636\!\approx\!0.98$, assuming an error
\begin{equation}
\sigma_{\mathrm{fit}}\!=\!f_{1}(\theta)\sigma_{\mathrm{gt}}\quad,\quad
\sigma_{\mathrm{gt}}\!=\!f_{0}\sigma_{\varepsilon}/\!\sqrt{2}\!\approx\!0.29
\end{equation}
for the individual shear estimate. 
This overall agreement with NFW is
consistent with shear profiles of clusters with comparable redshift and data
quality \citep{2006A&A...451..395C}. We discuss the NFW parameter values 
obtained by the fit and the radial range over which the NFW fit is valid 
(the middle and lower panels of Fig.~\ref{fig:profile}) in
Sect.~\ref{sec:disc}.

Gravitational lensing by a single axially symmetric deflector causes  
tangential alignment of the resulting ellipticities. Thus, the ellipticity 
cross-component $g_{\times}\!=\!\langle e_{\times}(\theta)\rangle$ corresponding
to a pure curl field around the centre 
should be consistent with zero at all $\theta$. The dotted line and diamonds in
the upper panel of Figure~\ref{fig:profile} show that $g_{\times}$ is indeed 
consistent or nearly consistent with zero in its error bars in all bins but the 
innermost $90\arcsec$.
This feature is, like the general shapes of both $g_{\mathrm{t}}$ and 
$g_{\times}$, quite robust against the choice of binning.
A tentative explanation for the higher $g_{\times}$ in the central bin 
might be additional lensing by 
the foreground mass concentration associated with the $z\!\approx\!0.25$
galaxies (cf.\ Sect.~\ref{sec:colours}), centred to the East of \cl00 .

To further investigate this hypothesis, we split up the ellipticity
catalogue into an eastern 
($\alpha_{\mathrm{J2000}}\!>\!\alpha_{\mathrm{CL0030}}$) and western 
($\alpha_{\mathrm{J2000}}\!<\!\alpha_{\mathrm{CL0030}}$) subset 
(with $50.0$~\% of the galaxies in each) and repeat the profile fitting for
both of them separately, as the influence of a possible
perturber at the position of G2 should be small compared to the eastern 
sub-catalogue.
In accordance with the mass distribution displayed in Fig.~\ref{fig:overlay}, in
which a higher and more extended surface mass density can be found west of the
centre of \cl00 ~than east of it, the $g_{\mathrm{t}}$ signal is more significant
in the sources lying to the West of the cluster than to the East. We find
$r_{200,\mathrm{W}}^{\mathrm{fit}}\!=\!1.82\pm0.22\,\mbox{Mpc}$,
$c_{\mathrm{NFW,W}}^{\mathrm{fit}}\!=\!2.1\pm1.2$, and
$r_{200,\mathrm{E}}^{\mathrm{fit}}\!=\!1.47\pm0.25\,\mbox{Mpc}$,
$c_{\mathrm{NFW,E}}^{\mathrm{fit}}\!=\!1.5\pm1.4$.
The cross components in the central bins of both subsets are similarly high
than in the complete catalogue with the eastern half also showing 
a high $g_{\times}$ in the second bin.
As the values for $r_{200}$ from the two sub-catalogues are consistent
given their uncertainties,
we find no clear indications for a significant impact of the foreground
structure. The inconspicuous lensing signal is consistent with the 
inconspicuous X-ray signal.

The deviation of $g_{\times}$ from zero by $\sim\!1.5\sigma$ in the
central bin, out of the $10$ bins we probe, is not unexpected and does thus 
not pose a severe problem for the interpretation of our results with respect
to $c_{\mathrm{NFW}}$ (Sect.~\ref{sec:cnfw}).

In a further test, we repeated the analysis centred on G1, the brightest 
cluster galaxy and found very similar results in terms of shapes of 
$g_{\mathrm{t}}$ and $g_{\times}$ and fit parameters.

\subsubsection{Likelihood analysis}

\begin{figure*}
\sidecaption
\includegraphics[angle=90,width=12cm]{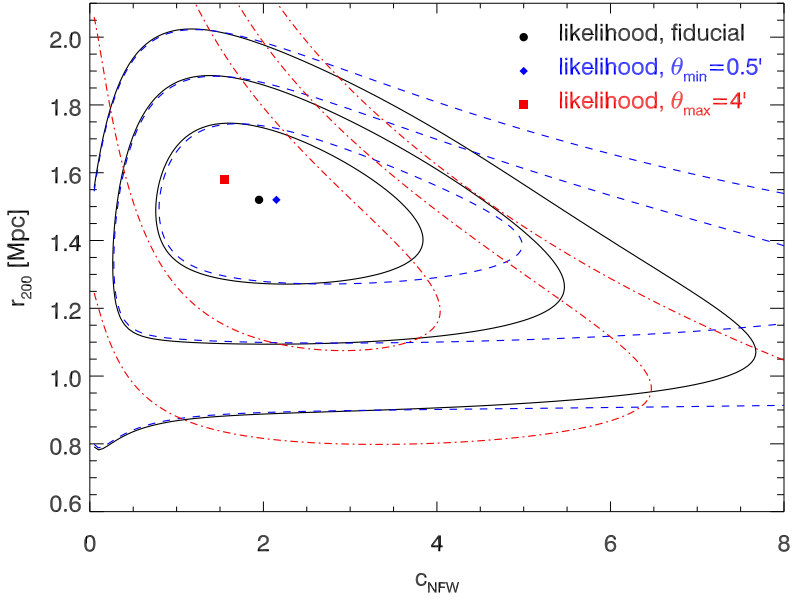}
\caption{Confidence contours in the NFW parameter space spanned by the 
virial radius $r_{200}$ and concentration $c_{\mathrm{NFW}}$, 
corresponding to confidence
levels of $99.73$\%, $95.4$\%, and $68.3$\%. Also given as symbols 
are the maximum likelihood cluster parameters for three different 
cases. They are: the fiducial model, using data in the complete range
$0\arcmin\!\leq\!\theta\!\leq\!15\arcmin$ (solid contours and dot); a model
where the central $30\arcsec$ are excised from the analysis (dashed contours
and diamond); and a model considering only data inside 
$0\arcmin\!\leq\!\theta\!\leq\!4\arcmin\!\approx\!r_{200}$
(dash-dotted contours and square).}
\label{fig:rccont}
\end{figure*}

While shear profiles serve well to investigate the agreement between a cluster
shear signal and a mass distribution like NFW, there are better methods to
infer model parameters, and hence the total cluster mass, than fitting 
techniques. Knowledge of the likelihood function
\begin{equation}
\mathcal{L}\!=\!\exp{\left(-\frac{1}{2}\left(\chi^{2}_{\mathrm{L}}(r_{200},
c_{\mathrm{NFW}})\!-\!\chi^{2}_{\mathrm{L}}(r_{200}^{\mathrm{min}},
c_{\mathrm{NFW}}^{\mathrm{min}})\right)\right)}
\end{equation}
allows us to quantify the uncertainties in the model parameters given the data
and -- an important advantage over fitting methods -- also their 
interdependence.
We evaluate the consistency between the tangential reduced shear 
$g_{\mathrm{t},i}(r_{200},c_{\mathrm{NFW}})$
predicted by an NFW model for the $i$-th sample galaxy and the 
tangential ellipticity component $\varepsilon_{\mathrm{t},i}$ from the data
by considering the function
\begin{equation}
\chi^{2}_{\mathrm{L}}\!=\!\sum_{i=1}^{N_{\mathrm{gal}}}{\frac{
\left|g_{\mathrm{t},i}(r_{200},c_{\mathrm{NFW}})\!-\!\varepsilon_{\mathrm{t},i}\right|^{2}}
{\sigma_{\mathrm{fit}}^{2}\left(1\!-\!\left|g_{\mathrm{t},i}(r_{200},c_{\mathrm{NFW}})\right|^{2}\right)^{2}}}
\end{equation}
\citep[for a derivation, see][]{2000A&A...353...41S} 
which we compute for a suitable grid of test 
parameters $r_{200}$ and $c_{\mathrm{NFW}}$ and determine the values 
$r_{200}^{\mathrm{min}}$ and $c_{\mathrm{NFW}}^{\mathrm{min}}$ for which 
$\chi^{2}_{\mathrm{L}}$ becomes minimal. 
The likelihood approach also allows us to introduce a more accurate noise 
estimator 
$\sigma^{2}_{\mathrm{SKE}}\!=\!\sigma_{\mathrm{fit}}^{2}(1\!-\!|g_{\mathrm{t},i}|^{2})^{2}$
for the individual shear estimate, taking into account the dependence of the 
noise on the shear value itself, as expected by the model.\footnote{Use of
this error model is denoted writing $\chi^{2}_{\mathrm{L}}$ instead of $\chi^{2}$.}

In Fig.~\ref{fig:rccont}, we present the regions corresponding to confidence
intervals of  $68.3$\%, $95.4$\%, and $99.73$\% in the 
$r_{200}$-$c_{\mathrm{NFW}}$-parameter space for three
radial ranges in which data galaxies are considered.
The solid curves denote the fiducial model 
with the complete $0\arcmin\!\leq\!\theta\!\leq\!15\arcmin$ range, giving
$r_{200}^{\mathrm{min}}\!=\!1.52^{+0.22}_{-0.24}\,\mbox{Mpc}$ and  
$c_{\mathrm{NFW}}^{\mathrm{min}}\!=\!2.0^{+1.8}_{-1.2}$. We will adopt these as the
fiducial results of our analysis (see Table~\ref{tab:param}), 
yielding a cluster mass of
$M_{200}(r_{200}^{\mathrm{min}})\!=\!7.2^{+3.6}_{-2.9}\times10^{14} \mathrm{M}_{\sun}$ 
(statistical uncertainties) by applying Eq.~(\ref{eq:mass}).
In the following, we will
discuss variations to this fiducial case (e.g.\ the other contours in
Fig.~\ref{fig:rccont}).
\begin{table} 
 \caption{Parameters resulting from NFW modelling of \cl00 ~for 
  models relying on different assumptions.
}
 \begin{center}
 \begin{tabular}{cccccc} \hline\hline
 Model & $r_{200}^{\mathrm{min}}/\mbox{Mpc}$ & $c_{\mathrm{NFW}}^{\mathrm{min}}$ & 
$M_{200}(r_{200}^{\mathrm{min}})^{\dagger}$ & $\mu^{\ddagger}$ \\ \hline
fiducial & $1.52_{-0.24}^{+0.22}$ & $2.0_{-1.2}^{+1.8}$ & 
$7.2\!\times\!10^{14} \mathrm{M}_{\sun}$ & $--$\\
$0.5\arcmin\!\leq\!\theta\!\leq\!15\arcmin$ & $1.52_{-0.24}^{+0.22}$ & $2.2_{-1.4}^{+2.8}$ & 
$7.2\!\times\!10^{14} \mathrm{M}_{\sun}$ & $1.00$ \\ 
$0\arcmin\!\leq\!\theta\!\leq\!4\arcmin$ & $1.58_{-0.50}^{+1.85}$ & $1.6_{-1.6}^{+2.4}$ & 
$8.1\!\times\!10^{14} \mathrm{M}_{\sun}$ & $1.12$ \\ \hline
$\max(|\varepsilon|)\!=\!1.0$ & $1.43_{-0.26}^{+0.24}$ & $1.9_{-1.3}^{+2.0}$ &
$6.0\!\times\!10^{14} \mathrm{M}_{\sun}$ & $0.83$ \\
$\max(|\varepsilon|)\!=\!10^{4}$ & $1.32_{-0.39}^{+0.29}$ & $1.5_{-1.3}^{+2.6}$ &
$4.7\!\times\!10^{14} \mathrm{M}_{\sun}$ & $0.65$ \\
centred on BCG & $1.51_{-0.24}^{+0.23}$ & $1.7_{-1.1}^{+1.9}$ &
$7.1\!\times\!10^{14} \mathrm{M}_{\sun}$ & $0.98$ \\
no contam. corr. & $1.50_{-0.24}^{+0.22}$ & $1.7_{-1.0}^{+1.6}$ &
$7.0\!\times\!10^{14} \mathrm{M}_{\sun}$ & $0.96$ \\
$f_{0}\!=\!0.88$ & $1.41_{-0.22}^{+0.19}$ & $1.9_{-1.2}^{+1.8}$ & 
$5.8\!\times\!10^{14} \mathrm{M}_{\sun}$ & $0.80$ \\
$f_{0}\!=\!1.13$ & $1.54_{-0.25}^{+0.23}$ & $2.0_{-1.2}^{+1.9}$ & 
$7.5\!\times\!10^{14} \mathrm{M}_{\sun}$ & $1.04$ \\
$\langle D_{\mathrm{ds}}/D_{\mathrm{s}}\rangle\!=\!0.30$ & $1.59_{-0.25}^{+0.24}$ &$2.1_{-1.3}^{+1.9}$ &
$8.3\!\times\!10^{14} \mathrm{M}_{\sun}$ & $1.14$ \\
$\langle D_{\mathrm{ds}}/D_{\mathrm{s}}\rangle\!=\!0.36$ & $1.45_{-0.23}^{+0.21}$ &$1.9_{-1.2}^{+1.8}$ & 
$6.3\!\times\!10^{14} \mathrm{M}_{\sun}$ & $0.87$ \\
\hline \hline \label{tab:param}
 \end{tabular}
 \begin{minipage}{80mm}
  \smallskip
  $^{\dagger}$ From Eq.~(\ref{eq:mass}). \\
  $^{\ddagger}$ Mass $\mu\!=\!M_{200}/M_{200}^{\mathrm{fid}}$ 
 in units of the fiducial cluster mass.
 \end{minipage}
 \end{center}
\end{table}

\begin{figure*}
\sidecaption
\includegraphics[angle=90,width=12cm]{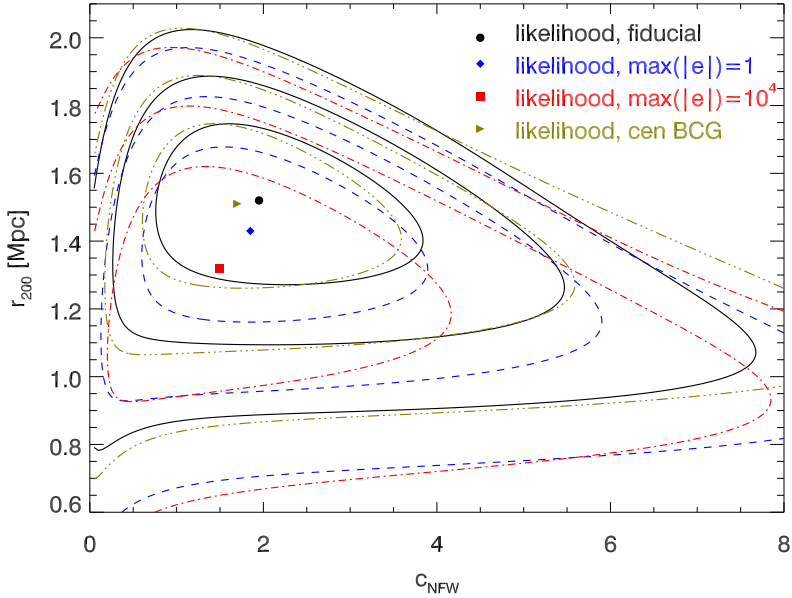}
\caption{Confidence contours and values of $r_{200}$ and $c_{\mathrm{NFW}}$ 
maximising the likelihood in
dependence of the maximum shear estimator $\max(|\varepsilon|)$ permitted in
the catalogue: Given are the fiducial case ($\max(|\varepsilon|)\!=\!0.8$,
solid contours and dot, see Sect.~\ref{sec:veri});
$\max(|\varepsilon|)\!=\!1.0$ (dashed contours and diamond); and
$\max(|\varepsilon|)\!=\!10^{4}$, equivalent to no cut (dot-dashed contours
and square). Triple dot-dashed contours and the triangle denote the results for
an otherwise fiducial model centred on the BCG of \cl00 .}
\label{fig:rc2}
\end{figure*}
\begin{figure*}
\sidecaption
\includegraphics[angle=90,width=12cm]{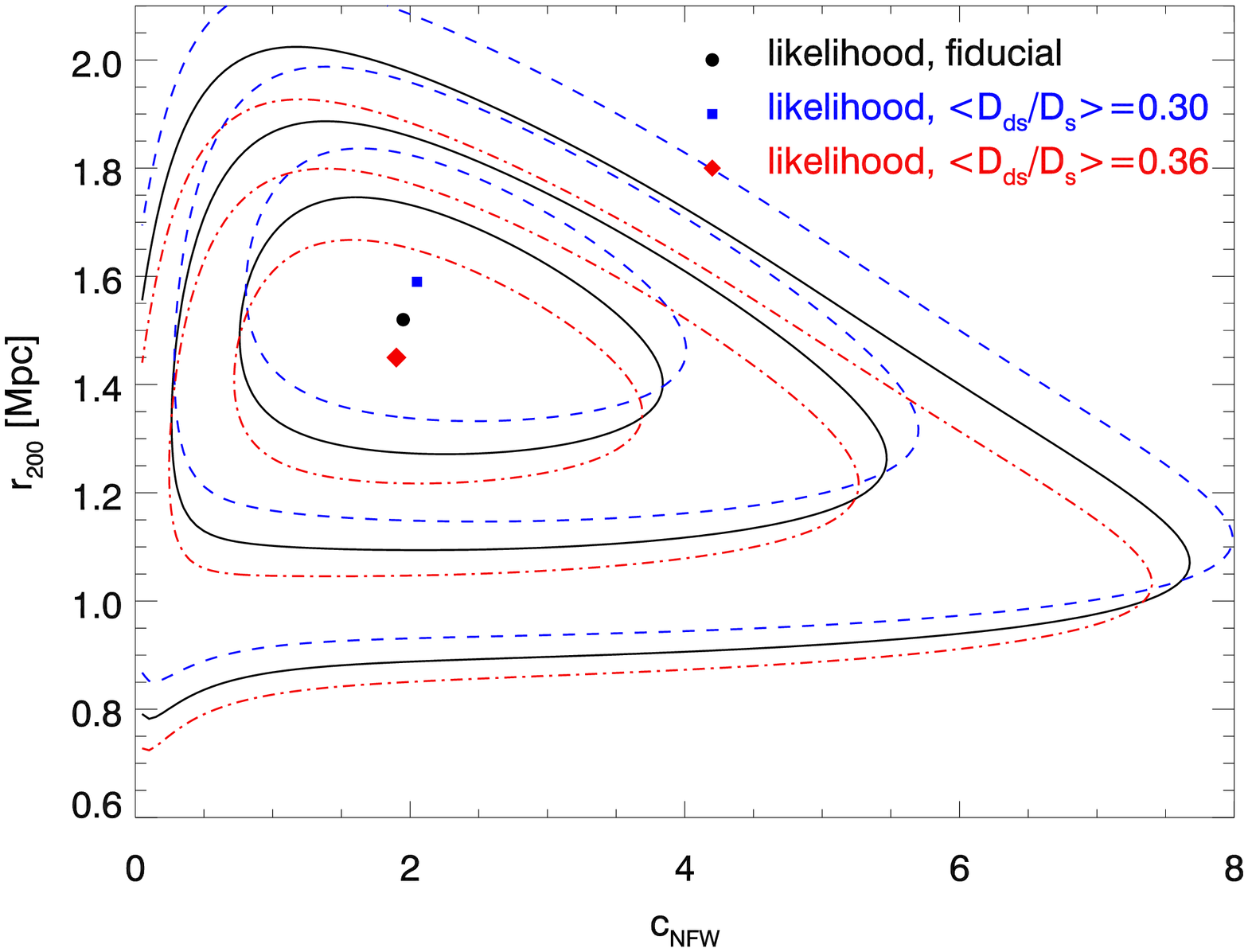}
\caption{Confidence contours and values of $r_{200}$ and $c_{\mathrm{NFW}}$ 
maximising the likelihood for
models of different geometric factor: The fiducial case,
$\langle D_{\mathrm{ds}}/D_{\mathrm{s}}\rangle\!=\!0.33$ (solid contours and dot) 
is compared to the low ($\langle D_{\mathrm{ds}}/D_{\mathrm{s}}\rangle\!=\!0.30$,
dashed contours and square) and high cases 
($\langle D_{\mathrm{ds}}/D_{\mathrm{s}}\rangle\!=\!0.36$, dot-dashed contours and
diamond) derived from the CFHTLS Deep fields in Sect.~\ref{sec:ddsds}.
}
\label{fig:dists}
\end{figure*}
\subsection{The Concentration Parameter} \label{sec:cnfw}

While our resulting $r_{200}$ seems reasonable for a galaxy cluster of the
redshift and X-ray luminosity of \cl00 , its concentration, despite the fact
that it is not well constrained by our data and cluster weak lensing in
general, seems low compared to the known properties of galaxy clusters:

\citet{2001MNRAS.321..559B} established a relation between mass and
concentration parameter from numerical simulations of dark matter haloes, 
using a functional form from theoretical arguments:
\begin{equation}
c_{\mathrm{NFW}}=\frac{c_{\mathrm{NFW},0}}{1+z}
\left(\frac{M_{\mathrm{vir}}}{M_{\ast}}\right)^{\alpha}
\label{eq:cm}
\end{equation}
with $c_{\mathrm{NFW},0}=14.5\pm6.4$ and $\alpha=-0.15\pm0.13$
for a pivotal mass $M_{\ast}\!=\!1.3\times 10^{13}\,h^{-1}\mathrm{M}_{\sun}$.
This, for $z=0.50$ and $M_{\mathrm{vir}}=7.2\times10^{14}\,\mathrm{M}_{\sun}$
gives $c_{\mathrm{NFW}}\!=\!3.7$.
\citet{2007MNRAS.379..190C}, analysing a sample of 62 galaxy clusters for which
virial masses and concentration parameters have been determined, and using the
same relation Eq.~(\ref{eq:cm}), find
$c_{\mathrm{NFW},0}=9.0$ and $\alpha=-0.13$, yielding
$c_{\mathrm{NFW}}=5.6_{-3.7}^{+7.3}$ for the virial mass and redshift of \cl00 .

This large interval is consistent within the error bars with our fiducial
$c_{\mathrm{NFW}}\!=\!2.0^{+1.8}_{-1.2}$ with
$\chi^{2}_{\mathrm{L}}/\nu_{\mathrm{dof}}\!=\!13413/13636$, 
but as the value itself remains
unusually small, we investigate it further. First, we test
$c_{\mathrm{NFW}}\!=\!4.0$, close to the value suggested by 
\citet{2001MNRAS.321..559B}, while fixing
$r_{200}^{\mathrm{fit}}\!=\!1.64\,\mbox{Mpc}$ and find 
$\chi^{2}_{\mathrm{L}}/\nu_{\mathrm{dof}}\!=\!13423/13637$ and the shear
profile of the resulting model (dash-dotted line in the middle panel of
Fig.~\ref{fig:profile}) to be clearly outside the error margin for the
innermost bin, demanding a significantly higher shear in the inner $90\arcsec$
than consistent with the measurements. With changes in $c_{\mathrm{NFW}}$
mainly affecting the modelling of the cluster centre, there is no such tension
in the other bins.
In the next step, we repeat the fit to the profile, now with
$c_{\mathrm{NFW}}\!=\!4.0$ fixed and $r_{200}$ as the only free parameter. 
The resulting best-fitting model yields
$r_{200}^{\mathrm{fit}4}\!=\!1.51\,\mbox{Mpc}$ (triple-dot dashed in the middle 
panel of Fig.~\ref{fig:profile}), still outside but close to the measured
$1\sigma$-margin of the data. As this fit gives
$\chi^{2}_{\mathrm{L}}/\nu_{\mathrm{dof}}\!=\!13418/13637$, we conclude that more
strongly concentrated models than the fiducial are indeed disfavoured.

Residual contamination by cluster galaxies reduces the measured concentration
parameter, as can be seen when ``switching off'' the contamination correction
factor (see Table~\ref{tab:param}). This is expected as contamination
suppresses the signal most strongly in the cluster centre.
Removing all galaxies at separations $\theta\!<\!0.5\arcmin$ from the likelihood
analysis, we indeed measure a higher 
$c_{\mathrm{NFW}}^{\mathrm{min}}\!=\!2.2^{+2.8}_{-1.4}$, but for the price of
larger error bars, as the same galaxies close to the cluster centre have the
highest constraining power on $c_{\mathrm{NFW}}$. As can be seen from the
dashed contours and the diamond in Fig.~\ref{fig:rccont}, excising the
$\theta\!<\!0.5\arcmin$ galaxies just stretches the confidence contours
towards higher $c_{\mathrm{NFW}}$, leaving
$r_{200}^{\mathrm{min}}\!=\!1.52_{-0.24}^{+0.22}\,\mbox{Mpc}$, and thus the 
inferred cluster mass unchanged (see also Table~\ref{tab:param}).

Replacing the contamination correction with a background selection down to the 
faintest magnitudes ($m_{\mathrm{faint}}\!=\!29$), removing a large
fraction of the ``red sequence-like'' galaxies in Fig.~\ref{fig:fcg}, also
yields a higher $c_{\mathrm{NFW}}^{\mathrm{fit}}\!=\!3.3\pm1.7$ in the shear profile 
fit, together with a slightly larger
$r_{200}^{\mathrm{fit}}\!=\!1.67\pm0.18\,\text{Mpc}$ and a less significant detection 
$S(\pmb{\theta}_{\mathrm{c}})\!=\!5.10$ 
than the fiducial case.
A further possible explanation for the low $c_{\mathrm{NFW}}$ 
due to additional lensing by the $z\!\approx\!0.25$ foreground
structure is rather unlikely (cf. Sect.~\ref{sec:fits}). 

\subsection{The Extent of the NFW Profile}

\citet{1997ApJ...490..493N} designed their profile to represent the mass
distribution of galaxy clusters in numerical simulations within the virial
radius. Thus, as theory provides no compelling argument to use it out to larger
radii, this practice has to be justified empirically.

In the lower panel of Fig.~\ref{fig:profile}, we show results for a toy model 
profile in which the shear signal drops faster than NFW outside $r_{200}$.
For simplicity, we chose the shear profile of a point mass, i.e.
\begin{equation}
g_{\mathrm{t,ext}}(\theta)\!=\!g_{\mathrm{t,NFW}}(\theta_{200})
\left(\frac{\theta_{200}}{\theta}\right)^{2}
\end{equation}
for $\theta\!>\!\theta_{200}$, the separation corresponding to $r_{200}$.
As in the middle panel of Fig.~\ref{fig:profile}, dashed, dot-dashed, and
triple dot dashed lines denote the fit to both $r_{200}$ and $c_{\mathrm{NFW}}$,
setting $c_{\mathrm{NFW}}\!=\!4.0$ for the same $r_{200}$, and fitting to   
$r_{200}$ for a fixed $c_{\mathrm{NFW}}\!=\!4.0$, respectively.
The truncation points $ \theta_{200}$ are marked by squares in 
Fig.~\ref{fig:profile}.
For the usual two-parameter model with
$\chi^{2}_{\mathrm{L,trunc}}\!-\!\chi^{2}_{\mathrm{L,NFW}}\!=\!0.80$, as for the other 
two models the truncated, the difference in goodness-of-fit between the 
truncated and pure NFW profiles is marginal. 

Secondly, we repeat the likelihood analysis for galaxies 
$0\arcmin\!\leq\!\theta\!\leq\!4\arcmin\!\approx\!\theta_{200}$ only. 
The dash-dotted contours and the square in Fig.~\ref{fig:rccont} for the 
resulting optimal parameters show the corresponding values. Here, 
$r_{200}^{\mathrm{min}}\!=\!1.58_{-0.50}^{+1.85}\,\mbox{Mpc}$ and
$c_{\mathrm{NFW}}^{\mathrm{min}}\!=\!1.6^{+2.4}_{-1.6}$
are more degenerate than in the fiducial case (cf.\ Table~\ref{tab:param}). 
We conclude that there is no evidence in the \cl00 ~data for a deviation of the
shear profile from NFW at $r\!>\!r_{200}$. Applying Occam's razor, we use this
profile for the whole radial range, but stress cautiously  that we cannot
preclude an \emph{underestimation} of the errors and, to a lesser extent,
a bias in the virial mass here.

\subsection{Shear Calibration} \label{sec:f0}

As already pointed out in Sect.~\ref{sec:veri}, the maximum shear estimator
$\max(|\varepsilon|)$ considered in the catalogue strongly affects averaged
shear observables. In Fig.~\ref{fig:rc2}, we quantify this dependence by 
comparing the confidence contours and best values for $r_{200}$ and 
$c_{\mathrm{NFW}}$ from the fiducial $\max(|\varepsilon|)\!=\!0.8$ catalogue
(solid contours and dot) to cases with $\max(|\varepsilon|)\!=\!1.0$ (dashed
contours and diamond) and $\max(|\varepsilon|)\!=\!10^{4}$ (dot-dashed contours and
square). The latter includes even the most extreme shear estimates\footnote{
Note that, although unphysical, shear estimates $\varepsilon\!>\!1$ in KSB
are to some extent justified when averaging over large ensembles.}.
The $\max(|\varepsilon|)$ cut, via the amplitude of the shear signal, mainly 
influences $r_{200}^{\mathrm{min}}$, reducing\footnote{The sign here is likely 
due to a statistical fluke; theory expects $r_{200}^{\mathrm{min}}$ to 
increase with a less strict $\max(|\varepsilon|)$.} it by $6$~\% ($13$~\%) for the
frequently used $\max(|\varepsilon|)\!=\!1.0$ and the extreme 
$\max(|\varepsilon|)\!=\!10^{4}$, respectively. In turn, the mass estimate would be
reduced by $17$~\% ($35$~\%), as can be seen from Table~\ref{tab:param}.

The influence on the mass estimate by the choice of $\max(|\varepsilon|)$ is
compensated by the shear calibration $f_{0}\!\neq\!1$ and one of the effects
which we account by considering different $f_{0}$.
Given the uncertainty $\sigma_{\mathrm{f}_{0}}\!=\!0.05$ (Sect.~\ref{sec:veri}), we repeat
the likelihood analysis with $f_{0}\!=\!1.13$. For the negative sign, the
signal dilution by foreground galaxies has to be taken
into account. Combining in quadrature the $18$~\% foreground dilution estimated from the
CFHTLS~D1 field (Sect.~\ref{sec:cfht}) with $\sigma_{\mathrm{f}_{0}}$, we arrive
at $f_{0}\!=\!0.88$ as the lower bound of the error margin.
The $+5$~\% ($-20$~\%) variation in $f_{0}$ translates into $+1.3$~\% ($-7.2$~\%) in 
$r_{200}^{\mathrm{min}}$, yielding again $\approx\!+4$~\% ($\approx\!-20$~\%) variation in $M_{200}$
(see Table~\ref{tab:param}).

\subsection{Combined Mass Error Budget} \label{sec:err}

Replacing the weak lensing centre in our fiducial model by the cluster's BCG
as the centre of the NFW profile, we find the resulting differences in
$r_{200}^{\mathrm{min}}$ and $c_{\mathrm{NFW}}^{\mathrm{min}}$ returned by the 
likelihood method, and hence in $M_{200}$, to be small (cf. triple dot-dashed
contours and triangle in Fig~\ref{fig:rc2}; Table~\ref{tab:param}). We
conclude the error on the chosen centre to be subdominant.

Variations in the geometric factor $\langle D_{\mathrm{ds}}/D_{\mathrm{s}}\rangle$
induce a similar scaling in $r_{200}^{\mathrm{min}}$ and 
$c_{\mathrm{NFW}}^{\mathrm{min}}$ as shear calibration does.
Using the error margin from the determination of the distance ratios from the
CFHTLS Deep fields (Sect.~\ref{sec:ddsds}), we produce likelihood contours for
$\langle D_{\mathrm{ds}}/D_{\mathrm{s}}\rangle\!=\!0.30$ (dashed lines and square in
Fig.~\ref{fig:dists}) and $\langle D_{\mathrm{ds}}/D_{\mathrm{s}}\rangle\!=\!0.36$
(dot-dashed contours and diamond). Comparing to the fiducial model (solid
contours and dot), we find an increase in $r_{200}^{\mathrm{min}}$ by $4.6$~\%
and by $14$~\% in $M_{200}$ for 
$\langle D_{\mathrm{ds}}/D_{\mathrm{s}}\rangle\!=\!0.30$ (a more massive lens is
needed for the same shear if the source galaxies are closer on average) and a
decrease by  $4.6$~\% in $r_{200}^{\mathrm{min}}$ and $13$~\% in $M_{200}$ for
$\langle D_{\mathrm{ds}}/D_{\mathrm{s}}\rangle\!=\!0.36$ (cf.\ Table~\ref{tab:param}).
%

An additional source of uncertainty in the mass estimate not discussed so far 
are triaxiality of galaxy cluster dark matter haloes and projection 
of the large-scale structure (LSS) onto the image.
\citet{2007MNRAS.374L..37K} and \citet{2007MNRAS.380..149C} showed with 
simulated clusters that masses of prolate haloes tend to get their masses 
overestimated in weak lensing while masses of oblate haloes are underestimated. 

Again owing to cosmological simulations, \citet{2005ApJ...629..781K} devised a
fitting formula for the largest-to-smallest axis ratio $\eta$ of a triaxial
haloes as a function of redshift and mass
\begin{equation}
\eta(M_{200},z)\!=\!\eta_{0} (1+z)^{\epsilon}\left(1-\zeta \ln\left(\frac{M_{200}}
{h\, 10^{15}\,\mathrm{M}_{\sun}}\right)\right)
\end{equation}
with $\epsilon\!=\!0.086$, $\zeta\!=\!0.023$, and $\eta_{0}\!=\!0.633$.
Inserting the values for \cl00 , we find $\eta\!=\!0.61$ and, like 
\citet{2009A&A...499..669D} whose lines we are following, derive the following
maximal biases from \citet{2007MNRAS.380..149C}: for a complete alignment of 
the major cluster axis with the line of sight mass is overestimated by $16$~\%,
while complete alignment with the minor axis results in a $10$~\%
underestimation. 

The projection of physically unrelated large scale structure can lead
to a significant underestimation of the statistical errors in $M_{200}$ and 
$c_{\mathrm{NFW}}$ \citep{2003MNRAS.339.1155H,2007MNRAS.379..317H}.
The simulations of \citet{2003MNRAS.339.1155H} yield an additional error of
$\pm1.2\,h^{-1}\!\times\!10^{14}\mathrm{M}_{\sun}\!=\!\pm1.67\times 10^{14}\mathrm{M}_{\sun}$ 
for a cluster in the mass range of \cl00 , and little redshift dependence for $z\!>\!0.2$.
Thus, we adopt this value as the systematic uncertainty due to large scale structure.

We define the systematice mass uncertainty $\sigma_{\mathrm{sys}}$
as the quadratic sum of the errors $\sigma_{\mathrm{cali}}$ from shear 
calibration, $\sigma_{\mathrm{geom}}$ from the geometric factor, 
$\sigma_{\mathrm{proj}}$ from projection, and $\sigma_{\mathrm{LSS}}$ from 
large-scale structure.\footnote{We remark, however, that strictly 
speaking $\sigma_{\mathrm{LSS}}$ qualifies as a statistical error.}
The total error, used in Fig.~\ref{fig:xray}, is then defined as the
quadratic sum also including $\sigma_{\mathrm{stat}}$:
\begin{equation}
\sigma_{\mathrm{tot}}^{2}=\sigma_{\mathrm{stat}}^{2}\!+\!
\sigma_{\mathrm{sys}}^{2}=\sigma_{\mathrm{stat}}^{2}\!+\!
\sigma_{\mathrm{LSS}}^{2}\!+\!\sigma_{\mathrm{proj}}^{2}\!+\!
\sigma_{\mathrm{geom}}^{2}\!+\!\sigma_{\mathrm{cali}}^{2}
\label{eq:err}
\end{equation}
We note that the statistical errors are already quite large and the
dominating factor in Eq.~(\ref{eq:err}). As its main result, this study arrives
at a mass estimate of 
$M_{200}\!=\!7.2^{+3.6+2.3}_{-2.9-2.5}\times 10^{14}\mathrm{M}_{\sun}$
for \cl00 , quoting \emph{separately} the statistical and systematic 
error as the first and second uncertainty.

\subsection{Comparison to X-ray masses}

\begin{figure*}
\sidecaption
\includegraphics[angle=90,width=12cm]{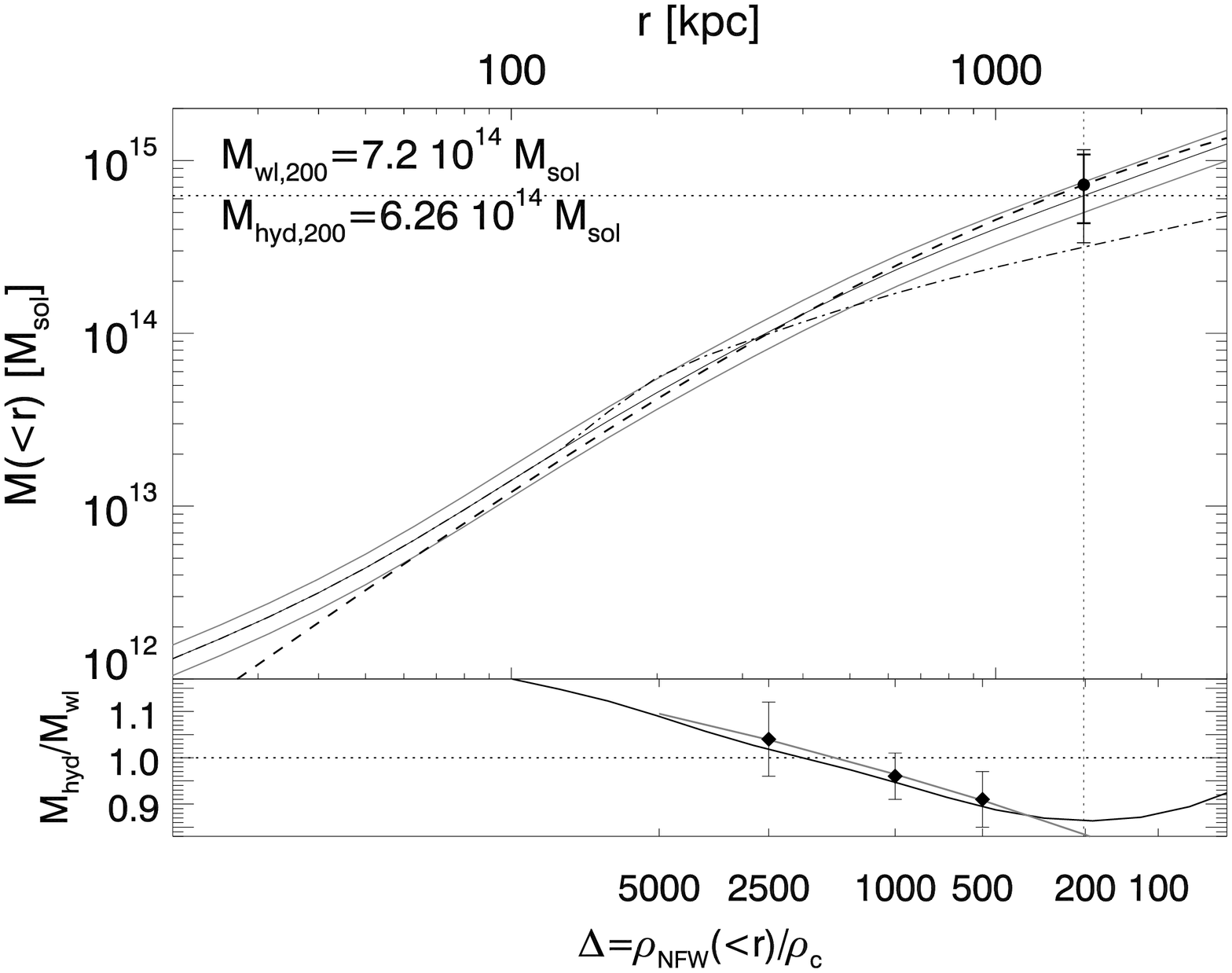}
\caption{Comparison of mass profiles of CL0030+2618. \newline
 \emph{Upper panel: }
The hydrostatic mass $M_{\mathrm{hyd}}(<\!r)$ derived
from the \textsc{Chandra} analysis 
(thick solid line). A constant ICM temperature is assumed and
the grey lines delineate the error margin derived from its error. 
The dash-dotted line gives the \textsc{Chandra} profile for
a more realistic temperature profile.
The dot with error bars and the dashed line denote the mass estimate and
profile $M_{\mathrm{wl}}(<\!r)$
from our weak lensing analysis, assuming an NFW profile. The thick error bars
show the statistical errors while thin bars include all components dicussed
in Sect.~\ref{sec:err}. 
\emph{Lower panel: }Ratio of X-ray to lensing mass as a function of radius
(black line).
The symbols and grey line show the $M_{\mathrm{hyd}}/M_{\mathrm{wl}}$ 
found by \citet{2010arXiv1001.0780Z} at three overdensity radii and their fitted relation.}
\label{fig:xray}
\end{figure*}

We will now compare this weak lensing mass to a mass profile drawn from the
\textsc{Chandra} analysis of \cl00 .
Under the assumption that the ICM is in hydrostatic equilibrium, the total
mass $M(<\!r)$ of a galaxy cluster within a radius $r$
can be derived as \citep[cf.][]{1988xrec.book.....S}:
\begin{equation}
M(<\!r)\!=\!\frac{-k_{\mathrm{B}}T_{\mathrm{X}}(r)r}{\mu m_{\mathrm{p}}G}
\left(\frac{\mathrm{d}\ln \rho_{\mathrm{g}}}{\mathrm{d}\ln r}
+\frac{\mathrm{d}\ln T_{\mathrm{X}}}{\mathrm{d}\ln r}\right)
\end{equation}
where $m_{\mathrm{p}}$ is the proton mass and $\mu$ the mean molecular mass.
In a first step, we treat the ICM temperature to be independent of the
radius and fix it to the \citet{2009ApJ...692.1033V} value of
$k_{\mathrm{B}}\langle T_{\mathrm{X}}\rangle\!=\!5.63\pm1.13\,\mbox{keV}$.
For the gas density $\rho_{\mathrm{g}}$, 
we use 
a (simplified) 
\citet{2006ApJ...640..691V} particle density profile
\begin{equation} \label{eq:v06prof}
\sqrt{n_{\mathrm{e}}n_{\mathrm{p}}}\!=\!n_{0}^{2}\frac
{\left(r/r_{\mathrm{c}}\right)^{-\alpha}}
{\left(1+r^{2}/r_{\mathrm{c}}^{2}\right)^{3\beta-\alpha/2}}
\frac{1}{\left(1+r^{\gamma}/r_{\mathrm{s}}^{\gamma}\right)^{\varepsilon/\gamma}}
\end{equation}
  \begin{table} 
   \centering
   \caption{Parameters and fit values of the \citet{2006ApJ...640..691V}
     ICM model for \cl00 .}
   \begin{tabular}{ccc}\hline\hline
     Parameter & Quantity & Fit Value \\ \hline\hline
    $n_{0}^{2}$ & pivot density & $3.784\times10^{-3}\,\mbox{cm}^{-3}$ \\
     $r_{\mathrm{c}}$ & core radius & $139\,\mbox{kpc}$ \\
     $r_{\mathrm{s}}$ & scale radius & $421\,\mbox{kpc}$ \\
     $\alpha$ & exponent & $0.5867$ \\
     $\beta$ & exponent & $0.4653$ \\
     $\varepsilon$ & exponent & $1.2293$
 \\ \hline\hline
   \end{tabular}
\label{tab:v06}
  \end{table}
with the parameters given in Table~\ref{tab:v06} and a fixed $\gamma\!=\!3$ 
and arrive at a
mass of $M_{\mathrm{hyd}}\!=\!(6.26\pm1.26)\times10^{14}\,\mathrm{M}_{\sun}$ 
at the virial radius of $r_{200}\!=\!1.52\,\mbox{Mpc}$ obtained in the lensing
analysis. We show the corresponding mass profile as the thick and its error
margin as the grey lines in the upper panel of Fig.~\ref{fig:xray}.

This value is in very good agreement with the weak lensing mass estimate
(dot with thick error bars for statistical and thin error bars for systematic
plus statistical uncertainties in Fig.~\ref{fig:xray}).
The consistency between the X-ray mass profile 
derived from $T_{\mathrm{X}}$ and
the (baryonic) ICM using Eq.~\ref{eq:v06prof} and the NFW
profile describing the combined dark and luminous matter densities
holds at all relevant radii $\gtrsim\!50\,\mbox{kpc}$ in a wide range
from the cluster core till beyond the virial radius.

Assuming an isothermal cluster profile, one likely overestimates the total
hydrostatic mass, as the ICM temperature is lower at the large radii
dominating the mass estimation around $r_{\mathrm{vir}}$. The competing effect
of the temperature gradient term in the hydrostatic equation is subdominant
compared to this effect of the temperature value.

Therefore, to estimate the systematic uncertainty arising from isothermality,
we consider a toy model temperature profile consisting of the flat core at 
$\langle T_{\mathrm{X}}\rangle$,
an power-law decrease at larger radii, and a minimal temperature 
$k_{\mathrm{B}}T_{0}\!=\!0.5\,\mbox{keV}$ 
in the cluster outskirts to qualitatively represent the features of an 
ensemble-averaged temperature profile:
\begin{equation}
k_{\mathrm{B}}T_{\mathrm{X}}(r)\!=\!\begin{cases}
k_{\mathrm{B}}\langle T_{X}\rangle & r\!\leq\!r_{\mathrm{i}} \\
p\, r^{q} & r_{\mathrm{i}}\!\leq\!r\!\leq\!r_{\mathrm{t}} \\
k_{\mathrm{B}}T_{0} & r\!\geq\!r_{\mathrm{t}} \end{cases}
\end{equation}
where we choose a core radius $r_{\mathrm{i}}\!=\!r_{200}/8$
\citep[as used in][]{2007A&A...461...71P}, a power-law slope $q\!=\!-0.4$
taken as a typical value found by Eckmiller et al.\
(in prep.), and fixing the truncation radius $r_{\mathrm{t}}$ and amplitude $p$
demanding continuity of $T_{\mathrm{X}}(r)$. 
The mass profile resulting from this temperature distribution is plotted in
Fig.~\ref{fig:xray} (upper panel) as the dash-dotted line, giving an estimate of the 
systematic uncertainty in the X-ray profile. 
Its value coincides with the lower end of the $1\sigma$ mass range for
$M_{\mathrm{wl}}$ at $r_{200}$, taking into account its systematic errors.
Another systematic factor in X-ray analysis is non-thermal pressure support,
leading to an underestimation of the X-ray mass by $\sim\!10$~\%
\citep[e.g.][]{2008A&A...482..451Z}. Taking into account all these
effects, we conclude a very good agreement of X-ray and weak lensing mass 
estimates of \cl00 , despite the potential perturbation by the line-of-sight
structure.  
%

In the lower panel of Fig.~\ref{fig:xray}, we show the ratio 
$M_{\mathrm{hyd}}/M_{\mathrm{wl}}$ of hydrostatic X-ray and weak lensing mass as
a function of radius. Although this quantity has a large error, our values
are in good agreement with the X-ray-to-lensing mass ratios  found by 
\citet{2010arXiv1001.0780Z} for a sample of relaxed clusters
for three radii corresponding to overdensities 
$\Delta\!=\!\rho_{\mathrm{NFW}}(<\!r)/\rho_{\mathrm{c}}\!=\!2500$, $1000$, 
and $500$
(black line).
We note that we recover well the relation $\frac{M_{\mathrm{hyd}}}{M_{\mathrm{wl}}}(\Delta)$
found by \citet{2010arXiv1001.0780Z} by fitting their cluster sample data (grey line).

\section{Summary and Conclusion}

With this study, we report the first results for the largest weak lensing
survey of X-ray selected, high-redshift clusters, the 
\emph{400d cosmological sample} defined by \citet{2009ApJ...692.1033V} 
and determine a weak lensing
mass for an interesting cluster of galaxies, \cl00 , which had not been studied
with deep optical observations before.
We observed \cl00 , along with other clusters of our sample, using the 
\megacam ~$\sim\!24\arcmin\times24\arcmin$ imager at the MMT, obtaining deep
$g'r'i'$ exposures. Employing an adaptation of the \citet{2005AN....326..432E}
pipeline, \texttt{THELI}, and the ``TS'' KSB shape measurement pipeline
presented by T.\ Schrabback in \citet{2006MNRAS.368.1323H}, we, for the first
time, measure weak gravitational shear with \megacam , showing its PSF
properties to be well suited for such venture.

The lensing catalogue of background galaxies is selected by a photometric
method, using $g'r'i'$ colour information. Despite similar number count
statistics, we find different photometric properties in our \megacam ~field
than in the CFHTLS Deep fields used to estimate the redshift distribution of 
the lensed galaxies.
The photometric measurements establish the galaxy we name G1, for which
\citet{1997MNRAS.285..511B} determined a redshift $z\!=\!0.516$ as the BCG of
\cl00 , ruling out a slightly brighter source found inconsistent in its
colours with the cluster redshift $z\!=\!0.50$. We find additional evidence
for the presence of a foreground structure at $z\!\approx\!0.25$ from 
photometry but find it does neither significantly affect the lensing nor the
X-ray mass estimate of \cl00 .

Having applied several consistency checks to the lensing catalogue
and optimising the aperture mass map of the cluster, we detect \cl00 ~at
$5.84\sigma$ significance. The weak lensing centre obtained by bootstrapping 
this map is in good agreement with the BCG position and the X-ray detections by
\textsc{Rosat}, \textsc{Chandra}, and \textsc{XMM-Newton}. Two tentative 
strong lensing arcs are detected in \cl00 .

Tangential alignment of galactic ellipticities is found to extend out to
$10\arcmin$ separation and well modelled by an NFW profile out to $>\!2r_{200}$.
The low concentration parameter found by least-squares fitting to the shear 
profile is confirmed by the likelihood method with which we determine \cl00
~to be parametrised by 
$r_{200}^{\mathrm{min}}\!=\!1.52^{+0.22}_{-0.24}\,\mbox{Mpc}$ and  
$c_{\mathrm{NFW}}^{\mathrm{min}}\!=\!2.0^{+1.8}_{-1.2}$. Modifying the likelihood
analysis for the fiducial case, we estimate systematic errors due to shear
calibration, the redshift distribution of the background galaxies, and the
likely non-sphericity of the cluster. Further, we confirm the best model to
change little if the BCG is chosen as cluster centre rather than the weak
lensing centre. We arrive at a virial weak lensing mass for \cl00 ~with 
statistical and systematic uncertainties of  
$M_{200}\!=\!7.2^{+3.6+2.3}_{-2.9-2.5}\times 10^{14}\mathrm{M}_{\sun}$, in 
excellent agreement with the virial mass obtained using \textsc{Chandra} and
the hydrostatic equation, 
$M_{\mathrm{hyd}}\!=\!6.26\pm1.26\times 10^{14}\mathrm{M}_{\sun}$.

Noting that statistical errors on the lensing mass are still high, we conclude
that high-quality data and well-calibrated analysis techniques are essential to
exploit the full available cosmological information from the mass function of
galaxy clusters with weak lensing.
Nevertheless, once lensing masses for all the $36$ clusters in the sample are
available, these statistical errors are going to be averaged out and reduced by
a factor of 6 by combining
all clusters when testing for cosmological parameters. Thus, understanding and
controlling systematic errors remain important issues.
We are going to proceed our survey with the analysis of further high-redshift
clusters from the \emph{400d cosmological sample} observed with \megacam .

\begin{acknowledgements}
HI owes thank to Tim Schrabback-Krahe and J\"org Dietrich for important
discussions and suggestions helpful for the advance of this study;
HI thanks Ismael Tereno and Rupal Mittal for much useful advice during the
day-to-day work; furthermore Mischa Schirmer, Karianne Holhjem, and Daniela 
Wuttke for showing up different perspectives on various aspects of the
analysis; Vera Jaritz for proofreading; 
and everybody who showed interest in its development.
We thank the staff at MMTO for help and hospitality during our
observing runs. The authors thank the anonymous referee for useful comments.
HI is supported by Deutsche Forschungsgemeinschaft through project B6
``Gravitational Lensing and X-ray Emission by Non-Linear Structures'' of
Transregional Collaborative Research Centre TRR 33 -- ``The Dark Universe''.
THR, YYZ, and DSH acknowledge support
by the Deutsche Forschungsgemeinschaft through Emmy Noether
research grant RE 1462/2 and by the BMBF/DLR through research
grant 50 OR 0601.
HH was supported by the European DUEL RTN, project MRTN-CT-2006-036133.
\end{acknowledgements}

\bibliographystyle{aa}
\bibliography{400dGCSWLP1-v7}

\appendix
\section{Details of Data Reduction} \label{sec:drdet}

In this Appendix, we provide a detailed view of the subtleties of the
data reduction outlined in Sect.~\ref{sec:datared}.

\subsection{Chips and Amplifiers} \label{sec:amp}

The MMT \megacam\ control software offers a number of options for the CCD
readout. As already mentioned in Sect.~\ref{sec:mmt}, 
there are 36 physical CCD chips each of them is equipped with
two output amplifiers, giving a readout of $1024\!\times4608\!$ (unbinned) 
pixels per amplifier. For our programme, we have
chosen to use all 72 amplifiers each reading out half a chip, thus reducing
readout time by a factor of two. As a result, \megacam\ raw images are
multi-extension \texttt{fits} files with 72 extensions. 

Owing to this, all run processing tasks are performed on the 72 subframes
individually. Files from the two chips of an amplifier are joined at the end of
the run processing prior to the astrometric calibration in order to increase
the usable surface for the astrometric procedures. 

\subsection{The ``Run Processing'' Stage} \label{sec:run}

\begin{itemize}

\item{\textbf{De-biasing: }
By stacking all bias frames taken within a suitable time interval
around the date of science observations, a \emph{master bias image} is 
constructed and
subtracted from all other frames.}  

\item{\textbf{Flatfielding: } \theli\ applies a
two-step process. First, science frames are divided by a
master sky flatfield frame.  In the second step,
the median of all science frames is calculated, discarding the positions at
which objects have been detected
by \textsl{SExtractor} \citep{1996A&AS..117..393B}.
Due to the dithering, \emph{for every pixel} in the field-of-view, these
``superflats'' contain signal from the sky background 
\emph{from slightly different positions on the sky}. 
Thus, the superflat provides a measure to compare the
response in different pixels.

Selecting the frames to contribute into the superflat to achieve
the optimal flatness of the background is the most time-consuming and 
work-intensive step in run processing, as inhomogeneities in individual 
frames will propagate into the superflat.
Imperfect photometric conditions and variable instrumental gains
are two common
reasons for science frames to be removed from the calculation of the
superflat. Very bright stars near target clusters, e.g.\ HIP\,9272 
(BD\,-00\,301) with $V\!=\!8.28$ at $2\arcmin$ distance from CL0159+0030,
exacerbate the situation.
Involving many iterations of the -- manual -- frame selection process, our
superflatfielding is effective in reducing the relative
background variation over the field in the superflatfielded exposures
to $<\!1.5$~\%, and to $<\!1.0$~\% for most exposures.

In the superflatfielding stage, the different sensitivities of the amplifiers
are determined and equalised, taking into account all exposures within the
\theli ~\textsl{run}. This we can do, because the relative sensitivities of
most of the amplifiers are constant most of the time. 
Gain equalisation is achieved by scaling each amplifier 
with an appropriate factor detailed in \citet[Sect.\ 4.7]{2005AN....326..432E}.
Some amplifiers, however,
experience gain fluctuation on short timescales of the order of days.
In such situation the same superflatfield frame can no longer provide the same
quality of flattening to all exposures
such that we had to split the $g'$-band data taken on Nov.\ 8, 2005 from the
rest of the exposures taken on Oct.\ 30, 2005 and Nov.\ 1, 2005, 
at the cost of a lower number of exposures contributing to the superflatfield
in each of the two sub-runs. 
}

\item{\textbf{De-fringing: }
In the bands where this is necessary,
the fringing pattern can be isolated from the high spatial frequencies
of the superflat and subtracted from the science frames.
On the other hand, we divide by the superflat containing the 
lower spatial frequencies which carry information about
the (multiplicative) ``flatfield'' effects.}

\item{\textbf{Satellite tracks: }
We identify satellite tracks by visual inspection when assessing frames for
superflat construction and mask pixels which
are affected in the given exposure. Masked pixels
(stored as a \texttt{DS9} region files) are set to zero in the construction
of the weight images.}

\item{\textbf{Weight images: }
Taking into account bad pixel information from the bias\footnote{Here,
  we also use dark frames, although they are not necessary for running
  \theli.}, flatfield, and superflatfield frames we construct a
weight image, i.e.  noise map, for each individual amplifier
and exposure in the \textsl{run}.  Our algorithm is not only sensitive
to cold and hot pixels but also to charge ``bleeding'' in the vicinity
of grossly overexposed stars.}
 
\end{itemize}

\subsection{The ``Set Processing'' Stage} \label{sec:set}

\begin{figure}
 \resizebox{\hsize}{!}{\includegraphics[angle=90]{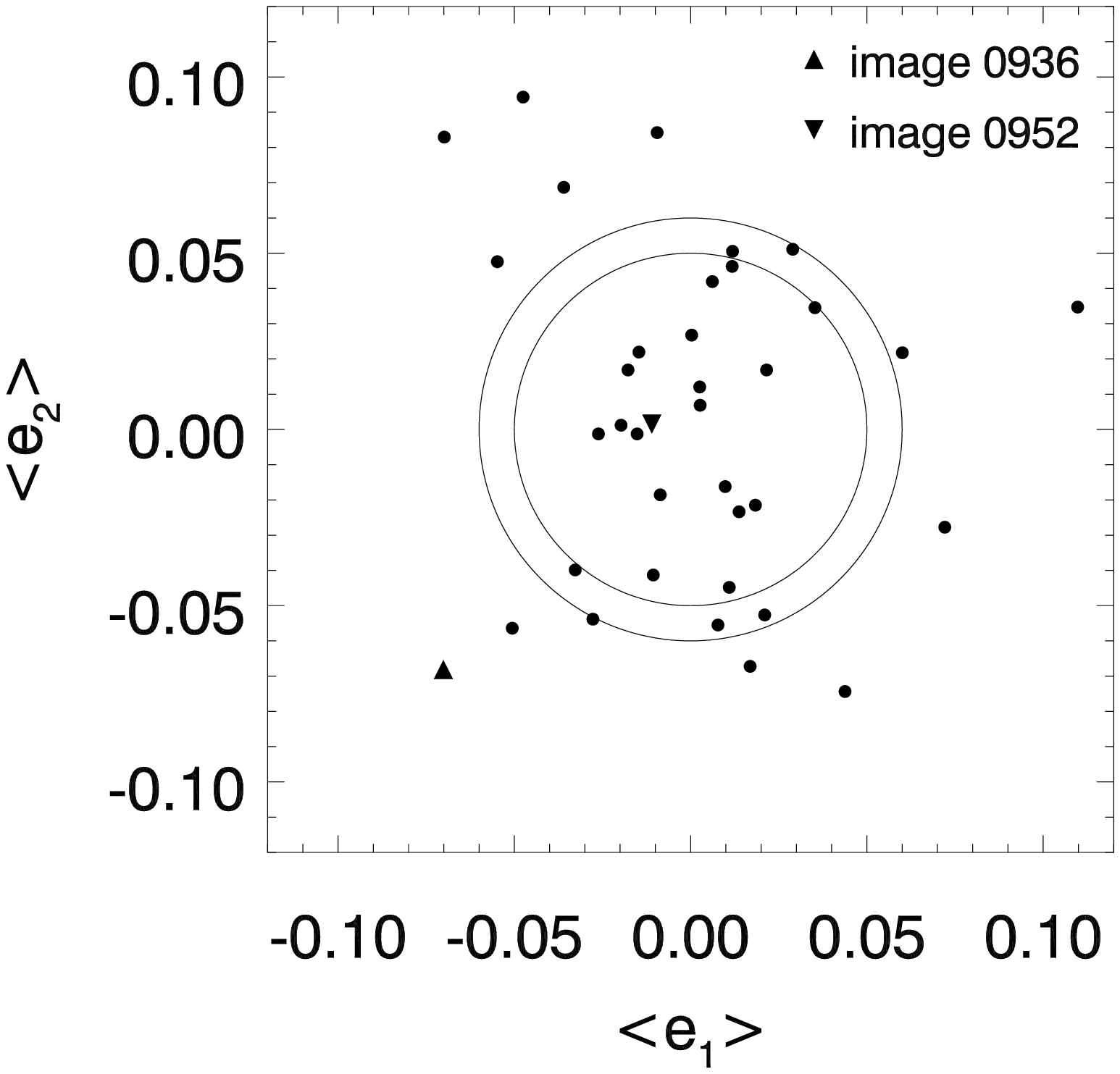}}
 \caption{Mean stellar anisotropies $\langle e_{1,2}\rangle$ found in the
  \textit{r'}-band frames of CL0030+2618 fulfilling the seeing condition
  $s<1.0\arcsec$. 
  All images with $\langle\left|e\right|\rangle<0.05$ are included
  in the final coaddition (inner circle), while those exceeding 
  $\langle\left|e\right|\rangle<0.06$ (outer circle) are always rejected.
  The decision for intermediary objects (see below) is based on visual 
  inspection.}
\label{fig:framesel}
\end{figure}

\begin{figure*}
 \includegraphics[width=0.62\textwidth,angle=90]{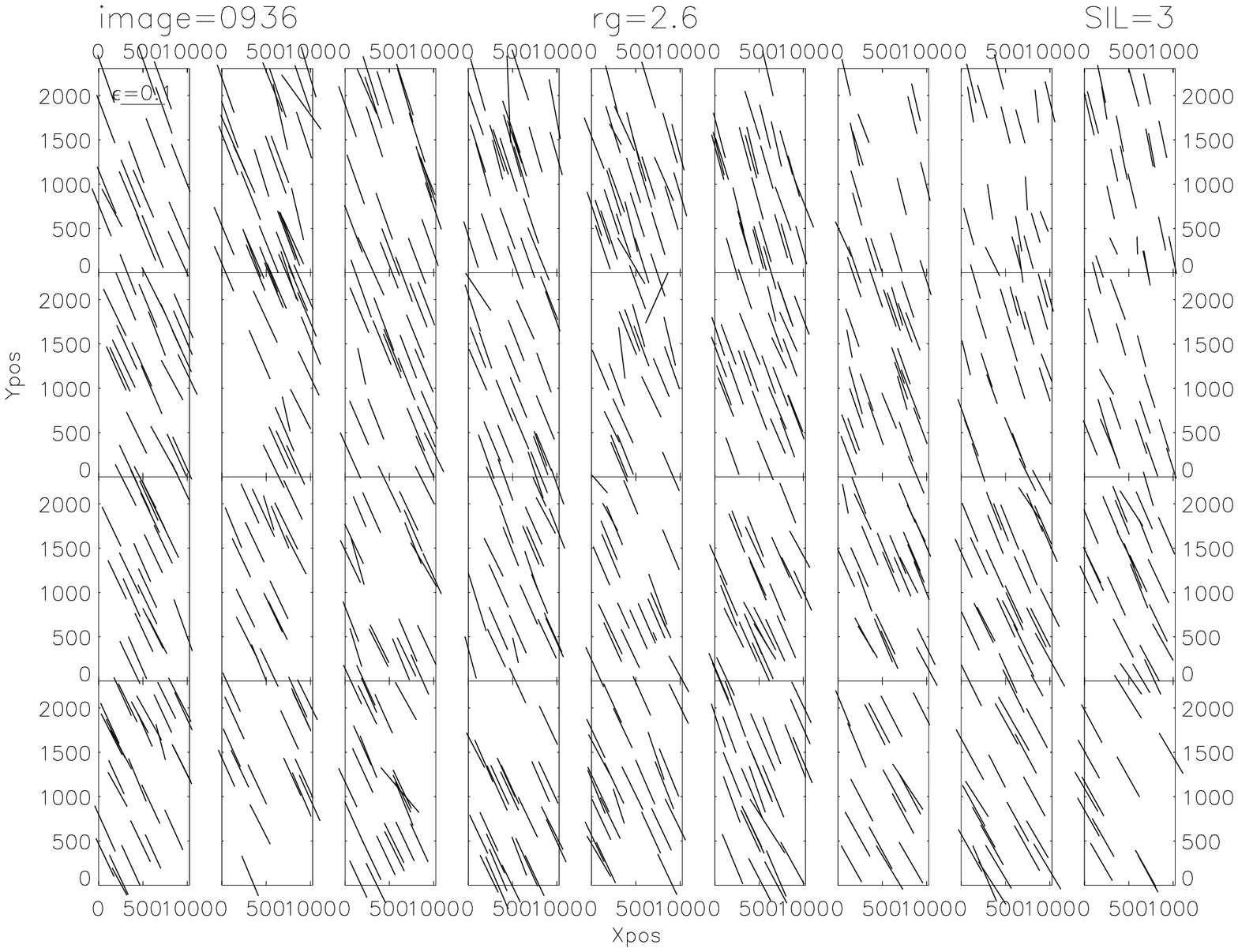}
 \includegraphics[width=0.62\textwidth,angle=90]{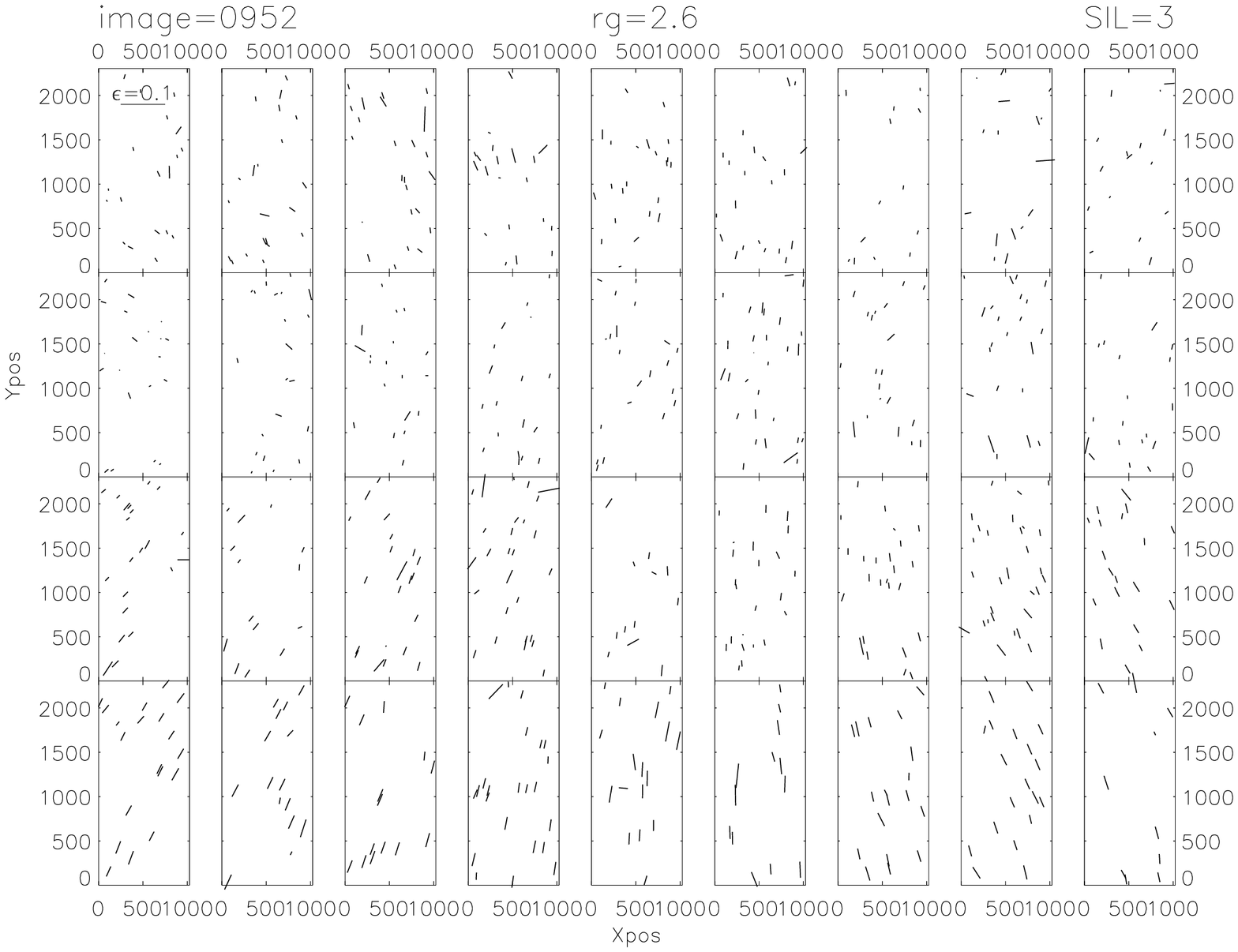}
 \caption{Spatial distribution of stellar anisotropies for an example 
  exposure of high overall PSF anisotropy. Shown are the sizes and orientations
  of the raw ellipticity $e$ for stars identified in the MMT \megacam 
  ~exposures of \cl00 ~labelled \texttt{0936} (\emph{upper panel}) and
  \texttt{0952} (\emph{lower panel}) in Fig.~\ref{fig:framesel}.
  While within each chip the $x$ and $y$ pixel
 axes are to scale the array layout is only schematic.}
\label{fig:indiv}
\end{figure*}

\begin{table*}
 \label{tab:photosol}
 \centering
 \caption{Coefficients of photometric calibration 
  defined by Eq.~(\ref{eq:photofit}) for the photometric nights
  used to calibrate the observations of \cl00 , CL0159+0030, and CL0809+2811.}
 \begin{tabular}{ccccccc}\hline\hline
   Filter & Obs.\ Date & $Z_{\mathrm{f}}^{\dagger}$ & $\beta_{\mathrm{f}}$ & colour index & $\gamma_{\mathrm{f}}$ & $n_{\mathrm{par}}^{\ddagger}$ \\ \hline
   $g'$ & 2005-10-30 & $27.27\pm0.02$ & $0.106\pm0.007$ & $g'\!-\!r'$ & $-0.14\pm0.02$ & 3 \\
        & 2005-10-31 & $27.15\pm0.02$ & $0.124\pm0.008$ & $g'\!-\!r'$ & $-0.08\pm0.01$ & 3 \\
        & 2005-11-01 & $27.35\pm0.02$ & $0.115\pm0.005$ & $g'\!-\!r'$ & $-0.21\pm0.02$ & 3 \\ \hline
   $i'$ & 2005-10-30 & $26.49\pm0.02$ & $0.127\pm0.004$ & $r'\!-\!i'$ & $-0.11\pm0.02$ & 3 \\
        & 2005-10-31 & $26.47\pm0.01$ & $0.122\pm0.002$ & $r'\!-\!i'$ & $-0.09\pm0.01$ & 3 \\
        & 2005-11-01 & $27.41\pm0.01$ & $0.119\pm0.002$ & $r'\!-\!i'$ & $-0.03\pm0.01$ & 3 \\ \hline
   $r'$ & 2005-10-30 & $26.95\pm0.02$ & $0.046\pm0.002$ & $g'\!-\!i'$ & $-0.10\pm0.02$ & 3 \\
        & 2005-10-31 & $26.90\pm0.01$ & $0.043\pm0.003$ & $g'\!-\!i'$ & $-0.05\pm0.01$ & 3 \\
        & 2005-11-01 & $26.96\pm0.01$ & $0.048\pm0.004$ & $g'\!-\!i'$ & $(-0.10)^{\S}$ & 2 \\
        & 2005-11-08 & $26.81\pm0.01$ & $0.046\pm0.003$ & $g'\!-\!i'$ & $(-0.10)^{\S}$ & 2 \\ \hline \hline
 \end{tabular}
 \begin{minipage}{120mm}
  \smallskip
  $^{\dagger}$ Normalised to an exposure time of $1\mbox{s}$ and 
    an airmass $a\!=\!0$.\\
  $^{\ddagger}$ Number of parameters used in the fit.\\
  $^{\S}$ Fixed to default value.
 \end{minipage}
\end{table*}

\begin{figure}
\resizebox{\hsize}{!}{\includegraphics[angle=90,width=0.9\textwidth]{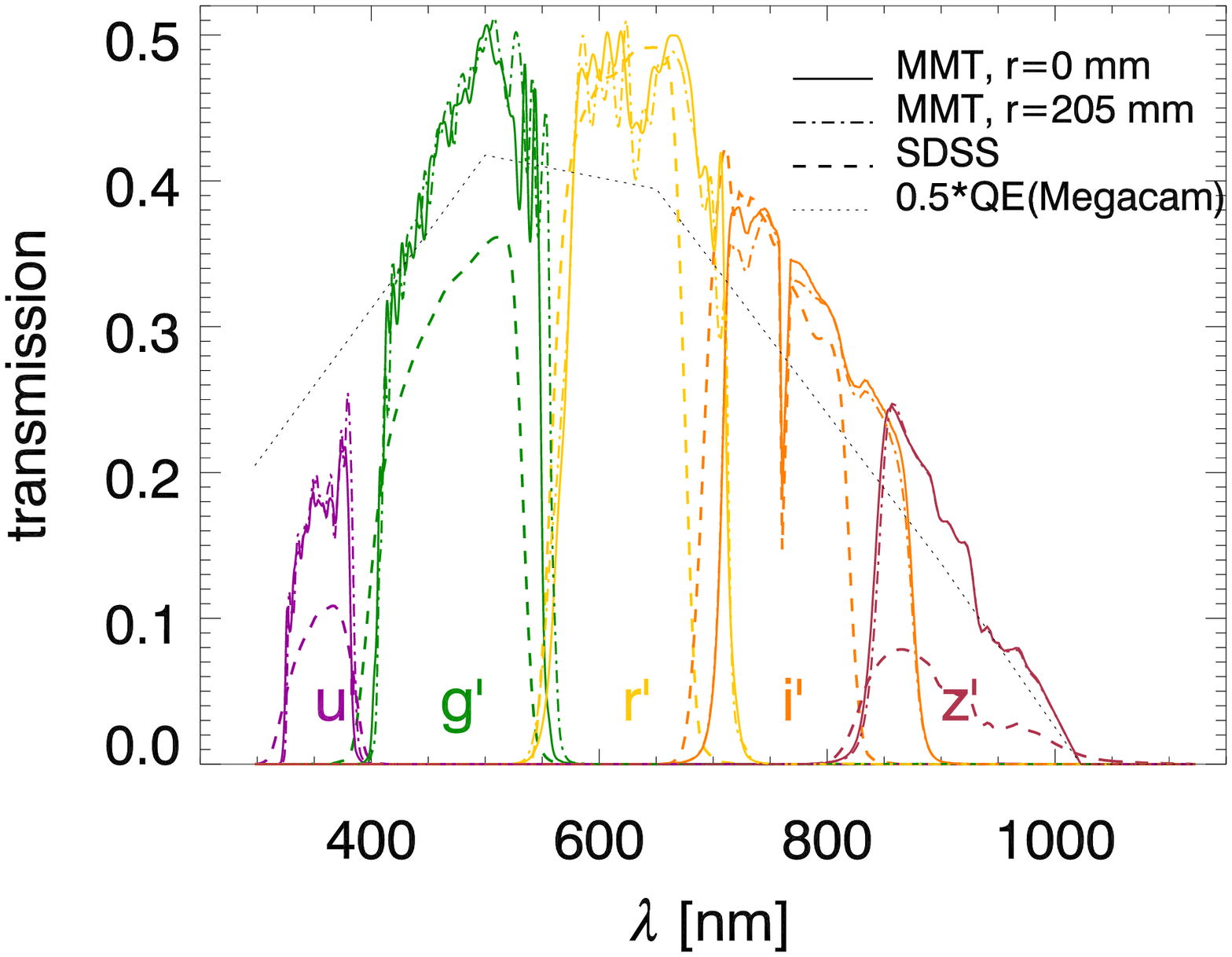}}
\caption{Comparison of the SDSS and \megacam ~filter systems. The plot
  shows the complete transmission curves for the $u'g'r'i'z'$
  filters of both systems as a function of wavelength, including the
  atmospheric transmissivity (as given for the SDSS site), the
  CCD quantum efficiency, and the actual effect of the filter, as
  measured in the laboratory.  The solid lines give sensitivities of
  \megacam ~filters for photons incident on the optical axis while the
  dash-dotted lines show the same quantity near the corner of the
  \megacam ~array.  Overplotted as dashed lines are the transmission
  curves defining the SDSS bandpass system.  The black, dotted curve
  shows the \megacam ~quantum efficiency we derive from the instrument
  specifications, scaled by one half to show it conveniently on the plot. 
  Note that we need to
  interpolate its values from only five points in the range
  $300~\text{nm}\!<\!\lambda\!<\!1000~\text{nm}$ and have to
  extrapolate outside it.  }
\label{fig:filters}
\end{figure}

\begin{itemize}

\item{\textbf{Astrometry: }
We perform the astrometric calibration of our data using the best
available catalogue as a reference. In case of overlap with SDSS Data Release
Six \citep{2008ApJS..175..297A} this
is the SDSS catalogue; otherwise we employ the shallower \texttt{USNO B1} 
catalogue \citep{2003AJ....125..984M}, as it is the densest all sky
astrometric catalogue.  The astrometric calibration 
is carried out by the TERAPIX software
\textsl{Scamp} \citep{2006ASPC..351..112B}, 
replacing the \textsl{Astrometrix} programme earlier
used within \theli. We find \textsl{Scamp} to be more robust 
than \textsl{Astrometrix} when
working on chips with a small field-of-view on the sky, as
given for \megacam ~(see Sect.~\ref{sec:datared}).
Compared to the otherwise similar design of
MegaPrime/MegaCam at the Canada-France-Hawaii Telescope the MMT \megacam ~chips
cover $\sim\!1/6$ of the solid angle on the sky, reducing the number of usable
sources for astrometry by a similar factor leading to less accurate and robust
astrometric solutions in case these are calculated on a chip-to-chip basis.

The most important innovation is, that while \textsl{Astrometrix}
determines an astrometric solution for each chip (\megacam ~amplifier in our
case) individually, \textsl{Scamp} recognises the amplifiers of one exposure
belonging together and can take into account information on the array
configuration, drastically reducing the effort to be invested into this
task. 

We provide these additional constraints by defining a
template for the same instrument configuration and filter. 
This template is drawn from
the observation of a dense field, i.e.\ a star cluster. This template
guarantees a sensible solution even with few ($\lesssim\!20$)
astrometric standard stars per chip, a condition frequently met with
\megacam ~in poor fields.

Furthermore,
by running \textsl{Scamp} on all frames in all filters for a given target
cluster with only one call to the software, we ensure consistency of the
astrometric solutions among the \theli ~\textsl{sets} corresponding to the
resulting stacks in different passbands.

For the combined data set of \cl00 , we achieve an accurate calibration with
a $1\sigma$ intrinsic accuracy of $0\farcs04$ of the sources detected with
\megacam ~and $0\farcs27$ with respect to the astrometric standard catalogue
\texttt{USNO B1}.
}

\item{\textbf{Relative photometry: }
Simultaneously with the astrometric calibration, the relative
photometric zeropoints of the frames are established by
\textsl{Scamp}.
In the first part of this two-step process, 
relative zeropoints are determined only from the
differences in flux found for the astrometric reference stars in
different exposures. 
These are  independent from the absolute photometric calibration detailed in
Sect.~\ref{sec:photo}.

In this first step, fluxes of the same object in different exposures are 
compared. 
For the coadded image
resulting from stacking to be well-calibrated, the
variation in relative zeropoints among the contributing frames needs to be
small. We decide to include only images which show a zeropoint less
than 0.1 magnitudes from the median zeropoint:
%
\begin{equation}
Z_{\mathrm{rel}}\!-\!\median(Z_{\mathrm{rel}})\!<\!-0.1
\end{equation}
In the second step, if the \emph{absolute} photometric calibration 
(Sect.~\ref{sec:photo}) has been applied already, we compute the
\emph{corrected zeropoints} defined in \citet[Eq.\ (2)]{2006A&A...452.1121H} 
of those individual frames we consider to be taken under photometric 
conditions. As detailed in \citet{2006A&A...452.1121H}, corrected zeropoints
are a useful consistency check, as they are the same for exposures obtained
in photometric conditions.  
}

\item{\textbf{Coaddition: } Conforming with \theli ~standard, \textsl{SWarp} 
is used to stack
(``coadd'') images. This, together with the \textsl{Scamp} astrometry, also
removes optical distortions, yielding a constant pixel scale in the coadded  
image. The final products of the \textsl{set} stage are the
coadded image and the corresponding weight image (Fig.~\ref{fig:coadd}).} 

\end{itemize}

\subsection{Image Selection} \label{sec:stack}

The success of a lensing analysis crucially depends on the data quality.
Because of the necessity to establish a common image coordinate system and to
rebin all data onto the new grid, the stacking process is a potential source of
biases to the shape information.\footnote{There is an ongoing debate whether
shapes should be measured on individual frames, instead.}
It is evident that the decision which frames should contribute to the shape
measurement is of great importance.
Apart from seeing and photometric quality which can be easily assessed while the
observation takes place, PSF anisotropy is a key factor as it can only be
corrected up to a certain degree.

We found that irrespective of the seeing, a significant fraction of the \megacam
~exposures from the observing runs in October 2004, October 2005, and January
2008, covering the four clusters \cl00 , CL0159+0030, CL0230+1836, and 
CL0809+2811 suffer from highly elliptic PSF which shows little variation over
the field-of-view. 

Inspecting the anisotropy stick plots like Fig.~\ref{fig:indiv} for a fair
fraction of all frames taken in these three above-mentioned runs, we come up
with the following criterion:
If the mean ellipticity in the stars is $\langle\left|e\right|\rangle\!<\!0.05$
(the smaller circle in Fig.~\ref{fig:framesel})
the variations over the field due to the properties of the optical system are
clearly discernible. These frames do not suffer from a large tracking error
and thus we include them in the analysis in any case.
On the other hand, exposures whose stars are more elliptical than
$\langle\left|e\right|\rangle\!>\!0.06$ (the larger circle in 
Fig.~\ref{fig:framesel}) show a stellar anisotropy which is mostly constant over
the field which we can attribute to MMT tracking errors of varying strength.
These frames are excluded from the lensing analysis.
In the intermediary case of $0.05\!<\!\langle\left|e\right|\rangle\!<\!0.06$ 
we decide on a case-to-case basis by inspecting the respective anisotropy stick
plot where frames in which the ``tracking error-like'' contributions seem
prevalent are excluded.

\subsection{The \megacam ~filter system} \label{sec:pcrel}

To establish the transformation between MMT and SDSS measurements, we need to
know the transmissivities of both instruments to a great detail. For \megacam ,
the instrument website\footnote{Overview:\\ 
\texttt{http://www.cfa.harvard.edu/mmti/megacam.html}; \\ filter data:\\
\texttt{http://www.cfa.harvard.edu/\textasciitilde bmcleod/Megacam/Filters/}
}
offers detailed laboratory transmission curves of the actual filters and a few
data points on the CCD's quantum efficiency. We average the tabulated 
quantum efficiency values over the 36 \megacam ~chips. 
Correspondingly, the SDSS collaboration provides data on the combined
sensitivity of its camera/filter 
system\footnote{\texttt{http://www.sdss.org/dr7/instruments/imager/}}.
Assuming the atmospheric absorption to behave the same on both sites, we can
directly compare the responses of the two instruments, as visualised in
Fig.~\ref{fig:filters}.

\subsection{Results of Photometric Calibration}

Photometric calibration is achieved by fitting Eq.~(\ref{eq:photofit}) to the 
instrumental magnitudes of the photometric standards (Sect.~\ref{sec:filters}).
For each filter, we chose a colour index 
in Eq.~(\ref{eq:photofit}) which has been proven a reliable
transformation in calibration of the Canada-France-Hawaii Telescope Legacy Survey
(CFHTLS) data which also uses a similar filter system\footnote{
http://www3.cadc-ccda.hia-iha.nrc-cnrc.gc.ca/megapipe/docs/filters.html)}.
These colour indices are given in Table~\ref{tab:photosol} which shows the
results for the fit parameters $Z_{\mathrm{f}}$, $\beta_{\mathrm{f}}$, and 
$\gamma_{\mathrm{f}}$ for the photometric nights used to calibrate the \cl00
~data (i.e.\ the datasets for \cl00 , CL0159+0030, and CL0809+2811; 
Sect.~\ref{sec:ccc}).

Comparing the colour terms $\beta_{\mathrm{f}}$ for the different nights, we
find considerable agreement of the values for all three bands, although the
error bars from fitting Eq.~(\ref{eq:photofit}) might underestimate the true
errors. While our determinations of $\beta_{\mathrm{r}}$ are all consistent
with each other, there is some tension between the $\beta_{\mathrm{g}}$ values.
In previous \megacam ~studies, \citep[Table~5]{2008ApJ...675.1233H} quote 
$\beta_{\mathrm{g}}\!=\!0.122\pm0.002$ and $\beta_{\mathrm{i}}\!=\!0.137\pm0.002$,
the first in agreement to our results, the latter significantly higher than our
value.
Furthermore, \citet{2008ApJ...688..245W} find 
$\beta_{\mathrm{g}}\!=\!0.091\pm0.068$, consistent with our values due to the
large error bar. 
We suggest that the large span in values for $\beta_{\mathrm{g}}$ might be
due to the dependence of the filter throughput as a function of distance to the
optical axis, which is strongest an this band. Further investigation is needed
to conclude about this issue. 

\section{The Background Catalogue}

\begin{figure}[h!]
 \resizebox{\hsize}{!}{\includegraphics[width=0.9\textwidth,angle=90]{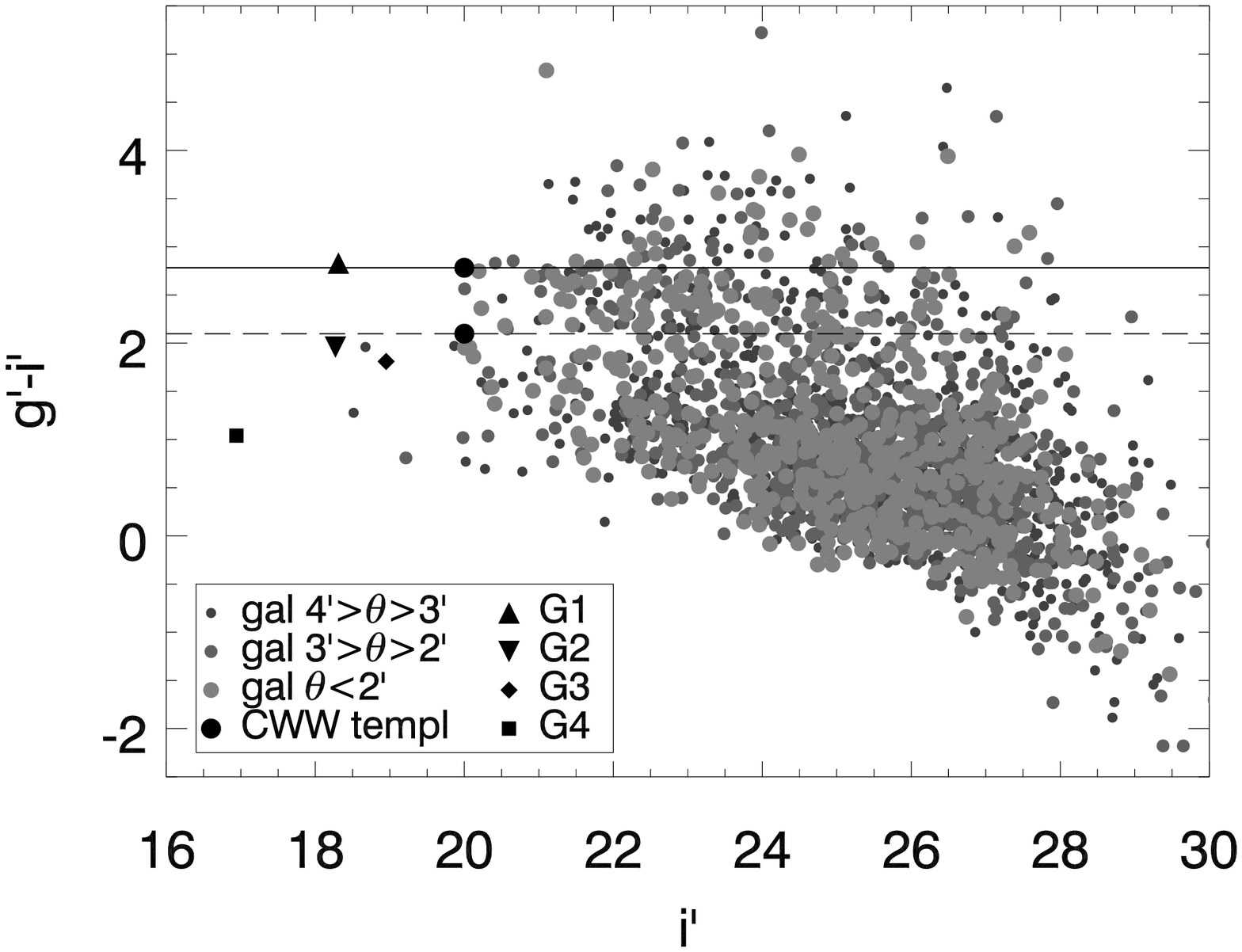}}
 \resizebox{\hsize}{!}{\includegraphics[width=0.9\textwidth,angle=90]{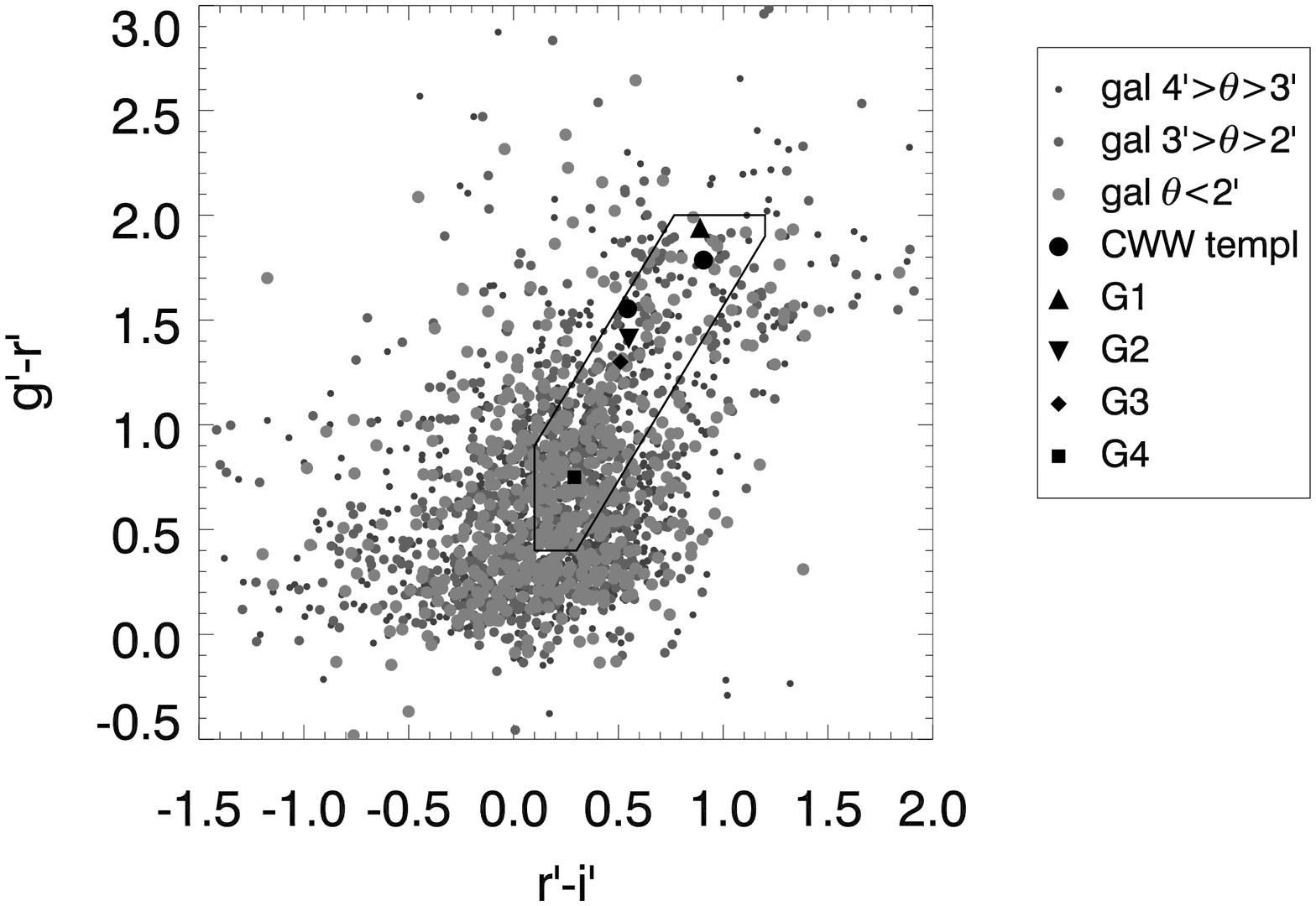}}
 \caption{\emph{Upper plot}: Colour-magnitude diagramme of KSB galaxies with a
  radial distance $\theta\!<\!4\arcmin$ from the centre of \cl00 . Symbol sizes
  and shades of grey denote galaxies from the galaxy shape catalogue 
  in different cluster-centric radial bins. The $g'\!-\!i'$ colours of 
  \citet[CWW]{1980ApJS...43..393C} template galaxies
  at $z\!=\!0.5$ (solid line and large dot at $i'\!=\!20$) and $z\!=\!0.25$ 
  (dashed line and large dot) are shown for comparison, as well as four notable
  bright galaxies detailed in Table~\ref{tab:gal}. 
\emph{Lower plot}: Colour-colour diagramme with the same objects. 
  The polygonal region delineating the locus of bright galaxies 
  (cf.\ Fig.~\ref{fig:ccselec}) is given for comparison.
}
 \label{fig:cmdccd}
\end{figure}

\begin{figure*}
\sidecaption
\includegraphics[width=12cm,angle=90]{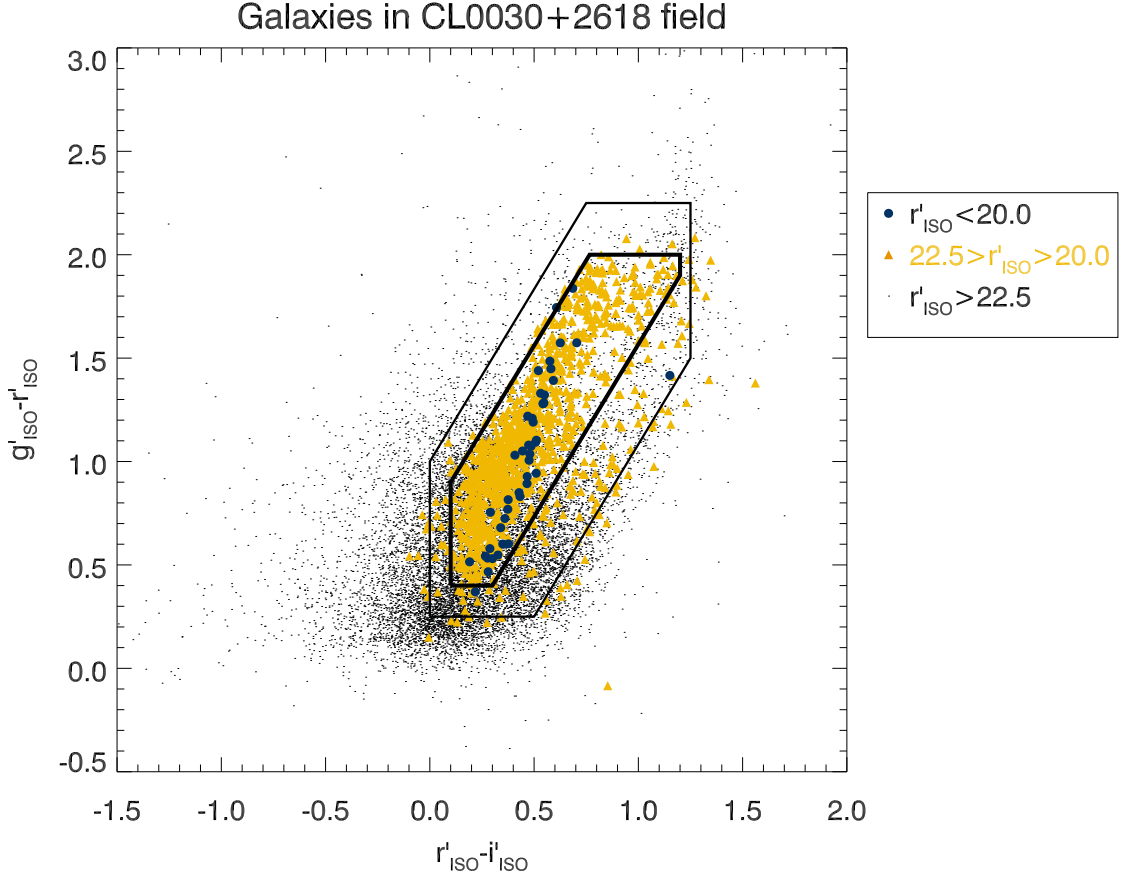}
\caption{Colour-colour selection of the lensing catalogue: plotted are
the $g'\!-\!r'$ vs.\ $r'\!-\!i'$ colours of the 
objects in the \emph{galaxy shape catalogue} (with 
cuts on $\min{(\tr{\tens{P}^{g}})}$, $\min{(\nu_{\mathrm{gal}})}$, and
$\max{(\left|(\varepsilon\right|)}$ already applied).
The galaxy sample is divided into three magnitude bins by the 
$m_{\mathrm{bright}}$ and $m_{\mathrm{faint}}$ parameters. All sources brighter 
than $m_{\mathrm{bright}}$ (largest dots) are removed in producing the final
lensing catalogue while all sources fainter than $m_{\mathrm{faint}}$ smallest dots) are kept. 
Intermediately bright galaxies with
$m_{\mathrm{bright}}\!<\!r'\!<\!m_{\mathrm{faint}}$ (medium-sized dots) mark the
transition between these two regimes. Only in this magnitude interval, the
selection into the final galaxy catalogue by colour indices applies: sources
outside the thick polygon bounding the region in which we find the brighter
galaxies and likely cluster members are included in the final catalogue.
See Table~\ref{tab:poly} for the definition of the polygon tracing the locus
of bright galaxies.
}
\label{fig:ccselec}
\end{figure*}

\begin{figure*}
\sidecaption
\includegraphics[width=12cm,angle=90]{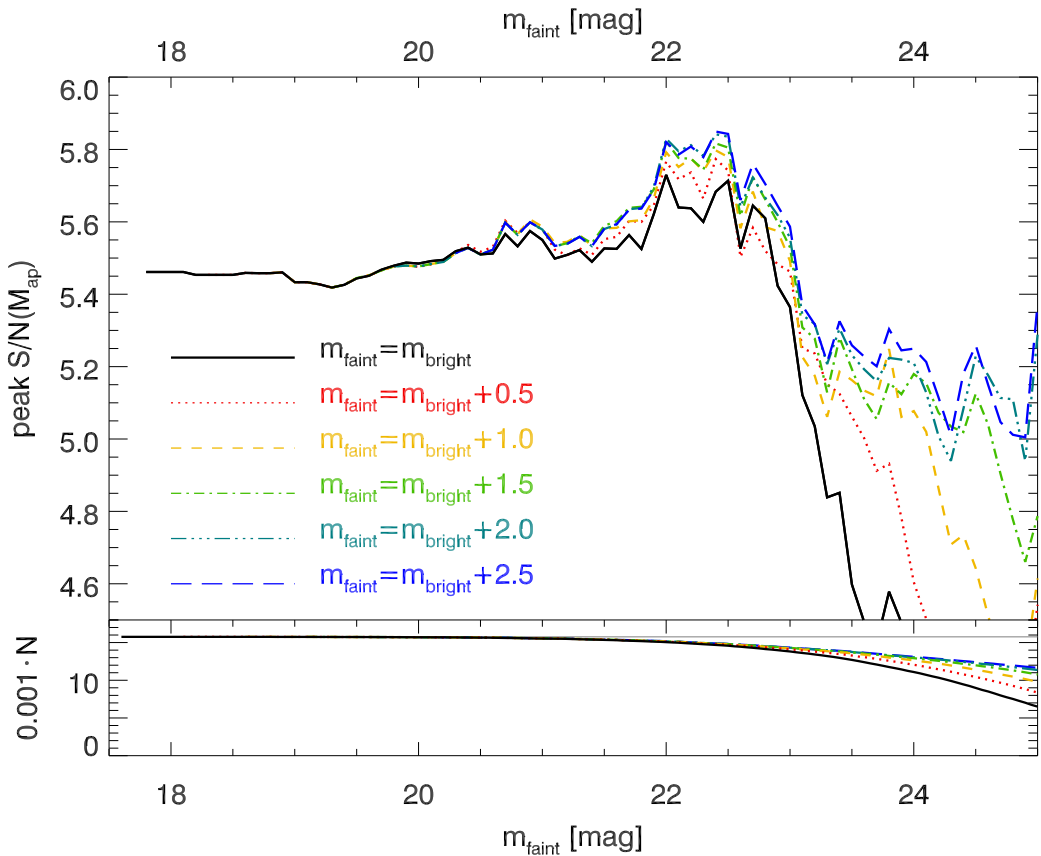}
\caption{\emph{Upper panel:}
The maximum $M_{\mathrm{ap}}$ signal-to-noise ratio found in the vicinity
of \cl00 ~as a function of the background selection introduced by 
$m_{\mathrm{faint}}$ and $m_{\mathrm{bright}}$. The solid line corresponds to
a magnitude cut $m_{\mathrm{bright}}\!=\!m_{\mathrm{faint}}$ while the 
dotted, dashed, dot-dashed, triple dot-dashed, and long-dashed lines 
denote background selections
by galaxy colours in increasingly wide intervals of
$m_{\mathrm{faint}}\!-\!m_{\mathrm{bright}}\!=\!\{0.5,1.0,1.5,2.0,2.5\}$ respectively.
Here, the smaller polygon in Fig.~\ref{fig:ccselec} is used, assuming a 
well-defined locus of cluster galaxies in colour-colour-space and, in turn, 
for a rather inclusive selection of galaxies.
\emph{Lower panel:} The number $N$ of galaxies in the shear catalogue as a 
function of $m_{\mathrm{bright}}$ and $m_{\mathrm{faint}}$. The horizontal line
gives $N\!\approx\!16000$ before applying any background selection for comparison.
The colours and line-styles denote the same cases as in the upper panel.}
\label{fig:ccss1}
\end{figure*}
\begin{table*} 
 \caption{Tested regions in colour-colour space inside which
galaxies with $m_{\mathrm{bright}}\!<\!r'\!<\!m_{\mathrm{faint}}$ are excised from
the lensing catalogue.} 
 \begin{center}
 \begin{tabular}{ccccccc} \hline\hline
 Polygon & $\min{(r'\!-\!i')}$ & $\max{(r'\!-\!i')}$ & $\min{(g'\!-\!r')}$ & $\max{(g'\!-\!r')}$ & $\min{(s)}$ & $\max{(s)}$ \\ \hline
 large & 0 & 1.25 & 0.25 & 2.25 & -1.0 & 0.583 \\
 small & 0.1 & 1.2 & 0.4 & 2.0 & -0.733 & 0.1 \\ \hline 
 \end{tabular}  \label{tab:poly}
 \end{center}
\end{table*}

\subsection{Photometric Analysis: The Red Sequence} \label{sec:redseq}
In clusters of galaxies at low and moderate redshifts, 
early-type galaxies, i.e.\ elliptical and spheroidal
systems tend to be more common than disk galaxies. 
Cluster galaxies are
observed to be deficient in gas and thus show little ongoing star formation
but are on average dominated by old, red stellar populations
\citep[e.g.,][]{1992MNRAS.254..601B}. 
In fact, cluster galaxies represent the reddest galaxies observed at a given
redshift with what are considered the most gas-depleted systems showing very
similar colours over a large range in magnitude \citep{1998ApJ...501..571G}.
Observationally, this \emph{cluster red sequence} is one of the currently most
prolific methods in detecting clusters of galaxies in the optical band
\citep[e.g.,][]{2000AJ....120.2148G,2005ApJS..157....1G}.

We consider the $(g'\!-\!i')$ vs.\ $i'$ 
colour-magnitude diagramme of the galaxies in the galaxy shape catalogue (i.e.\
before cuts to select sources for their lensing signal apply) close to the
coordinates of \cl00 ~in order to identify the red sequence of this $z\!=\!0.5$
cluster as the observed $g'$ and $i'$ passbands are on different sides of the 
Balmer break at the cluster redshift.

Having removed most extended galaxies early-on in the KSB pipeline, 
we do not expect to find the most prominent cluster members in the catalogue for
which shear estimates are determined. 
Indeed, the upper panel of Fig.~\ref{fig:cmdccd} shows a rather broad distribution
in $g'-i'$ colour of the galaxies at $\theta\!<\!4\arcmin$ from the \textsc{Rosat}
cluster centre. Nevertheless, we find an enhancement in the number of galaxies
extending from around $(g'\!-\!i')\!\approx\!2.8$ for the brighter ($i'\!\approx\!21$) 
to $(g'\!-\!i')\!\approx\!2.3$ for the fainter ($i'\!\approx\!27$) sources in our catalogue,
especially coming from a high number of galaxies very close ($\theta\!<\!2\arcmin$)
to the cluster centre.

The CWW80 template for an elliptical $z\!=\!0.50$ galaxy predicts
$(g'\!-\!i')\!\approx\!2.8$. 
This (solid line and large dot at $i'\!=\!20$ in the upper panel of 
Fig.~\ref{fig:cmdccd}) is in good agreement
with the bright end of our observed tentative cluster red sequence, indicating
we indeed detect the red sequence of \cl00 .\footnote{
For this argument, we can neglect the known slope of the red 
sequence due to the lower metallicity of the many dwarf galaxies among the
fainter cluster members \citep{1998ApJ...501..571G}.}
At $z\!=\!0.25$, the tentative redshift of the foreground structure, the same
template yields $(g'\!-\!i')\!\approx\!2.1$ 
(dashed line and large dot at $i'\!=\!20$ in the upper panel of Fig.~\ref{fig:cmdccd}).
The broad distributions in $g'\!-\!i'$ colours and the not very distinctive red
sequence of \cl00 ~are in agreement with the presence of a foreground group.

In the lower panel of Fig.~\ref{fig:cmdccd},  we show the $g'\!-\!r'$ colours
of the same central region galaxies as a function of their $r'\!-\!i'$ colours.
(Compare also Fig.~\ref{sec:colcolsel}.) In addition to the
main clump, there is a population of galaxies with both red $g'\!-\!r'$ and
$r'\!-\!i'$ colours following the locus of the bright galaxies in 
Fig.~\ref{fig:ccselec}. As expected, the CWW80 templates for 
$z\!\approx\!0.25$ and $z\!\approx\!0.50$ belong to the redder population, and
would for $m<m_{\mathrm{faint}}$ be excluded from the lensing catalogue by the
background selection (Sect.~\ref{sec:colcolsel}). 

\subsection{Background Selection by Galaxy Colours} \label{sec:colcolsel}
The selection of background galaxies is based on $r'\!-\!i'$ vs.\ $g'\!-\!r'$ 
colour-colour-diagrammes for galaxies of intermediate
magnitude works as follows: We identify the region in the 
colour-colour-diagramme populated by the brightest galaxies, a sample we
assume to be dominated by cluster ellipticals. 
As the cluster red sequence shows,
the colour of early-type systems in a cluster of galaxies varies only little
with magnitude. 
Indeed, as can be seen from Fig.~\ref{fig:ccselec}, the bright galaxies
observed in the \cl00 ~field show a well-defined relation between their
$r'\!-\!i'$ and $g'\!-\!r'$ colours.
Inferring that the fainter cluster members which
fall into the $m_{\mathrm{bright}}\!<\!r'\!<\!m_{\mathrm{faint}}$ interval on
average show the same colours as their brighter companions, we exclude
those intermediately bright galaxies \emph{with colours similar to
those of the brighter objects} while keeping those inconsistent with the
colours of the bright sample. 

Following a method introduced by \citet{2005A&A...437...49B} and 
\citet{2007A&A...471...31K}, we empirically define
two polygonal regions in the colour-colour diagramme, 
a ``small'', rather inclusive polygon and a ``large'' polygon for  
a more conservative selection
(thick and thin lines in Fig.~\ref{fig:ccselec}, respectively).
We test the influence of the colour-colour selection on the lensing signal
for those two cases. 
Table~\ref{tab:poly} gives the respective limits in $g'\!-\!r'$, $r'\!-\!i'$,
and the second-order colour index $s:=\frac{8}{3}r'\!-\!\frac{5}{3}i'\!-\!g'$
chosen to be parallel to the locus of the bright galaxies in 
Fig.~\ref{fig:ccselec}.

Figure~\ref{fig:ccss1} (upper panel)
shows the effect of the background galaxy selection 
on the $S$-statistics if the ``small'' polygon defined in Table~\ref{tab:poly}
is used for the intermediary bright galaxies. 
Here, the solid line denotes a pure magnitude cut at 
$m_{\mathrm{bright}}\!=\!m_{\mathrm{faint}}$ while the different line styles show
cases in which the colour-colour criterion acts in different intervals of
$m_{\mathrm{faint}}\!-\!m_{\mathrm{bright}}$.
First thing to note is that the $S$-statistics depends more sensitively on 
$m_{\mathrm{faint}}$ than on $m_{\mathrm{bright}}$, with its maximum in the range
$22.0\!<\!m_{\mathrm{faint}}\!<\!22.5$, irrespective of $m_{\mathrm{bright}}$.
The greater relative importance of $m_{\mathrm{faint}}$ does not come as a
surprise as, in the $r'\!<\!25$ magnitude range we study here, source counts
are rising steeply towards fainter magnitudes (Fig.~\ref{fig:counts}).

Secondly, we notice that the improvement in the $S$-statistics upon using
the best value of $m_{\mathrm{faint}}\!=\!22.5$, which we now adopt, over the
case of not applying photometric criteria to our catalogue
(corresponding to $m_{\mathrm{faint}}\!=\!17.6$) is small: 
$S\!=\!5.73$ for $m_{\mathrm{bright}}\!=\!m_{\mathrm{faint}}$ as compared to
$S\!=\!5.46$.
This may partly be explained by the small number of catalogue objects affected
by background selection. As can be seen by comparing the number of objects in
the lensing catalogue as a function of $m_{\mathrm{faint}}$ and 
$m_{\mathrm{bright}}$ in the lower panel of Fig.~\ref{fig:ccss1} with 
the $S$-statistics, as selection
starts removing (signal-diluting foreground) galaxies from the catalogue at
$m_{\mathrm{faint}}\!\gtrsim\!21.5$, the $S$-statistics begins to increase around
the same point. 
For instance, with a magnitude cut at $m_{\mathrm{faint}}\!=\!22.5$, the
remaining $92.5$~\% of the sources yield a $S\!=\!5.73$, while for a 
 $m_{\mathrm{faint}}\!=\!21.5$ magnitude cut, the remaining $97.3$~\% of the  
catalogue give $S\!=\!5.53$.

Similarly, the strong decrease in detection significance for 
$m_{\mathrm{faint}}\!\gtrsim\!22.7$ -- most pronounced for the 
$m_{\mathrm{bright}}\!=\!m_{\mathrm{faint}}$ case -- can be explained by a cut
at faint magnitudes rejecting an increasingly large number of signal-carrying
background galaxies. For the various $m_{\mathrm{bright}}\!<\!m_{\mathrm{faint}}$ 
cases, the higher signals for a given $m_{\mathrm{faint}}$ demonstrate that 
galaxies of intermediary magnitude 
with colours inconsistent with cluster ellipticals
are kept in the catalogue and contribute to the signal.

Repeating this analysis with the ``large'' polygon defined in 
Table~\ref{tab:poly}, we find the dependence of $S$ on $m_{\mathrm{bright}}$
for a given $m_{\mathrm{faint}}$ to be largely reduced. This can be explained 
by the restrictive choice of the ``large'' compared to the ``small'' polygon,
leaving only few galaxies of intermediary magnitude in the catalogue.

For the following analyses, we choose the ``small'' polygon and the 
parameter combination $m_{\mathrm{faint}}\!=\!22.5$, 
$m_{\mathrm{bright}}\!=\!20.0$, yielding the near-optimal\footnote{We prefer 
$m_{\mathrm{faint}}\!=\!22.5$ over the slightly better
$m_{\mathrm{faint}}\!=\!22.4$ because of the greater robustness of the 
$m_{\mathrm{faint}}\!=\!22.5$ cases with respect to changes in 
$m_{\mathrm{bright}}$.}
overall detection of the cluster: $S\!=\!5.84$. We also tested
catalogues with $m_{\mathrm{faint}}\!-\!m_{\mathrm{bright}}\!>\!2.5$, but did not
find any further increase in the $S$-statistics.

\subsection{Comparison to Photometric Redshift Surveys}

\begin{figure}
\resizebox{\hsize}{!}{\includegraphics[angle=90,width=0.9\textwidth]{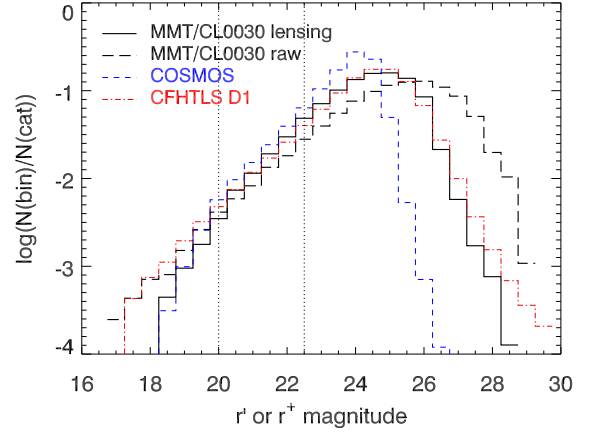}}
\caption{Source number counts in the \cl00 ~and exemplary photometric redshift
fields. Given are numbers of sources as fractions of the total number of
objects in the $r'$-band catalogue for the MMT \megacam 
~\cl00 ~raw (long-dashed curve) and lensing (before background selection; 
solid curve) catalogues as well as for the CFHTLS D1 field (dash-dotted 
curve). The dashed curve denotes the COSMOS $r^{+}$-band number counts. 
Vertical dotted lines indicate $m_{\mathrm{bright}}$ and $m_{\mathrm{faint}}$.}
\label{fig:counts}
\end{figure}
\begin{figure*}
 \sidecaption
 \includegraphics[width=0.95\textwidth]{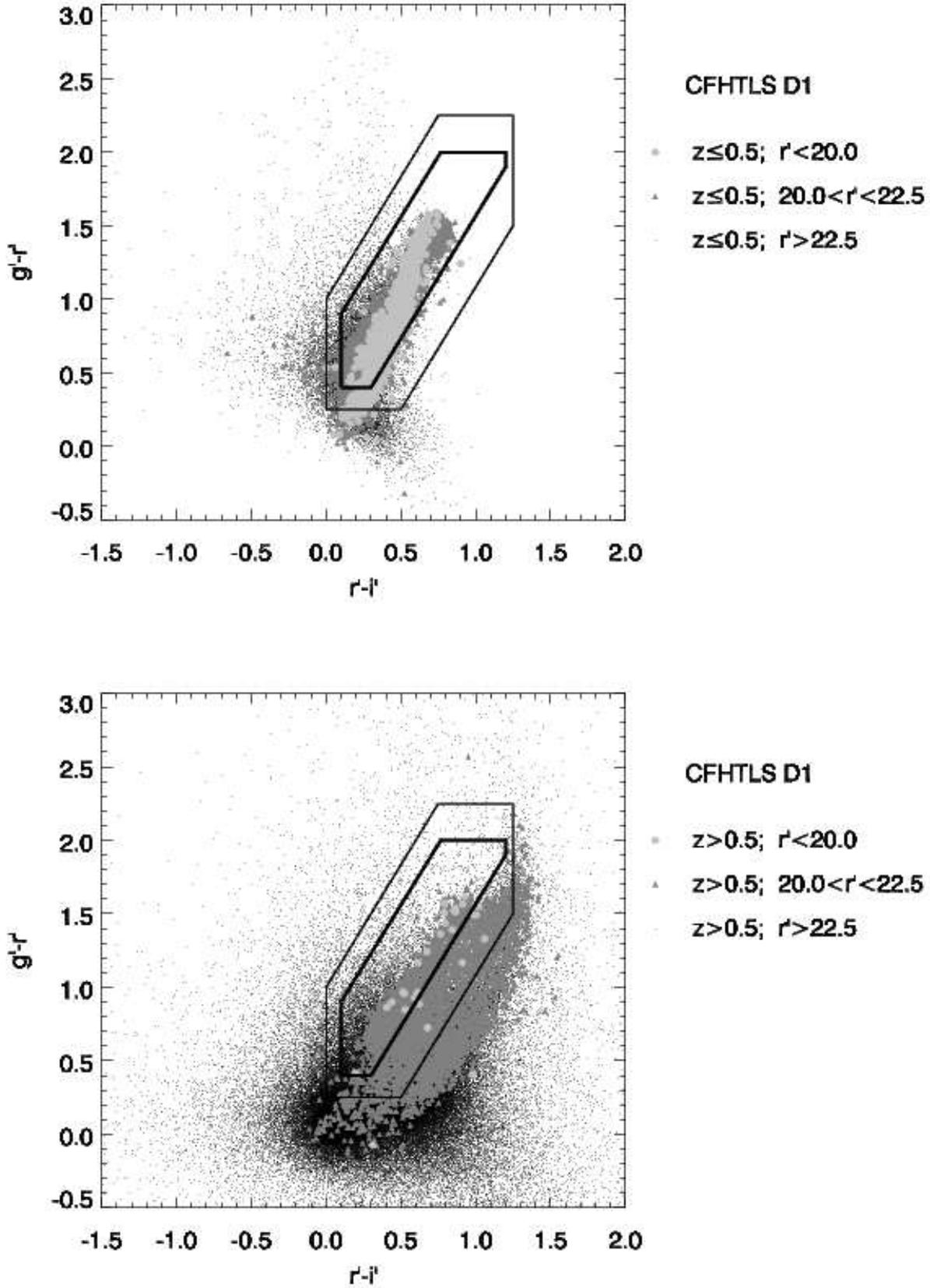}
 \caption{Colour-colour diagrammes of photo-$z$ galaxies 
in the CFHTLS~D1 field. Shown are the $r'\!-\!i'$ against $g'\!-\!r'$ colours
for foreground ($0.01\!<\!z\!\leq\!0.5$, upper panel) and background 
($z\!>\!0.5$, lower panel) galaxies, divided into the three magnitude bins
defined in Sect.~\ref{sec:colcolsel} and Fig.~\ref{fig:ccselec}: 
$r'\!<\!m_{\mathrm{bright}}\!=\!20.0$
(large symbols), $m_{\mathrm{bright}}\!<\!r'\!<\!m_{\mathrm{faint}}$ (medium-sized
symbols) and $r'\!>\!m_{\mathrm{faint}}\!=\!22.5$ (small symbols). Also shown are
the polygonals regions giving the locus of bright and intermediate galaxies in
the \cl00 ~field.
}
\label{fig:d1col}
\end{figure*}
In order to
check the significance of the optimal values empirically found for
$m_{\mathrm{bright}}$ and $m_{\mathrm{faint}}$ 
-- i.e., do they provide an effective distinction between galaxies at
redshift $z\!\leq\!0.5$ and those at $z\!>\!0.5$? --
and to estimate the geometric factor needed to convert gravitational shear 
into a mass estimate, we compare our data 
to two catalogues with known photometric redshift distributions,
the CFHTLS Deep~1 field \citep{2006A&A...457..841I} and the COSMOS survey
\citep{2009ApJ...690.1236I}.
In Fig.~\ref{fig:counts}, we compare the source number counts as a function of
magnitude of the MMT/\megacam ~catalogue of the \cl00 ~field
(before and after selection of high-quality shape objects, i.e. the unflagged
\texttt{SExctractor} objects compared to the galaxy shape catalogue)
to the CFHTLS D1 (MegaCam at CFHT, SDSS filter system)
and COSMOS photo-z sources. For the latter, the \textsc{Subaru} 
$g^{+}r^{+}i^{+}$ magnitudes similar to the SDSS filters have been used.
From the CFHTLS we use all unflagged sources classified as galaxies, detected
in all five bands ($u^{*}g'r'i'z'$) and with a photo-$z$ derived from at least
three bands whose $1\sigma$ error margin $\Delta z_{\mathrm{ph}}$ satisfies
$\Delta z_{\mathrm{ph}}/(1+z_{\mathrm{ph}})\!<\!0.25$.
Likewise, we use all unflagged sources classified as galaxies having an
unflagged photo-$z$ estimate in the COSMOS catalogue that are detected in the
\textsc{Subaru} $g^{+}r^{+}i^{+}$ and CFHT $i'$ passbands.

Fig.~\ref{fig:counts} illustrates how the various cuts in the KSB pipeline
remove faint galaxies from the catalogue, shifting the maximum 
$r'_{\mathrm{mh}}$ of the histogramme
from $r'_{\mathrm{mh}}\!=\!26.0\pm0.5$
to $r'_{\mathrm{mh}}\!=\!25.0\pm0.5$. We note that the CFHTLS~D1 shows a very 
similar histogramme over most of the relevant magnitude range 
$20.5\!<\!r'\!<\!27.0$, also peaking at $r'_{\mathrm{mh}}\!=\!25.0\pm0.5$. 
The other fields of the CFHTLS Deep Survey, D2 to D4, show a behaviour similar
to D1 and are omitted from Fig.~\ref{fig:counts} for the sake of clarity. 
The COSMOS photo-z catalogue, 
on the other hand, is shallower, with $r^{+}_{\mathrm{mh}}\!=\!24.0\pm0.5$, but
showing a number count function similar to the one in the \cl00 ~data at the
bright end.
Therefore, we use CFHTLS as a reference survey, estimating the
relations between galaxy colours and photometric redshift in the \cl00 ~data
from the D1 field and using all fields to derive the redshift distribution.  

\subsubsection{Photometric Properties} \label{sec:cfht}

First, we want to investigate the effect of the photometric cuts
optimising the aperture mass detection on the redshift distribution
of the CFHTLS~D1 catalogue.

In Fig.~\ref{fig:d1col}, we compare the $r'\!-\!i'$ and $g'\!-\!r'$ colours
of CFHTLS~D1 galaxies with photometric redshift 
$0.01\!<\!z_{\mathrm{ph}}\!\leq\!0.5$ (upper panel) and $z_{\mathrm{ph}}\!>\!0.5$
(lower panel) to the polygonal regions found from Fig.~\ref{fig:ccselec} 
containing all bright ($r'\!<\!20.0$) and most of  the intermediate 
($20.0\!<\!r'\!<\!22.5$) galaxies in the \cl00 ~field.
The bright and intermediate nearby ($0.01\!<\!z_{\mathrm{ph}}\!\leq\!0.5$)
galaxies indeed populate a similar region in the colour-colour diagramme like
their \megacam ~counterparts, albeit being slightly shifted towards bluer
$g'\!-\!r'$ colours. Thus, given its simplicity,
our background selection is quite efficient for the $r'\!<\!22.5$ foreground 
galaxies, removing $85$~\% of them from the CFHTLS~D1 catalogue. On the other 
hand, the number of bright ($r'\!<\!20.0$) background 
($z_{\mathrm{ph}}\!>\!0.5$) galaxies is negligible. Only $28$~\%
of the intermediary CFHTLS~D1 \emph{background} galaxies, redder in 
$r'\!-\!i'$ than the foreground sources but not in $g'\!-\!r'$, 
are removed by the selection criteria.

Concerning the faint ($r'\!>\!22.5$) galaxy population, 
the first observe that, despite the similar source counts 
(Fig.~\ref{fig:counts}), the colour distributions of faint sources in the 
CFHTLS~D1 and \cl00 ~fields differ qualitatively. 
Further investigations will be needed to relate this observation to a possible
cause in the data reduction pipeline.
This difference in the colour distributions affects the impact of the
background selection: In contrast to the $6.0$~\%
sources removed as foregrounds from the \cl00 ~catalogue, the size of the
CFHTLS D1 catalogue is reduced by only $0.8$~\%. 
(The rates differ little for the D2 to D4 fields.)

Second, we note the existence
of a significant fraction of $z_{\mathrm{ph}}\!\leq\!0.5$ galaxies
\emph{even to very faint magnitudes}: we find $15$~\% of the $r'\!>\!22.5$
sources and $8$~\% of the $r'\!>\!25.0$ sources to
be foregrounds to \cl00 , judging from their photo-$z$s. Consequently, our
background selection cannot identify these sources, leading to a contamination
of the lensing catalogue and a dilution of the lensing signal. 
\citet[their Fig.~16]{2006A&A...457..841I} and 
\citet[their Fig.~14]{2009ApJ...690.1236I} confirm the existence of this
population of faint galaxies at low $z_{\mathrm{ph}}$.
Although there certainly is a contribution by \emph{catastrophic outliers} 
to which a $z_{\mathrm{ph}}\!\leq\!0.5$ has been assigned erroneously, 
the comparison with spectroscopic redshifts 
\citep[their Fig.~12]{2006A&A...457..841I} indicates that most are indeed
faint nearby galaxies.

Hence, applying the background selection to the whole catalogue,
the rate of $z_{\mathrm{ph}}\!\leq\!0.5$ galaxies only drops from $18.2$~\%
to $17.6$~\%. This indicates a similar level of residual contamination in the 
\cl00 ~background catalogue (Sect.~\ref{sec:fcg}), given its redshift 
distribution follows the one in CFHTLS~D1. 
We account for the shear dilution caused by foreground galaxies as a source of
systematic uncertainty. To this end, we measure $18.0$~\%
galaxies at $z_{\mathrm{ph}}\!\leq\!0.5$ in the background-selected CFHTLS~D1
catalogue, once the $2.2\!<\!g'\!-\!i'\!<\!3.0$ sources, already covered in the
correction factor for \emph{cluster} galaxies (Sect.~\ref{sec:fcg}) are 
excised. We consider this $18.0$~\% uncertainty in the systematic error
derived from shear calibration effects (Sect.~\ref{sec:f0}).
%

\subsubsection{Redshift Distribution} \label{sec:ddsds}

\begin{figure}
 \resizebox{\hsize}{!}{\includegraphics[angle=90,width=0.9\textwidth]{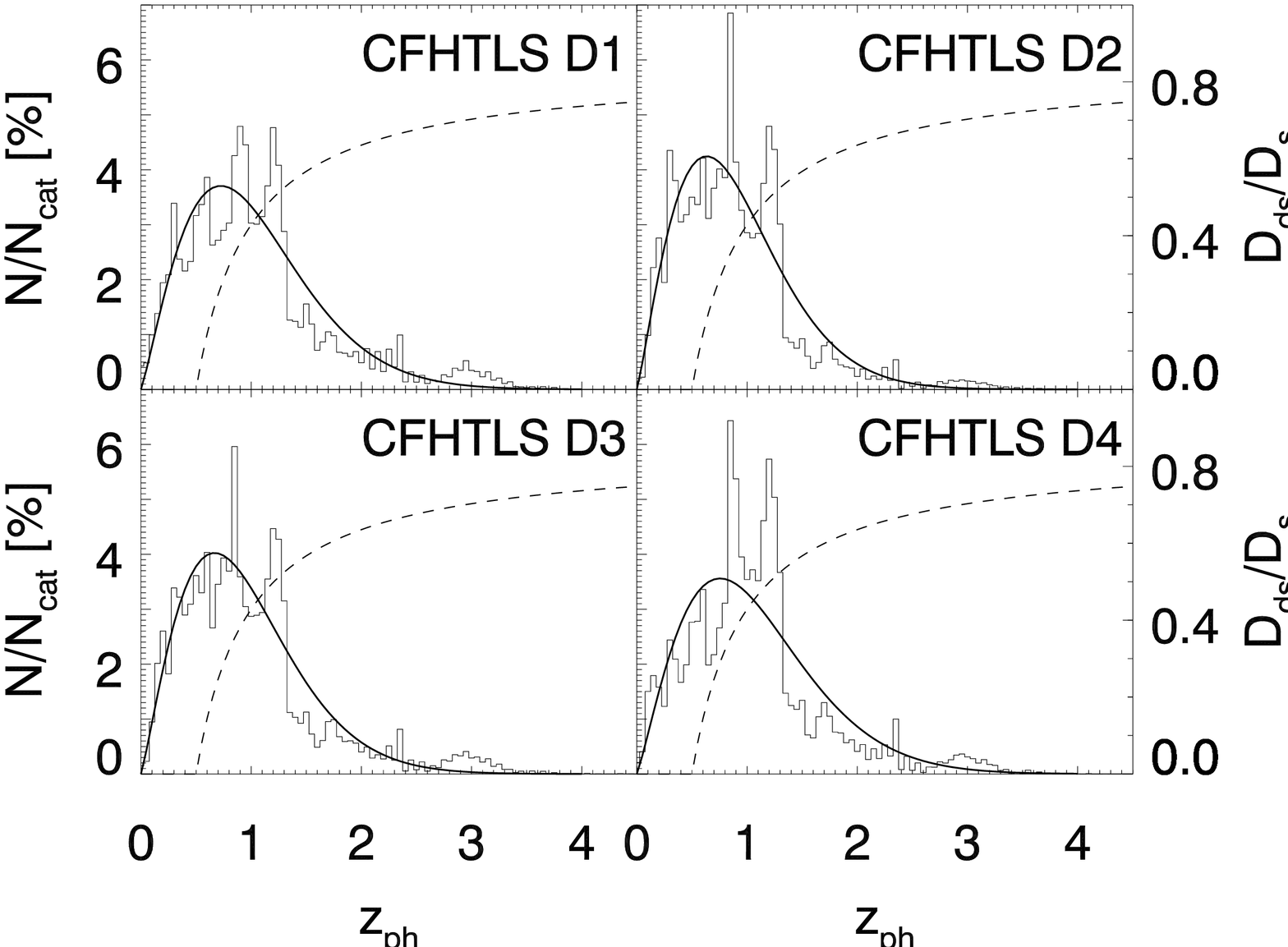}}
 \caption{Photometric redshift distributions of the CFHTLS~D1 to D4 fields
  after application of the photometric cuts defined in Sect.~\ref{sec:colcolsel}
  (histogrammes) and \citet{2001A&A...374..757V} best fits to these (solid 
  lines). The function 
  $D_{\mathrm{ds}}(z_{\mathrm{s}};z_{\mathrm{d}}\!=\!0.5)/D_{\mathrm{s}}(z_{\mathrm{s}})$ is denoted by dashed lines.}
 \label{fig:zdist}
\end{figure}
\begin{table}
 \caption{Best fit parameters $z_{0}$, $A$ (fixed), and $B$ (fixed) 
 of Eq.~(\ref{eq:vw}) to the CFHTLS D1 to D4 redshift distributions.}
 \begin{center}
 \begin{tabular}{cccccc} \hline\hline
  Field & $z_{0}$ & $A$ & $B$ & $\median(z_{\mathrm{ph}})$ & $\langle D_{\mathrm{ds}}/D_{\mathrm{s}}\rangle$ \\ \hline
  D1 & $0.87$ & $(1.15)$ & $(1.5)$ & $0.91$ & $0.345$ \\
  D2 & $0.76$ & $(1.15)$ & $(1.5)$ & $0.79$ & $0.297$ \\
  D3 & $0.80$ & $(1.15)$ & $(1.5)$ & $0.83$ & $0.316$ \\
  D4 & $0.90$ & $(1.15)$ & $(1.5)$ & $0.95$ & $0.358$ \\ \hline \hline
 \label{tab:zdist}
 \end{tabular}
 \end{center}
\end{table}
We use the redshift distribution in the CFHTLS Deep Fields
to estimate $\langle D_{\mathrm{ds}}/D_{\mathrm{s}}\rangle$, 
the catalogue average of the ratio of angular diameter distances between 
deflector and source and source and observer. In the absence of 
(spectroscopic or photometric) redshifts of the individual galaxies,
this essential quantity has to be determined from fields with a known redshift
distribution.

In Fig.~\ref{fig:zdist}, we show the binned photometric redshift distributions
we find for the CFHTLS~D1 to D4 fields after having applied the same photometric
cuts as to the \cl00 ~data. The apparent spikes seen at certain redshifts in
all the four fields are artifacts caused by the photo-$z$
determination. Because of those, we prefer calculating 
$\langle D_{\mathrm{ds}}/D_{\mathrm{s}}\rangle$ using a fit to the 
$z_{\mathrm{ph}}$-distribution.
We choose a functional form introduced by \citet{2001A&A...374..757V}:
\begin{equation}
p_{\mathrm{z}}(z_{\mathrm{ph}})\!=\!\frac{B}
{z_{0}\Gamma\left(\frac{1+A}{B}\right)} 
\left(\frac{z_{\mathrm{ph}}}{z_{0}}\right)^{A}
\exp\left(-\left(\frac{z_{\mathrm{ph}}}{z_{0}}\right)^{B}\right)
\label{eq:vw}
\end{equation}
where $z_{0}$ is a typical redshift of the sources, and $A$ and $B$ shape
parameters governing the low-redshift regime and the exponential drop-off
at high redshifts. The prefactor including the Gamma function renders
$p_{\mathrm{z}}(z_{\mathrm{ph}})$ a normalised probability distribution.
We fit the  binned redshift distributions in the range
$0\!\leq\!z_{\mathrm{ph}}\!\leq\!4$, fixing $B\!=\!1.5$ for reasons of
robustness to the default value suggested by \citet{2001A&A...374..757V}.
Next, $A\!=\!1.15$ is fixed too, to the value preferred for three of the four 
fields. The final results are summarised in Table~\ref{tab:zdist}.
Note that $D_{\mathrm{ds}}(z_{\mathrm{s}};z_{\mathrm{d}}\!=\!0.5)/D_{\mathrm{s}}(z_{\mathrm{s}})$
varies substantially over the range 
$0.8\!\lesssim\!\median(z_{\mathrm{ph}})\!\lesssim\!1.0$ spanned by the median 
redshifts of the fits to D1 to D4 (see Fig.~\ref{fig:zdist}). 
We now calculate the
average distance ratio for each field by integrating this function with the
redshift distribution over all redshifts larger than $z_{\mathrm{d}}\!=\!0.5$:
\begin{equation}
\Bigg\langle\frac{D_{\mathrm{ds}}}{D_{\mathrm{s}}}\Bigg\rangle\!=\!
\int_{z_{\mathrm{d}}}^{\infty}{\!\!p_{\mathrm{z}}(z)
\frac{D_{\mathrm{ds}}(z;z_{\mathrm{d}})}{D_{\mathrm{s}}(z)} \mathrm{d}z}
\end{equation}
For the mass estimation of \cl00 , we use the average and standard deviation
$\langle D_{\mathrm{ds}}/D_{\mathrm{s}}\rangle\!=\!0.33\pm0.03$ of the distance
ratios obtained for the four CFHTLS fields (see Table~\ref{tab:zdist}) as
fiducial value and uncertainty margin for the distance factor of our \megacam
~background sample. These values are consistent with the results for
$\langle D_{\mathrm{ds}}/D_{\mathrm{s}}\rangle$ computed directly from 
the histogrammes in Fig.~\ref{fig:zdist}.

\end{document}